\documentclass{aa}
\bibpunct{(}{)}{;}{a}{}{,} % to follow the A&A style
\usepackage[english]{babel}
\usepackage{multirow}
\usepackage{graphicx}
\usepackage{subcaption}
\usepackage{xspace}
\usepackage{xcolor}
\usepackage{booktabs}
\usepackage{txfonts}

\setcitestyle{notesep={ }}

\usepackage{hyperref}

\newcommand{\ms}{\ensuremath{\mathrm{m\,s^{-1}}}\xspace}
\newcommand{\kms}{\ensuremath{\mathrm{km\,s^{-1}}}\xspace}

\newcommand{\vsini}{\ensuremath{v \sin{i}}\xspace}

\begin{document}

\title{The CARMENES search for exoplanets around M dwarfs}

\subtitle{Radial velocities and activity indicators from cross-correlation functions with weighted binary masks\thanks{Table \ref{tab:rvabs} is available in electronic form at the CDS via anonymous ftp to cdsarc.u-strasbg.fr (130.79.128.5) or via \url{http://cdsweb.u-strasbg.fr/cgi-bin/qcat?J/A+A/}}}

\author{M.~Lafarga \inst{1,2},
        I.~Ribas \inst{1,2},
        C.~Lovis \inst{3},
        M.~Perger \inst{1,2},
        M.~Zechmeister \inst{4},
        F.\,F.~Bauer \inst{5},
        M.~K\"urster \inst{6},
        M.~Cort\'es-Contreras \inst{7},
        J.\,C.~Morales \inst{1,2},
        E.~Herrero \inst{1,2},
        A.~Rosich \inst{1,2},
        D.~Baroch \inst{1,2},
        A.~Reiners \inst{4},
        J.\,A.~Caballero\inst{7},
        A.~Quirrenbach \inst{8},
        P.\,J.~Amado \inst{5},
        J.\,M.~Alacid \inst{7},
        V.\,J.\,S.~B\'ejar \inst{9,10},
        S.~Dreizler  \inst{4},
        A.\,P.~Hatzes \inst{11},
        T.~Henning \inst{6},
        S.\,V.~Jeffers \inst{4},
        A.~Kaminski \inst{8},
        D.~Montes \inst{12},
        S.~Pedraz \inst{13},
        C.~Rodríguez-López \inst{5},
        J.\,H.\,M.\,M.~Schmitt \inst{14}
}

\institute{
        Institut de Ci\`encies de l'Espai (ICE, CSIC), Campus UAB, C/ de Can Magrans s/n, 08193 Cerdanyola del Vall\`es, Spain\label{ice}
        \and
        Institut d'Estudis Espacials de Catalunya (IEEC), C/ Gran Capit\`a 2-4, 08034 Barcelona, Spain\label{ieec}
    \and
        Observatoire Astronomique de l'Universit\'e de Gen\`eve, 51 Ch. des Maillettes, 1290 Versoix, Switzerland
        \and
    Institut f\"ur Astrophysik, Georg-August-Universit\"at, Friedrich-Hund-Platz 1, 37077 G\"ottingen, Germany
    \and
        Instituto de Astrof\'isica de Andaluc\'ia (IAA-CSIC), Glorieta de la Astronom\'ia s/n, 18008 Granada, Spain
        \and
        Max-Planck-Institut f\"ur Astronomie, K\"onigstuhl 17, 69117 Heidelberg, Germany
        \and
        Centro de Astrobiolog\'ia (CSIC-INTA), ESAC, Camino Bajo del Castillo s/n, 28692 Villanueva de la Ca\~nada, Madrid, Spain
        \and
        Landessternwarte, Zentrum f\"ur Astronomie der Universt\"at Heidelberg, K\"onigstuhl 12, 69117 Heidelberg, Germany
        \and
        Instituto de Astrof\'isica de Canarias, V\'ia L\'actea s/n, 38205 La Laguna, Tenerife, Spain
        \and
        Departamento de Astrof\'isica, Universidad de La Laguna, 38026 La Laguna, Tenerife, Spain
        \and
        Th\"uringer Landesstenwarte Tautenburg, Sternwarte 5, 07778 Tautenburg, Germany
        \and
        Departamento de Astrof\'isica y Ciencias de la Atm\'osfera, Facultad de Ciencias F\'isicas, Universidad Complutense de Madrid, 28040 Madrid, Spain
        \and
        Centro Astron\'onomico Hispano Alem\'an, Observatorio de Calar Alto, Sierra de los Filabres, E-04550 G\'ergal, Spain
        \and
        Hamburger Sternwarte, Gojenbergsweg 112, 21029 Hamburg, Germany
}

\titlerunning{RVs and activity indicators from CCFs with weighted binary masks}

\authorrunning{Lafarga et al.}

\date{Received, 29 November 2019; Accepted, 04 March 2020}

\abstract
{For years, the standard procedure to measure radial velocities (RVs) of spectral observations consisted in cross-correlating the spectra with a binary mask, that is, a simple stellar template that contains information on the position and strength of stellar absorption lines. The cross-correlation function (CCF) profiles also provide several indicators of stellar activity.}
{We present a methodology to first build weighted binary masks and, second, to compute the CCF of spectral observations with these masks from which we derive radial velocities and activity indicators. These methods are implemented in a python code that is publicly available.}
{To build the masks, we selected a large number of sharp absorption lines based on the profile of the minima present in high signal-to-noise ratio (S/N) spectrum templates built from observations of reference stars. We computed the CCFs of observed spectra and derived RVs and the following three standard activity indicators: full-width-at-half-maximum as well as contrast and bisector inverse slope.
}
{We applied our methodology to CARMENES high-resolution spectra and obtain RV and activity indicator time series of more than 300 M dwarf stars observed for the main CARMENES survey.
Compared with the standard CARMENES template matching pipeline, in general we obtain more precise RVs in the cases where the template used in the standard pipeline did not have enough S/N.
We also show the behaviour of the three activity indicators for the active star YZ CMi and estimate the absolute RV of the M dwarfs analysed using the CCF RVs.
}
{}


\maketitle
%
%-------------------------------------------------------------------

\vspace*{0cm}

\section{Introduction}

The Doppler spectroscopy or radial velocity (RV) technique is one of the main methods to detect, confirm, and characterise planetary companions.
It pioneered the discovery of the first exoplanet orbiting a main sequence star \citep{mayor1995Jup51Peg}.
Since then, Doppler spectroscopy and photometric searches for transits have become the most successful techniques to detect exoplanets, with about 900 and 3000 discoveries, respectively, of the more than 4000 exoplanets confirmed to date\footnote{\url{http://exoplanet.eu/} as of November 2019}.

Early RV searches for exoplanets mainly focused on solar-type stars and, due to their intrinsic faintness, only included a handful of M-dwarf stars. Out of approximately the first hundred exoplanet detections, only a few were found orbiting M dwarfs \citep{marcy1998gj876, delfosse1998gj876, marcy2001gj876, butler2004gj436}.
Nevertheless, in the last two decades, interest has rapidly grown regarding M dwarfs as exoplanets hosts, which now account for over 200 exoplanets orbiting about 100 hosts \citep{martinez-rodriguez2019exomoons}.
Compared to more massive and hotter FGK-type stars, the star-to-planet mass ratio is lower for M dwarfs and their habitable zones (HZs) are located closer in. Therefore, the gravitational pull of an HZ-orbiting planet is stronger for M dwarfs, implying RV signals with larger amplitudes. Also, orbital periods are shorter, thus reducing the monitoring timescales. These circumstances make the detection of small, rocky exoplanets feasible with the current instrumentation.
Moreover, M dwarfs are the most abundant stars in the solar neighbourhood, which makes them the most common potential exoplanet hosts.
The main issue with M dwarfs is that they can exhibit high levels of stellar activity.
This activity manifests itself in the form of features in the stellar surface that create distortions on the spectral line profiles used to measure RVs, which complicates the detection of exoplanets.
They are also relatively fainter than more massive stars and emit most of their bulk energy in the near-infrared.

The traditional method of measuring Doppler shifts in spectra is to directly cross-correlate them with a template.
This is known as the cross-correlation function (CCF) method \citep{queloz1995spectroscopylowSNR, baranne1996elodie}.
Binary masks are commonly used templates and they represent a very simple model of the stellar spectrum. They consist of a set of boxcar functions (a rectangle-shaped function that has a constant value over a specific range and is zero elsewhere) centred at the minimum wavelength of a large number of absorption lines that are present in the stellar spectrum.
In fact, early echelle spectrographs used physical masks, that is, spatial filters with real holes at the positions of absorption lines \citep{baranne1979Coravel}, thus explaining the name.
Depending on how much RV information \citep{bouchy2001photonnoise} each line contains, the boxcar functions can have different weights, which improve the RV measurements \citep{pepe2002coralieSaturns}. Deep and narrow lines, that is, lines with the steepest profiles, have more information than shallow and broad lines, hence their weight in the mask is larger.

The CCF method has been the standard approach to obtain precise RV measurements of FGK-type stars.
Their spectra display a number of resolved and unblended absorption lines across the visible wavelength range, which are ideal features to be represented by a weighted binary mask.
M dwarfs also show a large number of absorption lines. However, due to their lower temperatures, the number of lines is so high that most of the features are blended, making their identification and, hence, the construction of a mask, more challenging.
Different methods consisting in a least-squares matching of a full spectrum template to the observations \citep{anglada-escude2012harpsTerra, astudillo-defru2015harps3Mdwarfs, zechmeister2018serval}, rather than only using a selected set of absorption lines, have proven to work better for M-dwarf stars 
since they use more Doppler information contained in the spectrum.
Recently, different methods have been proposed with the goal of improving RV extraction, such as using individual absorption lines \citep{dumusque2018indivline} or approaches using Gaussian processes \citep{rajpaul2020rvmethod}.

The CCF continues to be useful for studying the stellar activity of stars, including M dwarfs.
The CCF profile gathers information from each individual spectral line selected in the mask, which acts as a kernel, and converts this into an average high signal-to-noise ratio (S/N) line profile.
Therefore, average changes in the individual line profiles due to stellar activity are reflected in changes in the CCF shape. Several parametrisations of the CCF profile describing its width, depth, or asymmetry are commonly used proxies of stellar activity.
Another benefit is that one can estimate the absolute RV value of the observed stars using RV measurements obtained from the CCF.
CCF masks are usually calibrated to an absolute RV scale, while template matching techniques provide RVs that are relative to the template used, which is not necessarily on an absolute scale. Of course, one could calibrate the template against an absolute reference, but the most straightforward way to do that would be by cross-correlation.

In this article, we study the CCF and its activity indicators in M-dwarf stars in the visible and near-infrared spectral range.
We use spectral data obtained with the CARMENES instrument \citep[Calar Alto high-Resolution search for M dwarfs with Exo-earths with Near-infrared and optical Echelle Spectrographs,][]{quirrenbach2016CARMENES, quirrenbach2018CARMENES}, 
located at the 3.5 m telescope at Calar Alto Observatory in Almería, Spain.
It consists of a pair of cross-dispersed, fibre-fed echelle spectrographs with different wavelength coverage: The visual (VIS) channel covers the spectral range $\lambda$ = 520--960\,nm at a resolution of $R=94\,600$, and the near-infrared (NIR) channel covers the range $\lambda$ = 960--1710\,nm at a resolution of $R=80\,400$.
The CARMENES survey (also guaranteed-time observations -- GTO program) has been ongoing since 2016. It monitors over 300 M dwarfs across all spectral subtypes with the main goal of detecting orbiting exoplanets \citep{reiners2018carmenes324}.

In Sect. \ref{sec:ccftheory} we present an overview of the CCF method and its parameters.
Then we explain the methodology we follow to build binary masks in Sect. \ref{sec:maskmethod} and to compute CCFs and their main parameters in detail in Sect. \ref{sec:ccfmethod}.
Next, in Sect. \ref{sec:carmenesgtomaskccf} we apply these methods to the M dwarf observations of the CARMENES GTO sample. There, we also comment on the masks and CCFs obtained for different targets, compare the RVs obtained with the default CARMENES template matching pipeline, show the behaviour of the CCF activity indicators for an active star in the sample, \object{YZ CMi}, and estimate the absolute RV of the CARMENES sample stars using the values derived from their CCFs.
Finally in Sect. \ref{sec:conclusion} we summarise our work.

%--------------------------------------------------------------------

\section{Cross-correlation with a weighted binary mask}
\label{sec:ccftheory}

Measuring the radial velocity of an observed spectrum via cross-correlation relative to a binary mask implies Doppler-shifting the mask to cover a range of velocities when in search for the best match\footnote{For convenience and to ensure the conservation of flux in the spectra, we did not rebin the spectra, but we worked with spectral bins sizes corresponding to the detector pixels. Since the wavelength scale is non-uniform due to the variation of the dispersion along each echelle order and instrumental distortions, this means that each line in the mask had to be shifted individually to the pixel position corresponding to its wavelength. For details, see Sect. \ref{sec:ccfmethoddetails}}. At each velocity step, the spectrum is multiplied by the mask. Since the mask has a zero value everywhere except at the positions of the selected absorption lines, only the spectrum pixels that totally or partially overlap with a mask line (non-zero values) contribute to the CCF value.
For a given velocity shift $v$, the cross-correlation of a spectrum consisting of $n$ pixels with a mask with $m$ lines is computed according to the following equation
\begin{equation}
\label{eq:ccf}
\mathrm{CCF}\left(v\right) = \sum_{l=1}^{m} \sum_{x=1}^{n} w_l \, f_x \, \Delta_{x\,l}\left(v\right),
\end{equation}
where $f_x$ is the flux of the spectrum at pixel $x$, $w_l$ is the weight of the mask line $l$, and $\Delta_{x\,l}$ is the fraction of pixel $x$ covered by the mask line $l$ when the mask is shifted by $v$.
This equation corresponds to the cross-correlation function (CCF) of two discrete functions. Their product is computed pixel by pixel, and the term $\Delta_{x\,l}$ is employed to take partial overlaps between the pixels of the spectrum and the mask into account. The mask lines are defined as having a width of 1 pixel, which corresponds to the minimum interval containing relevant information. The minimum of the CCF corresponds to the Doppler shift of the stellar spectrum with respect to the mask.
An analytic function, most frequently a Gaussian, is fitted to the CCF profile for a robust estimate of the position of the minimum and to characterise its profile.

Compared to template matching algorithms, the CCF method can be considered to be a simplistic least-squares minimisation in which we minimise the product of the observed spectrum and model (i.e. the CCF mask); afterwards, we fit the CCF with a function, such as a Gaussian, to estimate the Doppler shift.
This is equivalent to computing the CCF between the spectrum and the mask previously convolved with a Gaussian \citep{baluev2015rm}.
The only parameter in the model is the RV shift applied to the mask. Other corrections, such as normalisations or slope changes in the spectrum continuum, which are usually forward-modelled in template matching procedures, are not simultaneously modelled when computing the CCF.\  But rather, they are adjusted a priori.
Despite the number of simplifications, the CCF method is capable of yielding precise results.
Another difference is that in template matching schemes that use co-added templates, the noise of each of the individual spectra used is present in the template. This creates self-biases towards zero velocity due to the noise correlating with itself, specially if the S/N of the template is low \citep{vankerkwijk1995hd77581vela}, which should not affect the CCF method.

As mentioned above, the CCF represents an average of the profiles of all the stellar absorption lines that are present in the mask used in velocity space.
The different processes in the stellar atmosphere that affect the average position and shape of the individual spectral lines are reflected in the CCF. Therefore, the shape of the CCF contains information on the line profile distortions caused, for example, by stellar activity features present in the photosphere, such as dark spots or faculae.

Parameters, such as the full-width-at-half-maximum (FWHM) and the contrast of the CCF, which are obtained from the Gaussian fit are used to measure temporal changes. Correlations of these parameters with the RV, or the presence of the same periodic signal in the RV and these parameters, may indicate the presence of activity-induced variations in the RVs. They can also trace instrumental variations, such as focus changes, or observational effects, such as barycentric broadening.
The bisector of the CCF is used to analyse its asymmetry, which is also an indicator of changes caused by stellar activity for lines of a different strength,  correlated with the atmospheric depth of their formation \cite[e.g.][]{gray2009thirdsignatureconvection}.
Different ways to quantify the bisector have been proposed.
A commonly used metric is the bisector inverse slope (BIS), which was introduced by \cite{queloz2001noplanet}. It is defined as the difference between the average velocity of the top region of the CCF (e.g. from 60 to 90\%) and the average velocity of the bottom region (e.g. from 10 to 40\%).
Different fractions of the CCF can be used to compute the BIS, such as the $\mathrm{BIS^+}$, which uses the 80--90\% and 10--20\% regions, and the $\mathrm{BIS^-}$, which uses the 60--70\% and 30--40\% regions as defined in \cite{figueira2013lineprofilevariations}. These alternative definitions have been shown to provide more significant results than the usual regions for some specific cases.

Other ways to quantify the asymmetry have also been proposed.
In \cite{boisse2011disentangling}, the asymmetry was measured as the difference between the RVs obtained by fitting a Gaussian to the upper and lower part of the CCF. This indicator, known as velocity span, $V_{\mathrm{span}}$, seems to be more robust than the BIS at a low S/N.
\cite{figueira2013lineprofilevariations} also defined two different indicators.
Instead of fitting a Gaussian function to the CCF, they fitted a bi-Gaussian function (two half-Gaussians with different FWHM, each modelling one side of the CCF). This fit of a bi-Gaussian gives an estimate of the RV (the minimum of the CCF) that is different than the usual Gaussian fit, so the difference between these two RVs, $\Delta V$ \citep[first used in][]{nardetto2006lineasymmetry}, can be used as an asymmetry measure. In some cases, the amplitude of this indicator is larger than that of the BIS, allowing for the detection of correlations with the RV with a smaller amplitude.
The second indicator, $V_{\mathrm{asy}}$, which was later modified by \cite{lanza2018lineprofilediagnostics}, compares the RV information content (the total flux and slope) between each side of the CCF to measure its asymmetry, and it is also found to provide stronger correlations with the RV than other indicators.
\cite{simola2018ccfskewnormal} also fitted a non-symmetric function to the CCF, in this case a skew normal distribution (a Gaussian with a skewness parameter), to account for asymmetry in the CCF profile. The parameters derived from this fit seem to be more sensitive to stellar activity.
Even though different asymmetry parametrisations are proposed in the literature, here, we focus on the original BIS definition because it is the most commonly used asymmetry metric (e.g. it is provided by the HARPS/HARPS-N data reduction software) and we do not find significant differences with the other definitions presented. A more in-depth study of the different asymmetry parametrisations is out of the scope of this work.

%--------------------------------------------------------------------

\section{Weighted binary mask creation method}
\label{sec:maskmethod}

To compute a CCF, we need to obtain the appropriate weighted binary mask.
The main issue is to select a set of absorption lines in the spectra that are appropriate for unbiased radial velocity determination.
The lines that we find in the stellar spectrum both mainly depend on the instrument used (i.e. its wavelength coverage and resolution) and the target properties (i.e. its effective temperature and projected rotational velocity).
In the case of M dwarfs, the spectrum contains a large number of lines and is rich in molecular bands. The lines are often not clearly separated and, in many cases, they can be severely blended.

Since we want to compute CCFs of M dwarfs observed with the CARMENES instrument, we chose to directly use the large number of available observations from the GTO survey to identify and select lines \citep{reiners2018carmenes324}.
We could have also used a high S/N, high-resolution synthetic stellar templates, or information from line lists, but initial tests showed that using our observations yielded clearly better results and the procedure is more straightforward.
The number of observations available for all the CARMENES survey targets is sufficient to build high S/N templates by co-adding them.

In the sections below, we describe the procedure we followed to build a mask that is suitable to compute the CCFs.
The methods to build masks and compute CCFs with their moments described in 
the current and next sections have been implemented in the code RACCOON (Radial velocities and Activity indicators from Cross-COrrelatiON with masks). 
Currently, the code works for CARMENES and HARPS/HARPS-N \citep{mayor2003harps,cosentino2012harpsn} data, and it is publicly available\footnote{\url{https://github.com/mlafarga/raccoon}}.

\subsection{High S/N stellar spectrum template}

We employed the CARMENES survey observations as the starting point to select the lines.
To minimise the effect of the photon noise of individual spectra, we created a high S/N spectral template by co-adding all of the different observations available of a specific star.
The templates were obtained using the SERVAL software \citep{zechmeister2018serval}, which is the main pipeline used to analyse RVs in the CARMENES survey.
SERVAL creates a high S/N stellar template by co-adding observations and then it uses this template to compute a least-squares fit to each of them from which the RV time series is obtained iteratively.
Before co-adding, the observations are corrected for the corresponding barycentric motion of the Earth and any other known drift so that the stellar lines are optimally aligned.
The templates have a similar format as the observations, and echelle orders are considered individually.
The CARMENES NIR orders have a middle gap because of the NIR detector configuration (it is a mosaic of two detectors with a middle gap). Due to this, throughout all this work, we treated the NIR half-orders as single ones.

We then normalised the template orders by dividing each of them by a fit to their `continuum'.
To measure the continuum of each order, we selected the pixels with the maximum flux over windows of a few tens to a few hundreds of pixels in width to avoid overfitting as there are numerous absorption lines present in the spectrum.
When selecting these pixels, we avoided spectral regions containing very strong lines or lines that can show strong chromospheric emission (such as \ion{He}{I} D$_3$, the \ion{Na}{I} D doublet, H$\alpha$, or the Ca infrared triplet) as well as regions with strong telluric features to prevent biases in the polynomial fit.
The selected pixels were used to fit a second-order polynomial, and finally the template spectrum was divided by it.

\subsection{Excluded regions} \label{sec:maskexcludedregions}

Since we use observations obtained from the ground to create the template, some regions of the spectra are affected by telluric contamination. Absorption and emission features coming from the Earth's atmosphere on top of the stellar spectrum need to be excluded.
The position of the telluric lines on the detector (pixel frame) is very stable.
However, the position of the stellar spectral features on the detector depends on the relative velocity between the observer and the target, which is determined by the orbital and rotation motion of the Earth and by the absolute radial velocity of the target.
The relative stellar velocity changes throughout the year, so the stellar spectrum shifts across the detector depending on the observation date.
Except for targets with large radial velocity amplitudes (e.g. binaries), the main contribution to this velocity shift comes from the Earth orbital motion (barycentric Earth radial velocity, BERV), which can be up to approximately $\pm\,30\,\kms$, depending on the sky position of the target.
To be sure any contribution from the Earth's atmosphere at any moment of the year is removed, we disregarded wavelengths encompassing an enlarged region around each telluric feature and not only the telluric line itself. To define this region, we used the maximum BERV of the observations of the target used to build the template, which can be significantly smaller than $30\,\kms$, and we broadened each telluric line by subtracting and adding this maximum BERV to the blue and red line limits, respectively.
In the VIS range, we masked the regions around telluric lines that are deeper than 5\%. In the NIR range, the threshold was 4\% in the \textit{Y} and \textit{H} bands and 2\% in the \textit{J} band, where the stellar lines are less abundant and shallower \citep[see][for details]{nagel2019tdtm}.
Instead of completely masking telluric contaminated regions, we could attempt to correct shallow telluric features with techniques such as the one presented in \citet{nagel2019tdtm}, which would increase the available spectral range. However, this is beyond the scope of this work and we leave it for a future paper.
We also masked out regions close to chromospheric lines that can show a strong emission component in active stars, as was done before when normalizing the template orders.

\subsection{Mask lines}

\subsubsection{Line identification}

We used the normalised spectrum template to identify absorption features.
Firstly, we searched for all local minima present in the template (except in the excluded regions mentioned above), which indicate the centre of the absorption lines.
We considered the region around these minima (the pixels in the region going from the adjacent maxima at each side of a minimum, see the grey shaded region in Fig. \ref{fig:masklinecharacterize}) to represent the absorption lines of the stellar spectrum. Most of these spectral features do not come from a single elemental transition, but they are the result of line blending.

\subsubsection{Line characterisation}

\begin{figure}[]
\centering
\includegraphics[width=\linewidth]{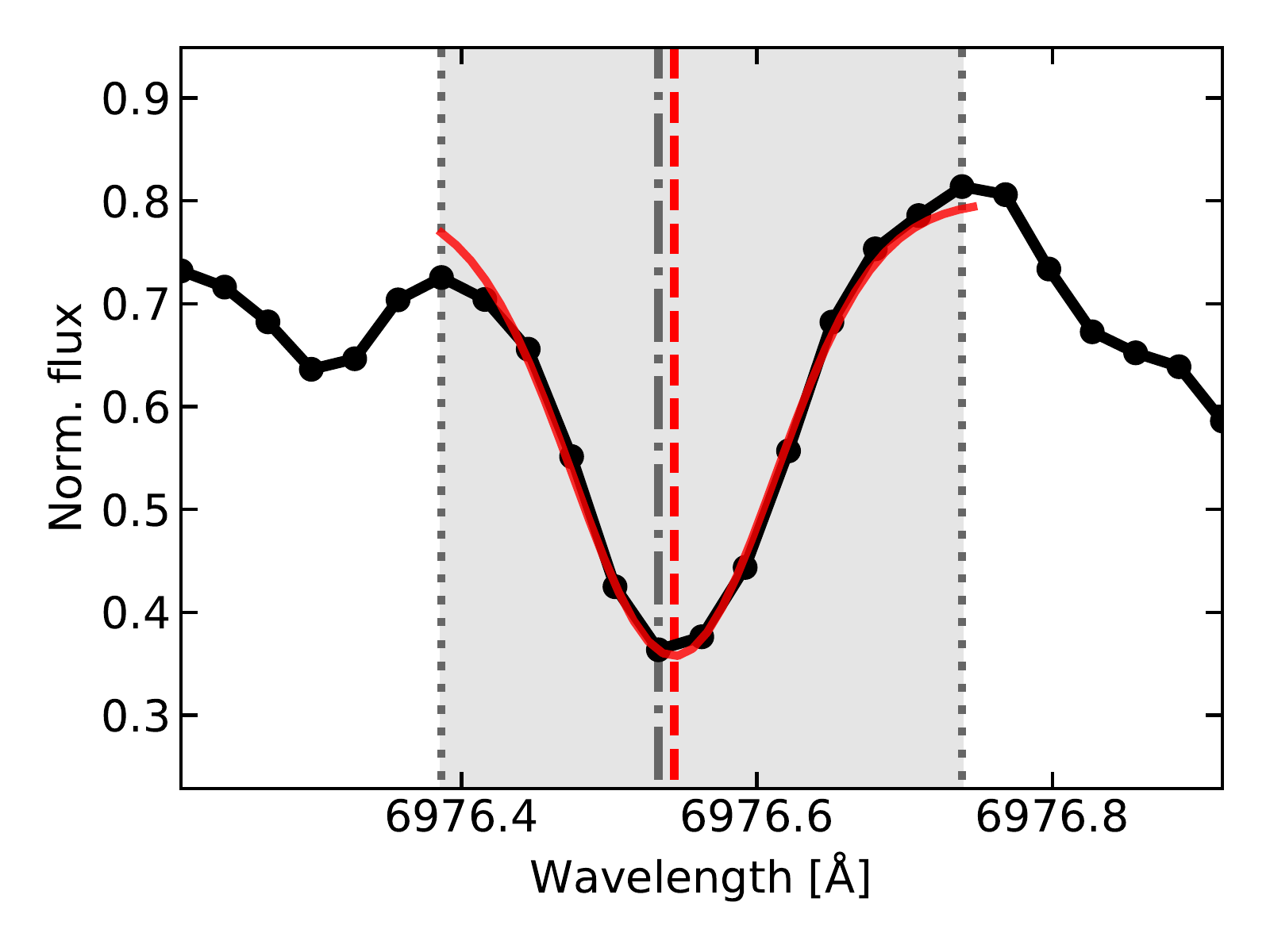}
\caption{Minima characterisation. The black dots and solid line are the template data showing an absorption feature. The shaded light grey region defines the pixels corresponding to the feature shown, which go from the local maxima (dotted grey vertical lines) at each side of the minimum (dotted dashed grey vertical line). The solid red line is the best Gaussian fit, and the dashed red vertical line indicates its minimum.}
\label{fig:masklinecharacterize}
\end{figure}

For each identified minimum or line, we measured a set of parameters in the template to quantitatively characterise its shape.
To have a more reliable estimation of these parameters, we fitted a Gaussian function to each spectral minimum
\begin{equation}
G(\lambda) = d + a \exp\left( - \frac{\left( \lambda - b \right)^2}{2c^2} \right).
\end{equation}
Figure \ref{fig:masklinecharacterize} shows an example of a line in a spectrum template with the best-fit Gaussian function. The pixels used in the fit are the ones between the two maxima at each side of the absorption feature.

To measure the central wavelength of each line, we used the position of the minimum of the Gaussian fit $b$. This gives a more robust and precise estimate than the wavelength corresponding to the local minimum of the spectrum template, which depends on the spectral sampling.
The depth of each line is given by the flux at the Gaussian minimum, $a+d$, so small values correspond to deep lines.
To measure the width of each line, we used the FWHM of the Gaussian fit, $2c\sqrt{2\ln\left(2\right)}$.
We computed the contrast or prominence of the line as the difference between the flux at the edges of the line (the adjacent local maxima) and the flux at the minimum of the line (what we defined as the line depth). This gives two values for the contrast; one corresponds to each side of the minimum, whose difference can be used to study the asymmetry of the line.

\subsubsection{Line selection}

\begin{figure*}[h]
\centering
\includegraphics[width=\linewidth]{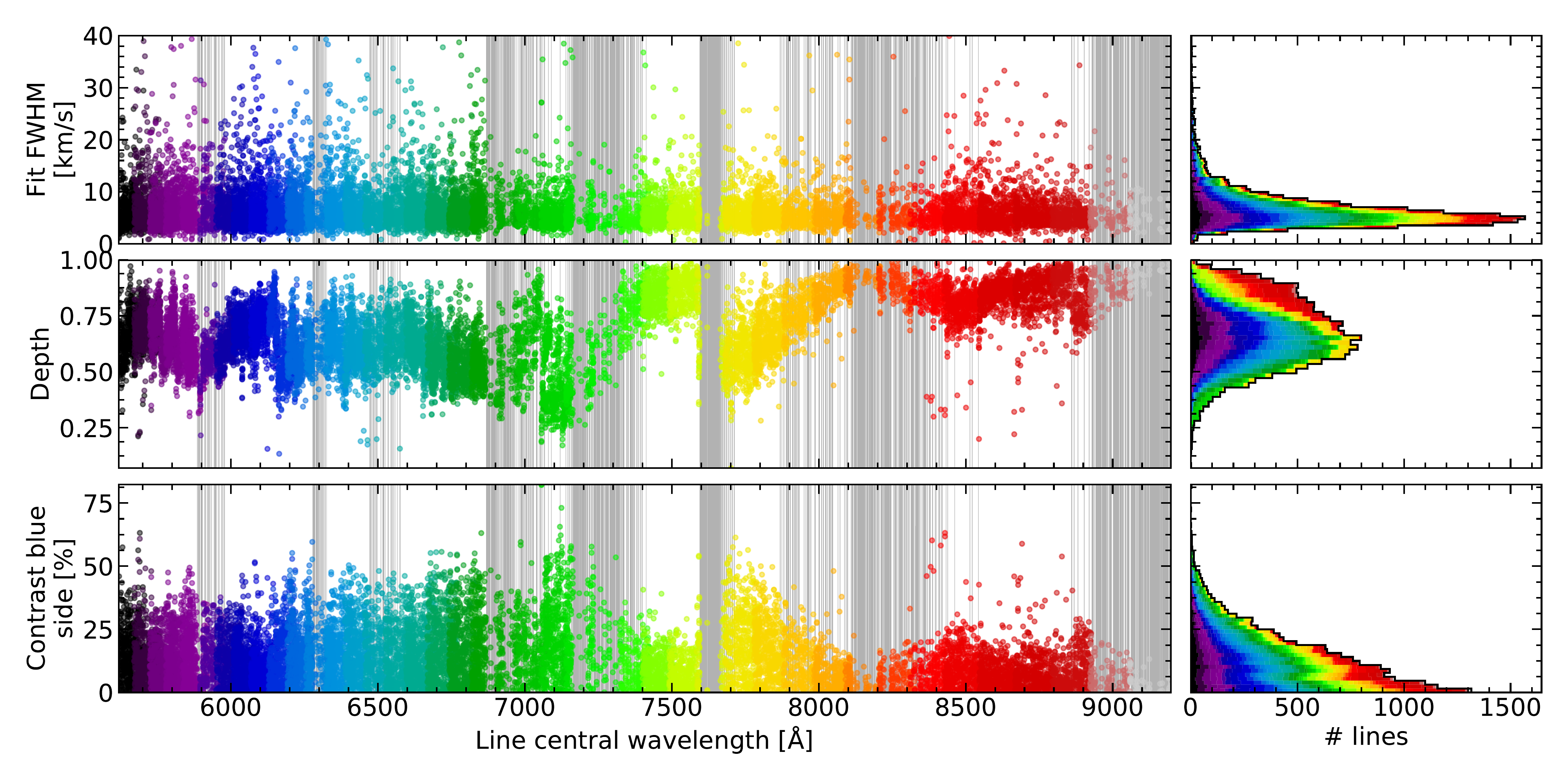}
\caption{Line parameters (\emph{top}: FWHM, \emph{middle}: depth, and \emph{bottom}: contrast) as a function of the line central wavelength ({\em left}) and the number of lines as a function of the parameters, where the values of each order have been stacked ({\em right}) for \object{Luyten's star} (Karmn J07274+035, M3.5\,V, low \vsini).
In all panels, colours correspond to different spectral orders. Lines in order-overlap regions, which are wider in the blue, are duplicated.
The shaded grey areas in the left panels correspond to regions contaminated by telluric features, broadened by $\pm30\,\kms$.
The black lines in the right panels correspond to the distribution of the lines in all orders.
}
\label{fig:masklineparamdistrib}
\end{figure*}

Figure \ref{fig:masklineparamdistrib} shows the distribution of the parameters we defined above for the lines identified in the VIS template of the slowly-rotating ($\vsini\leq2\kms$), M3.5\,V, Luyten's star (GJ~273, Karmn~J07274+052). We obtained similar values for low \vsini targets across all spectral subtypes.
As mentioned above, we excluded minima close to very strong lines and those that fall in regions with strong telluric contamination. We also excluded minima that do not contain enough pixels to be properly characterised (due to noise or too weak lines) and minima with profiles that highly deviate from a Gaussian (mainly blends for which the Gaussian fit gives unphysical values).

Based on these parameters, we defined some threshold values to select the lines that will become part of the final mask.
We tried to select as many lines as possible so that the CCF has a high S/N, allowing us to derive its parameters with the best precision. At the same time, we want to avoid weak lines that contribute more noise than signal.

We tested several masks created using different parameter combinations (depth, width, contrast, asymmetry) and different selection values for each parameter.
To analyse the performance of the masks, we used them to compute the CCFs of the observed spectra of some targets with low activity levels and low \vsini, and we checked which mask yielded the smallest RV scatter.
This analysis shows that using different combinations of parameters or slightly different selection values did not significantly change the masks or the CCF parameters obtained.
The final parameters and values we adopted are shown in Table \ref{tab:maskcarmenesparams}.

\begin{table}[]
\centering
\caption{Line parameter limits used to select the final lines of our low \vsini masks.}
\begin{tabular}{ccc}
\hline\hline
FWHM range & Minimum contrast & Depth quantile \\[0pt]
[\kms] & [\%] & [\%] \\
\hline
$2 - 30$ & 6 & 60 \\
\hline
\end{tabular}
\label{tab:maskcarmenesparams}
\end{table}

We selected lines with FWHM within a certain range. The FWHM values had to be increased for targets with larger $v\sin i$, up to few hundreds of \kms for the most extreme cases, the ones with \vsini~$\sim$~30--40\,\kms.
In doing this, we made sure that most of the lines that were included in the CCF have a similar width while avoiding the narrowest lines, which tend to correspond to weak noisy lines that were not removed in previous steps, and the broadest lines, due to strong, unsuitable lines or blends. The removed lines do not provide significant RV information and they tended to increase the RV scatter.
To avoid lines that are too asymmetric and to further remove weak lines, we found that the simplest method was to select lines for which the contrast at both sides is larger than a minimum value.

We also selected lines depending on their depth. However, for this parameter we did not use a single threshold value.
The spectra of M dwarfs do not show a clear continuum due to the large number of features present; so across the spectral range we observe large differences in the line depths, with much deeper lines in regions containing strong molecular bands.
This makes it difficult to just use a single cut value for the whole spectral range.
Since we are using a normalised high S/N template, the depth values of the lines range from 1, for the shallowest lines, to 0 for the deepest ones.
In some orders, all of the lines can be very shallow, with the deepest ones reaching values of only 0.8, for example,  while other orders contain very deep lines, reaching values of 0.2, for instance.
To minimise this bias, we rescaled the depth values so that, for each order, we considered that the depth goes from 1 to the average of the 10\% deepest lines of the order.
On this new scale, we selected lines with depths smaller than the 60\% quantile of the new depth range.

We found that values around the ones shown in Table \ref{tab:maskcarmenesparams} allow us to select enough lines over the whole spectral range to obtain a CCF with a good S/N and precise parameters. Also, masks with these parameters worked well for most of the M dwarfs of the CARMENES survey sample.
However, since the lines in the mask are weighted according to their RV information, changing these values to include more or less weak or wide lines did not change the CCF profiles significantly because these lines have lower weights (see next section).

Moreover, it is possible to fine-tune these values to a specific target and select optimum lines to study the velocity or the activity signals present in the spectrum.
It has been shown that it is possible to classify spectral lines according to their sensitivity to stellar activity and, hence, according to their contribution to the activity signal in the final RVs \citep{wise2018activelines, dumusque2018indivline, lisogoroskyi2019activityalphacenb, cretignier2019indivline}. Using masks with lines selected to be more or less affected by activity should result in CCF profile parameters that are more or less sensitive to stellar activity variations.

\subsection{Final mask}

\subsubsection{Position and weight}

The final mask was comprised of the lines we selected following the approach explained above.
Each line has a `position', given by the minimum of the Gaussian fit in wavelength space, and a `weight', which we calculated as the contrast divided by the FWHM of the line.
In this way, the lines with the steepest profiles (i.e. the ones with more RV content) are the ones that contribute to the CCF the most.

When building the mask, all features are defined to have zero width (Dirac $\delta$ functions), independently of their measured values. At the time of computing the CCF of a specific observation, we forced each of the mask features to have the width of one pixel, which depend on the instrument and the position of the mask on the spectrum (see Sect. \ref{sec:ccfmethod} for details).

\begin{figure*}[]
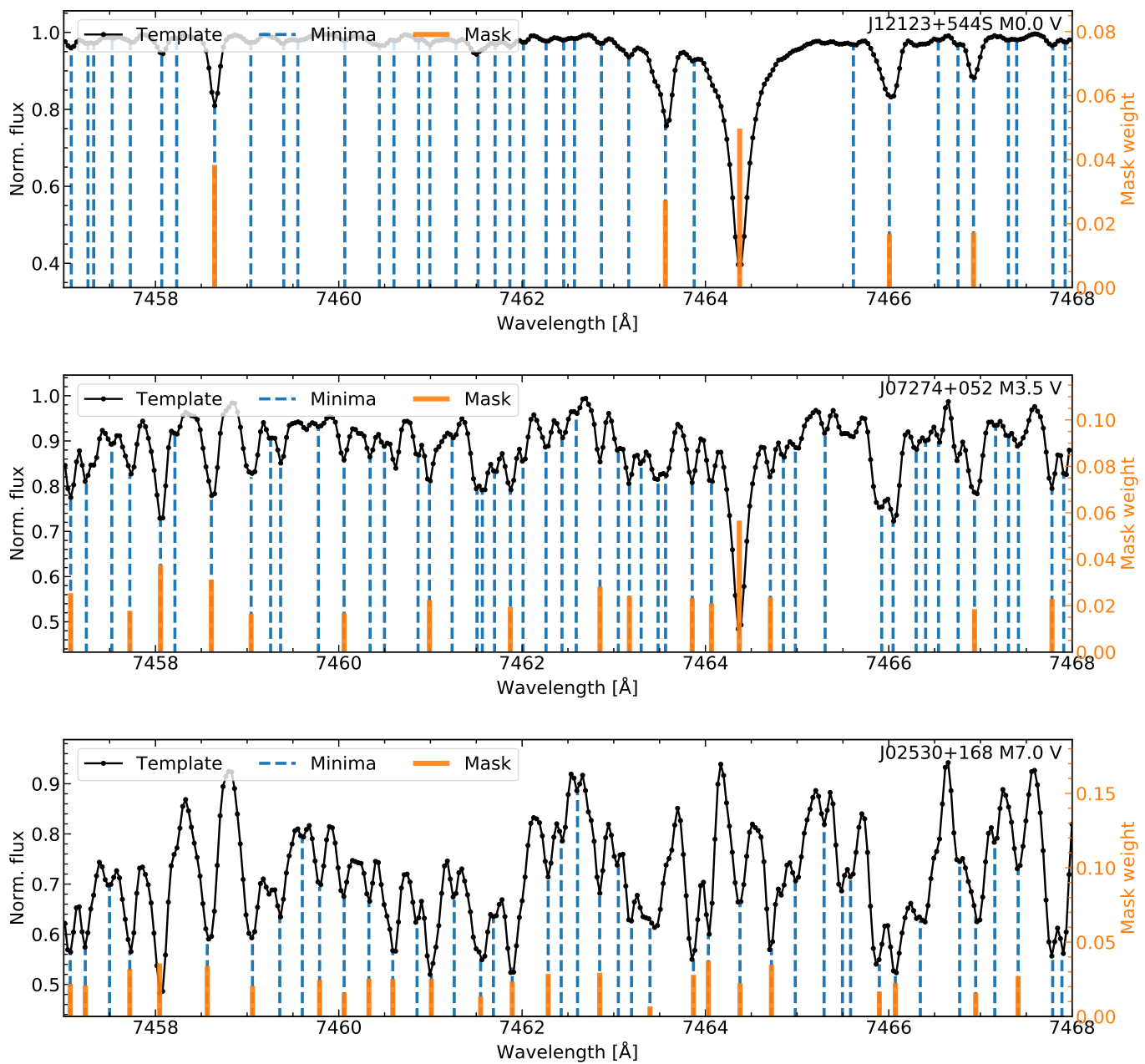

\centering
\begin{subfigure}[b]{1.0\linewidth}
\includegraphics[width=\linewidth]{{{J12123+544S_M0.0_V_tpl_mask_mincleanrv0_w07457-07468}}}
\end{subfigure}
\\
\begin{subfigure}[b]{1.0\linewidth}
\includegraphics[width=\linewidth]{{{J07274+052_M3.5_V_tpl_mask_mincleanrv0_w07457-07468}}}
\end{subfigure}
\\
\begin{subfigure}[b]{1.0\linewidth}
\includegraphics[width=\linewidth]{{{J02530+168_M7.0_V_tpl_mask_mincleanrv0_w07457-07468}}}
\end{subfigure}
\caption{Section of a normalised spectrum template used to build masks (solid black data points and line), minima found in the template (dashed blue line), and mask containing the minima selected based on some conditions (solid orange lines) for three stars of different spectral type and low \vsini:
\object{HD 238090} (M0.0\,V; GJ~458A, Karmn~J12123+544S; {\em top panel}), 
\object{Luyten's star} (M3.5\,V; GJ~273, Karmn J07274+052; {\em middle panel}), 
and \object{Teegarden's star} (M7.0\,V; Karmn J02530+168; {\em bottom panel}).
The three targets are Doppler-shifted to the same reference frame.
}
\label{fig:masklineselected}
\end{figure*}

In Fig. \ref{fig:masklineselected} we show a section of the normalised templates of stars of spectral types M0.0\,V, M3.5\,V, and M7.0\,V. We also plot the local minima found and the final mask lines selected following the criteria explained here, along with their corresponding weight.

\subsubsection{Order merging}

Since CARMENES is an echelle spectrograph, the observed spectra and also the high S/N template created from these observations are split in different spectral orders across the detector. The bluest orders have some overlap 
at the order ends, so there are some duplicated lines in our list.
To have a single final mask covering the entire spectral range, we merged the lines in overlapping order regions, which occur in the VIS channel and in the bluest NIR orders.
We identified lines that are the same in two consecutive overlapping orders by comparing their central wavelengths.
If the lines have central wavelengths that are close enough to be the same line (we chose a maximum separation of 0.05 \AA), we took the mean of their positions and their weights, and these are the values that are included in the mask.
If a line in an overlapping region is not found in both orders that overlap, we did not include it in the final mask.

\subsubsection{Absolute velocity}\label{sec:maskrvabs}

We used the SERVAL template made by adding observations to build the mask so the mask wavelengths obtained are shifted to the RV of the star.
To remove this contribution, we could just Doppler-shift the line wavelengths by the opposite of the star's velocity.
But since these velocities are known to different accuracies, instead, we used a synthetic stellar spectrum as a reference. We cross-correlated the mask with a PHOENIX synthetic spectrum \citep{husser2013phoenix}, which we assumed to define the zero-velocity frame, and we corrected the shift of the mask lines by the velocity obtained from the CCF.
We used models with solar metallicity $\mathrm{[Fe/H]}=0.0$, surface gravity $\log g=5.00,$ and effective temperature closest to the one of the star whose observations we used to build the template obtained from \cite{schweitzer2019carmenesMR}.

%--------------------------------------------------------------------

\section{CCF and parameter computation}\label{sec:ccfmethod}

\subsection{Spectrum preparation: Flux and wavelength correction}

As before, we assumed that we have a high-resolution echelle spectrum, which is divided into different orders, and each order consists of a list of data points. One data point represents all of the detector pixels summed perpendicular to the dispersion direction, with a central wavelength and a flux value.

The CARMENES science spectrum is extracted relative to the spectrum of a flat lamp taken during the daily calibrations \citep{zechmeister2014fox}. %halogen lamp
We divided the spectrum flux by an instrument response function (provided by the standard CARMENES reduction pipeline together with the reduced spectrum) to remove the SED of the flat lamp.
Due to the way the spectrum was extracted, its flux values do not reflect the S/N of the original image.
To weigh each order according to its original S/N, we rescaled the flux of each order so that its mean counts are equal to the square of the original S/N.
When other spectral extraction pipelines are used \citep[e.g. HARPS DRS,][]{mayor2003harps}, the extracted spectral orders need to be corrected for the blaze response of the spectrograph grating.
In addition, we applied a small correction to ensure a constant overall spectral energy distribution and thus minimise biases arising from different extinction values, for example \citep{berdinas2016systematicsccf}. These rescaling values were calculated from the spectrum with a higher S/N in the time series. By doing this, we altered the original S/N of the orders, but we avoided systematic biases in the CCFs.

In the bluest orders of the CARMENES VIS channel, the blue end of the order is noisier than the rest of the wavelengths. We found that removing the first 200-300 pixels improves the RV precision obtained.

We Doppler-shifted the spectrum to correct for the barycentric motion of the Earth and any instrumental drift measured so that the RV shift derived from the spectrum comes solely from the star.
In doing so, the centre of the CCF  is always located close to the absolute RV of the target.

\subsection{Mask preparation: Line selection for spectra at different epochs}

When computing the CCF of a series of observations of a target taken at different times of the year, we need to make sure that we are using exactly the same mask lines at each epoch to avoid measurement biases. Since the stellar spectrum shifts across the detector during the year according to the relative velocity between the observed star and the Earth, the lines at the detector edges are not always visible.
Therefore, to make sure that we were using exactly the same mask lines at any epoch, we removed the mask lines with wavelengths smaller than the wavelength of the blue cutoff plus the maximum BERV of the target observations, and all the lines with a wavelength larger than the red cutoff minus the maximum BERV.
Furthermore, since we displaced the mask across the spectrum to compute the CCF, some mask lines appear and disappear at the order cutoffs depending on the Doppler shift applied to the mask.
So we also removed the lines that, depending on the mask position, do not always fall within the spectral order limits. In other words, for each order, we only kept the mask lines that are between $\lambda_{\mathrm{min}}$ and $\lambda_{\mathrm{max}}$
\begin{align}
\lambda_{\mathrm{min}} &= \lambda_1 \left( \mathrm{BERV_{max}} - v_{\mathrm{min}} \right) / c \\
\lambda_{\mathrm{max}} &= \lambda_n \left( - \mathrm{BERV_{max}} - v_{\mathrm{max}} \right) / c, \notag
\end{align}
where $\lambda_1$ and $\lambda_n$ are the minimum and maximum wavelength of the order, respectively, $\mathrm{BERV_{max}}$ is the maximum barycentric shift of the observations considered, $v_{\mathrm{min}}$ and $v_{\mathrm{max}}$ are the minimum and maximum shift applied to the mask (i.e. the CCF velocity range), and $c$ is the speed of light.

When creating a mask from observations of a specific star, we already removed the stellar lines that fall on regions affected by tellurics at any time of the year.
However, telluric lines overlap with different stellar lines depending on the absolute radial velocity of the star. One needs to further remove the mask lines that can be affected by tellurics if the absolute velocities of the star used to create the mask and the target star are different.
We followed the same procedure as carried out when excluding regions during the mask creation process (Sect. \ref{sec:maskexcludedregions}, we broadened the telluric features by the maximum BERV of the target observations and removed the mask lines that overlap with them).
Additionally, we needed to take into account the displacement of the mask across the spectrum. Some mask lines overlap with telluric-contaminated regions for some of the Doppler shifts applied, so we also removed them.
After this process, we obtained a set of lines per order that, regardless of the Doppler shift applied, are not affected by tellurics and are visible in the spectrum at any epoch.

\subsection{CCF location in RV space}

Prior to computing the CCF of each order, we need to define the RV range and the steps that we use to Doppler-shift the mask, which determine the span and sampling of the CCF.
We need to ensure the full coverage of the CCF span (its minimum and part of the wings), but also, we should not extend far into the continuum to save on computing time.
To quickly locate the position of the CCF centre in RV space and estimate its width, we computed a test CCF spanning a large RV range of $\sim 200\,\kms$, using coarse steps of $\sim 1\,\kms$ in the spectral order with the highest S/N.
When both the target's absolute RV and \vsini are known, this first step can be skipped.

\begin{table}
\centering
\caption{Average pixel sampling per spectral element (SE), resolving power, and pixel size in velocity units.}
\begin{tabular}{cccc}
\hline\hline
Instrument & Sampling $s$ & Resolving & RV step \\
& [pix/SE] & power     & [\ms]   \\
\hline
CARMENES VIS & 2.5 &  94\,600 & 1268 \\
CARMENES NIR & 2.8 &  80\,400 & 1332 \\
\hline
\end{tabular}
\tablefoot{Sampling and resolving power from \cite{quirrenbach2018CARMENES}.
}
\label{tab:instr}
\end{table}

The sampling of the CCF velocity grid should be given by the average pixel size in velocity units, $\Delta v$, which can be obtained from the resolving power $R$ of the instrument and the pixel sampling per spectral element.
The resolving power at a given wavelength $\lambda$ can be expressed as $\lambda$ divided by the minimum difference that can be distinguished at this wavelength, $\delta\lambda$, which is typically the size of a spectral element. By applying the Doppler equation, we get the size of a spectral element in velocity units $\delta v$
\begin{equation}
R = \frac{\lambda}{\delta\lambda} = \frac{c}{\delta v}
\Rightarrow \delta v = \frac{c}{R}
\Rightarrow \Delta v = \frac{c}{R \cdot s},
\end{equation}
where $c$ is the speed of light.
By dividing this quantity by the average pixel sampling per spectral element, $s$, we get the size of a pixel in velocity units $\Delta v$, which gives us the step size by which we have to shift the mask. The average values for the CARMENES VIS and NIR arms are shown in Table \ref{tab:instr}.
Another way to find this step size is to directly compute the average width of each spectrum pixel in velocity units.
It is important to use the correct sampling to ensure that the CCF flux reflects the actual number of photons of the spectrum and, therefore, its S/N. This is used to estimate the statistical uncertainty on the RV. For example, if we were using a denser sampling (smaller RV steps), we would count the flux in the same pixels more than once, resulting in biased RV errors.

By using this RV step, we obtain a well-sampled CCF for which a Gaussian fit provides its minimum, width and depth, with their corresponding uncertainties. However, this type of sampling is not sufficiently dense to measure the bisector correctly and compute the BIS. For this, we used a denser sampling of $\sim 250\,\ms$. As with the RV, in order to estimate the uncertainty of the BIS, we needed to use the CCF computed with the correct sampling so that it is consistent with the original S/N of the spectrum (see Sect. \ref{sec:ccfmethodbis} for details).

\subsection{CCF computation details}\label{sec:ccfmethoddetails}

The calculation of the CCF as a function of the Doppler-shift $v$ was performed by applying Eq. \ref{eq:ccf} to each of the spectral orders separately.
The product of the two signals, that is, the spectrum and mask, was computed in pixel space, so both the spectrum and mask have to be expressed in pixel units.
Since we did not rebin or merge the spectral orders, each data point (wavelength and flux) corresponds to one pixel, so the spectrum is already in pixel space. The values of each data point correspond to the centre of the pixel.
Our mask lines are Dirac $\delta$ functions (zero width) expressed in wavelength, so we mapped each one to the corresponding position in the spectral order, interpolating between the wavelengths of the spectral pixels.
We then gave each of them a width of 1 pixel by adding or subtracting 0.5 pixels to the central value to obtain the pixel boundaries. In this way, we transformed our $\delta$ mask lines into bins of 1-pixel width.

As expressed in Eq. \ref{eq:ccf}, if a spectrum pixel overlaps (totally or partially) with a mask line (non-zero region of the mask), we multiply the value of the flux in that pixel by the value of the mask line weight, thus taking the fraction of the pixel overlapping by the mask into account.
If the overlap is not exact, we do not take the flux value of the centre of the pixel. Instead, we would linearly interpolate the flux between consecutive pixels, and take the interpolated flux value at the centre of the overlap region between the pixel and the mask. In this way, we would obtain a smoother CCF profile and still maintain the observed number of photons.

We computed the CCF on spectral orders that have been corrected for the blaze function of the echelle spectrograph to avoid biases due to the blaze slope.
However, this does not take into account the difference in S/Ns along the order, from lower S/N edges to a peak S/N at the centre.
Thus, a blaze-corrected spectrum does not preserve the flux and, therefore, its CCF does not immediately reflect the original S/N of the spectrum, which we need in order to estimate the uncertainty of the RV estimate.
To correct for this, we should multiply all the pixels of an spectral absorption line by the value of the blaze function at the centre of the line. In this way, we approximately recover the original flux of the line without reintroducing the biasing effects caused by the slope of the blaze.
However, since we did not know exactly which pixels correspond to a line in the spectrum, we instead multiplied the mask lines by the corresponding blaze value.

\subsection{Final co-added CCF}

We co-added the CCFs of the individual orders to obtain a final CCF that represents the full spectrum. Since we computed the order CCFs using the same RV grid, we could directly sum the flux at each RV step
\begin{equation}\label{eq:ccfsum}
\mathrm{CCF}(v) = \sum_{o} \mathrm{CCF}_o(v),
\end{equation}
where $\mathrm{CCF}_o(v)$ is the CCF of order $o$ and $\mathrm{CCF}(v)$ is the final co-added CCF.

Depending on the instrument and the star observed, not all the orders contain the same number of useable lines or have a sufficient S/N.
For the CARMENES observations, some orders are severely affected by tellurics, which is reflected in a low number of lines in the mask and poor CCFs. Furthermore, for very late M dwarfs, the bluest orders show a markedly degraded S/N.
The CCFs of very low S/N orders (S/N $\leq\,10$) or orders with few usable lines ($\leq\,10$ lines) do not significantly contribute to the final CCF and may actually be counterproductive.
Our tests show that discarding such poor orders may improve the profile of the final CCF, in the sense of obtaining better precision on the derived parameters.

We did not need to apply any weight to the CCFs of the different orders because the flux level of each of them is already proportional to the S/N of the original spectrum order, the number of lines used, and their quality, which are the parameters that we would use to asses the quality and, therefore, weight of each CCF.
However, the final flux also reflects the values of the mask line weights. These weights depend on the RV-content of each line, but the absolute values are arbitrary and should not affect the final flux of the CCF. Therefore, we applied a normalisation.
We only considered the mask lines that actually contributed to the co-added CCF, that is, the lines kept after removing telluric-contaminated regions and order ends of the orders used in the co-adding.
We computed the average of the weights of these mask lines and divided the flux of the co-added CCF by it.
In this way, the flux of the CCF reflects the true flux of the pixels used (scaled according to their RV-content) and the absolute weight of the mask lines is irrelevant.

%--------------------------------------------------------------------

\subsection{CCF parameters and errors}

\begin{figure}[]
\centering
\includegraphics[width=\linewidth]{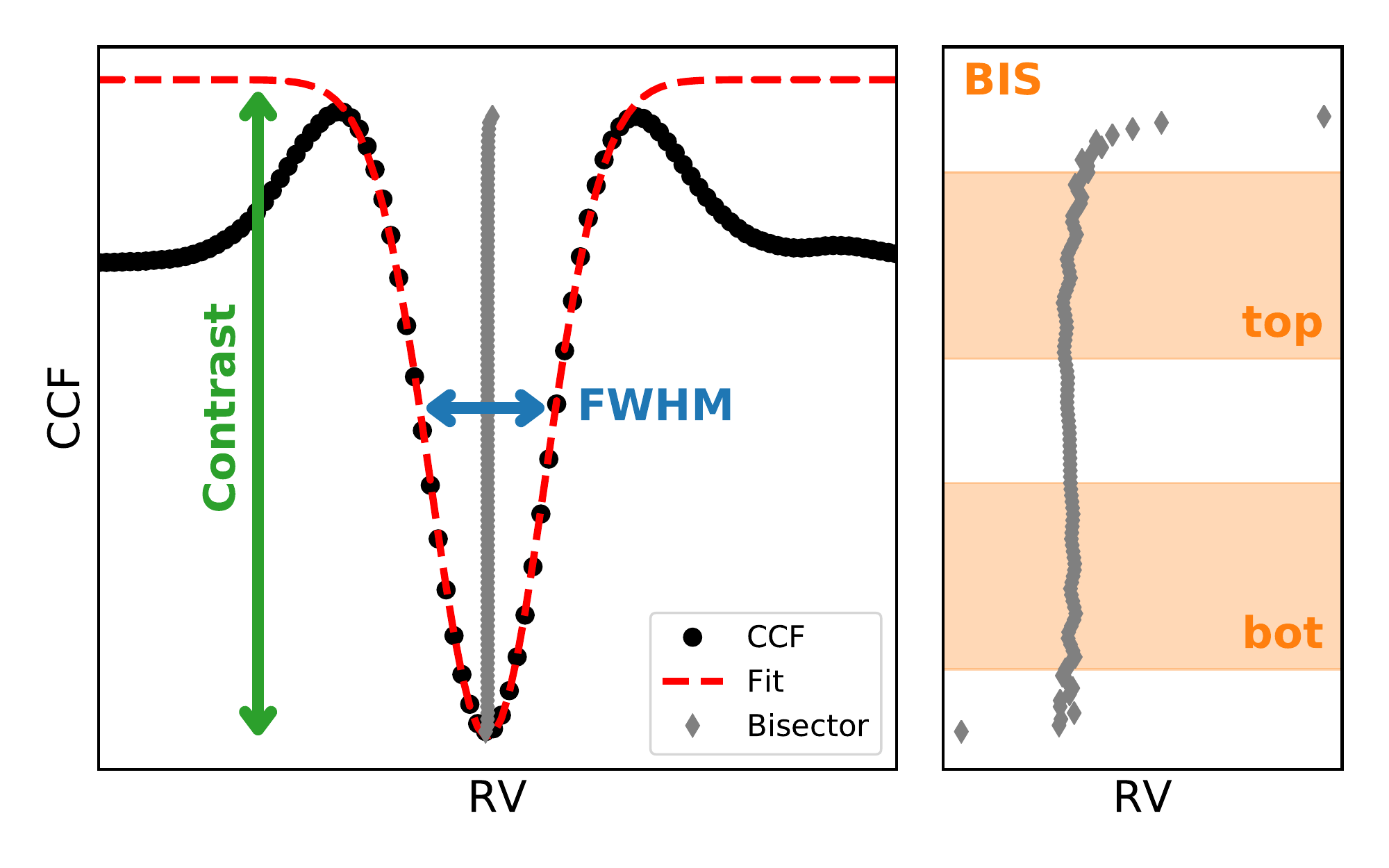}
\caption{Illustration of the CCF activity parameters measured in this work.
We plotted a typical CCF of an M dwarf (black dots) and the best Gaussian fit (red dashed line) together with the FWHM and contrast derived from the fit ({\em left}). We also plotted the bisector of the  CCF (grey diamonds) and a zoom to the central region of the CCF to show the bisector shape ({\em right}). We indicate the top and bottom regions used to compute the BIS.}
\label{fig:ccfparams}
\end{figure}

To study the CCF profile, we fitted a Gaussian function to the central region (from maximum to maximum; see Fig. \ref{fig:ccfparams}) of the co-added CCF.
From this function, we derived the RV shift of the spectrum with respect to the mask, the FWHM, and the contrast of the CCF profile.
We also analysed its asymmetry by studying its bisector.
Here, we describe the estimation of these four parameters and their uncertainties.

%-------------------------------------------------

\subsubsection{CCF minimum: RV}\label{sec:ccfmethodrvmin}

We estimated the velocity difference between the observed spectrum and the mask as the position of the minimum of the best Gaussian fit to the CCF. For time series RV work, we are interested in the relative velocities and, therefore, the differential values between the target and the mask are already appropriate.
An intuitive method in order to estimate the RV uncertainty would be to use the formal error of the fit. However, the CCF is not a perfect Gaussian function as it represents the convolution of thousands of lines with different profiles. Therefore, the formal error of the fit overestimates the true uncertainty as the Gaussian model is not adequate. To obtain a realistic value for the uncertainty, we consider the total RV content in the CCF assuming photon noise.
The capability of the CCF in providing a precise measure of the RV depends on its steepness and depth. A high S/N, deep, and narrow CCF is much more constraining than a low S/N, shallow, and wide CCF. In turn, this is defined by the S/N of the original spectrum by the number of lines used and by their profiles.
Descriptions of the estimation of photon-limited errors in RV measurements from spectral lines or CCFs can be found in
\cite{butler1996dopplerprecision}, \cite{bouchy2001photonnoise}, and \cite{boisse2010sophieJupiter}, for example.
Here we detail the procedure we followed, which is based on the methodologies of the previous references.

The first step is to compute the uncertainty of the flux on each sampling point of the CCF of each order. This uncertainty is propagated from the uncertainties of the individual pixels that are added when computing the CCF.
We consider that the uncertainty on the flux of a pixel $x$ in the spectral order $o$ is given by its photon noise $\sigma_{{\rm PN},x}$ and readout noise $\mathrm{RON}$ as follows:
\begin{equation}
{\sigma_f}_{ox}^2 = \left(\sigma_{{\rm PN},x}\right)^2 + \mathrm{RON}^2 = f_{ox} + \mathrm{RON}^2.
\end{equation}
Assuming Poissonian statistics, the photon noise of the pixel $x$ is calculated as the square root of the total flux in the pixel, $f_{ox}$, where we use the flux value before rescaling the orders to a constant overall spectral energy distribution, that is, the value representing the observed S/N.

The uncertainty of the CCF of order $o$ at velocity $v$, $\mathrm{CCF}_{o}\left(v\right)$, is then the square root of the quadratic sum of the uncertainties of all the pixels used when computing the CCF at that velocity.
For partial pixels, the flux is weighted by the fraction of pixel that overlaps with the line $l$, which is similar to Eq. \ref{eq:ccf}
\begin{equation}
\sigma^{2}_{\mathrm{CCF},o}\left(v\right) =
\sum_{l=1}^{m} \sum_{x=1}^{n} {\sigma_f}_{ox}^2 \Delta_{x\,l}\left(v\right).
\end{equation}
When computing the CCF of spectra extracted with a blaze function, the flux of pixel $x$ is also normalised to the original flux level by multiplying it by the interpolated blaze value, corresponding to the pixel of the mask line $l$, so that the photon noise reflects the original S/N of the data.\ This is the same process as with the spectrum (see Section  \ref{sec:ccfmethoddetails}).

The uncertainty of each RV sampling point of the co-added CCF, $\sigma^{2}_{\mathrm{CCF}}\left(v\right)$, is given by the quadratic sum of the uncertainties of all the orders used
\begin{equation}
\sigma^{2}_{\mathrm{CCF}}\left(v\right) =
\sum_{o} \sigma^{2}_{\mathrm{CCF},o}\left(v\right),
\end{equation}
which is the error propagation of Eq.~\ref{eq:ccfsum}.

From the value of the uncertainty of the CCF flux ($\sigma_{\mathrm{CCF}}\left(v\right)$, the error on the y-axis), we estimate the velocity uncertainty of each CCF point ($\sigma\left(v\right)$, the x-axis of the CCF) by using the derivative of the CCF as
\begin{equation}
\sigma\left(v\right) = \sigma^{2}_{\mathrm{CCF}}\left(v\right)  \left( \frac{\mathrm{d}\mathrm{CCF}\left(v\right)}{\mathrm{d} v} \right) ^ {-1}.
\end{equation}

The uncertainty of the final RV is then given by the velocity errors of all the points of the CCF
\begin{equation}
\sigma_{\mathrm{RV}} =
\left(\sqrt{\sum_v\frac{1}{\sigma\left(v\right)^2}}\right)^{-1}.
\end{equation}
For this procedure to provide realistic uncertainties, it is essential to employ the CCF sampled in RV steps corresponding to the average pixel size.

%-------------------------------------------------

\subsubsection{CCF profile variations: FWHM and contrast}

To determine the width of the CCF, we used the FWHM of the best-fit Gaussian, which is measured in velocity units (typically \kms).
The contrast of the CCF is the amplitude of the Gaussian divided by its baseline level. We inverted the sign of the amplitude, which is negative, and multiplied it by 100 so that the contrast values are positive and given in a percentage.
We considered the formal errors of the fit as the uncertainties for these two quantities.

%-------------------------------------------------

\subsubsection{CCF profile variations: Bisector (asymmetry)}\label{sec:ccfmethodbis}

To quantify the asymmetry in the CCF, we used the standard bisector inverse slope as defined in \cite{queloz2001noplanet}.
We measured the bisector of the CCF in the same interval in which we fitted the Gaussian, that is, the range comprised within the maxima at each side of the global minimum.
We linearly interpolated the CCF profile and took the mid points between the left and right side profiles at about 100 different evenly spaced depths. This yields the bisector as a function of the depth $d$
\begin{equation}\label{eq:bisector}
\mathrm{Bisector}(d) = \left( v ( \mathrm{CCF_{left}}=d ) + v ( \mathrm{CCF_{right}}=d ) \right) / 2,
\end{equation}
where $v ( \mathrm{CCF_{left}}=d )$ and $v ( \mathrm{CCF_{right}}=d )$ are the interpolated velocities at depth $d$ at the left and right sides of the CCF.
We then consider the following two regions: one at the top of the CCF, from 90 to 60\% of the baseline level, and one at the bottom, from 40 to 10\% (see orange regions in Fig. \ref{fig:ccfparams}). We computed the average RV of the points in these two regions, and the difference between them, $\mathrm{RV_{top}}$ and $\mathrm{RV_{bot}}$, yields the value of the BIS
\begin{align}
\mathrm{BIS} &= \mathrm{RV_{top}} - \mathrm{RV_{bot}} \\
                         &= \langle \mathrm{Bisector}(d) \rangle_{d=60-90\%} - \langle \mathrm{Bisector}(d) \rangle_{d=10-40\%}. \notag
\end{align}
As mentioned above, using the CCF computed with the correct RV sampling (i.e. where the RV steps are the size of the average pixel width in velocity units) does not allow us to correctly measure the bisector. Instead, we used a CCF computed with a denser sampling, with steps of 250 \ms.

Regarding the uncertainty, since the BIS is computed from the (interpolated) points of the CCF, we started with the velocity uncertainties of each point of the CCF as computed above (Sect. \ref{sec:ccfmethodrvmin}).
As was done with the CCF flux when computing the bisector, we linearly interpolated the RV errors of the actual CCF data points to the $\sim100$ points, corresponding to the evenly spaced depths used to define the bisector.
For each bisector point, the error $\sigma_{\rm bisector}(d)$ comes from the interpolated uncertainties at each side of the CCF by error propagation of Eq. \ref{eq:bisector}.
We then computed the uncertainty on the top and bottom regions of the bisector as the mean of all the points considered in each region divided by the squared root of the number of points $n_{\rm top}$ or $n_{\rm bot}$
\begin{align}
\sigma_{\rm top} &= \langle \sigma_{\rm Bisector}(d) \rangle_{d=60-90\%} / \sqrt{n_{\rm top}} \\
\sigma_{\rm bot} &= \langle \sigma_{\rm Bisector}(d) \rangle_{d=10-40\%} / \sqrt{n_{\rm bot}}. \notag
\end{align}
Finally the BIS uncertainty is the square root of the quadratic sum of the uncertainties on the top and bottom regions
\begin{equation}
\sigma(\mathrm{BIS}) = \sqrt{\sigma_{\rm top}^2 + \sigma_{\rm bot}^2}.
\end{equation}

%--------------------------------------------------------------------

\section{Application to the CARMENES survey}\label{sec:carmenesgtomaskccf}

\subsection{Sample}

We followed the procedure explained in the previous sections to build a set of masks using CARMENES survey observations to compute the CCFs for most of the targets in the sample.
The CARMENES survey sample \citep{alonsofloriano2015carmenesinlowres, cortescontreras2017carmenesinhighres, schweitzer2019carmenesMR, reiners2018carmenes324} includes about 350 M dwarfs of subtypes from M0.0 to M9.0.
Most of the targets are early and mid M dwarfs, and only less than 20 stars have a spectral type equal or later than M6.0.
Most of the stars have low projected rotational velocities, $\vsini \leq 10\, \kms$, although some show \vsini values of a few tens of \kms.
The sample includes both inactive and active stars for early and mid spectral types, while most of the late spectral type targets show high levels of activity.

In total, we computed CCF profiles and parameters of 323 targets in the VIS, excluding those with $\vsini \gtrsim 60\,\kms$ and spectroscopic binary systems \citep[already studied in][]{baroch2018carmenesBinaries}. In the NIR, we computed CCFs for 305 targets. Since there is less RV content in the NIR than in the VIS, we could not get reliable masks and CCFs for the faintest targets and the ones with $\vsini \gtrsim 25\,\kms$; we therefore excluded them from this analysis.
We used different masks for targets of different spectral types, but  this is especially the case for targets of different \vsini.
In the following, we explain the tests we performed to select the final set of masks used for different types of targets, which we divide into slow and fast rotators.

\subsection{Slowly-rotating stars}

\begin{table*}
\centering
\caption{Targets used to build the masks employed to compute the CCFs of the low \vsini stars with their spectral types, rotational velocities, and magnitudes.\tablefootmark{a}}
\begin{tabular}{llccccccccc}
\hline\hline
Karmn & Name & Sp. & $v\sin i$\tablefootmark{b} & $J$\tablefootmark{c} & \multicolumn{2}{c}{VIS template} & VIS mask & \multicolumn{2}{c}{NIR template} & NIR mask \\
\cline{6-7}\cline{9-10}
& & type & [km/s] & [mag] & \# obs & S/N & \# lines & \# obs & S/N & \# lines \\
\hline
J12123+544S & \object{GJ 458A} & M0.0\,V & $\leq 2.0$ & 6.88 & 110 & 1189 & 2019 & 105 & 1423 & 157 \\
J11033+359  & \object{GJ 411}  & M1.5\,V & $\leq 2.0$ & 4.20 & 310 & 1907 & 3762 & 300 & 2247 & 166 \\
J19169+051N & \object{GJ 752A} & M2.5\,V & $\leq 2.0$ & 5.58 & 129 & 1194 & 4332 & 125 & 1381 & 201 \\
J07274+052  & \object{GJ 273}  & M3.5\,V & $\leq 2.0$ & 5.71 & 733 & 1090 & 5334 & 721 & 1663 & 283 \\
J13229+244  & \object{GJ 3779} & M4.0\,V & $\leq 2.0$ & 8.73 & 113 &  701 & 5423 & 110 & 1022 & 269 \\
J20260+585  & \object{GJ 1253} & M5.0\,V & $\leq 2.0$ & 9.03 & 183 &  667 & 6136 & 182 &  991 & 443 \\
J10564+070  & \object{GJ 406}  & M6.0\,V &        2.9 & 7.09 &  79 &  517 & 4382 &  76 & 1078 & 353 \\
J02530+168  & \object{Teegarden's star} & M7.0\,V & $\leq 2.0$ & 8.39 & 245 &  732 & 5387 & 244 & 1528 & 389 \\
\hline
\end{tabular} 
\tablefoot{
\tablefoottext{a}{We also show the number of observations used to build the templates, the S/N of a reference order (order 82 in the VIS and order 58 in the NIR), and the number of lines of the final mask in the VIS (from order 108 to order 68, 5623--9068 $\AA$) and NIR channels (from order 63 to order 36, 9618--17118 $\AA$).}
\tablefoottext{b}{\cite{reiners2018carmenes324}}, 
\tablefoottext{c}{\cite{skrutskie20062mass}}.
}
\label{tab:maskcarmenestargets}
\end{table*}

\begin{figure*}[]
\centering
\begin{subfigure}[b]{\linewidth}
\includegraphics[width=\textwidth]{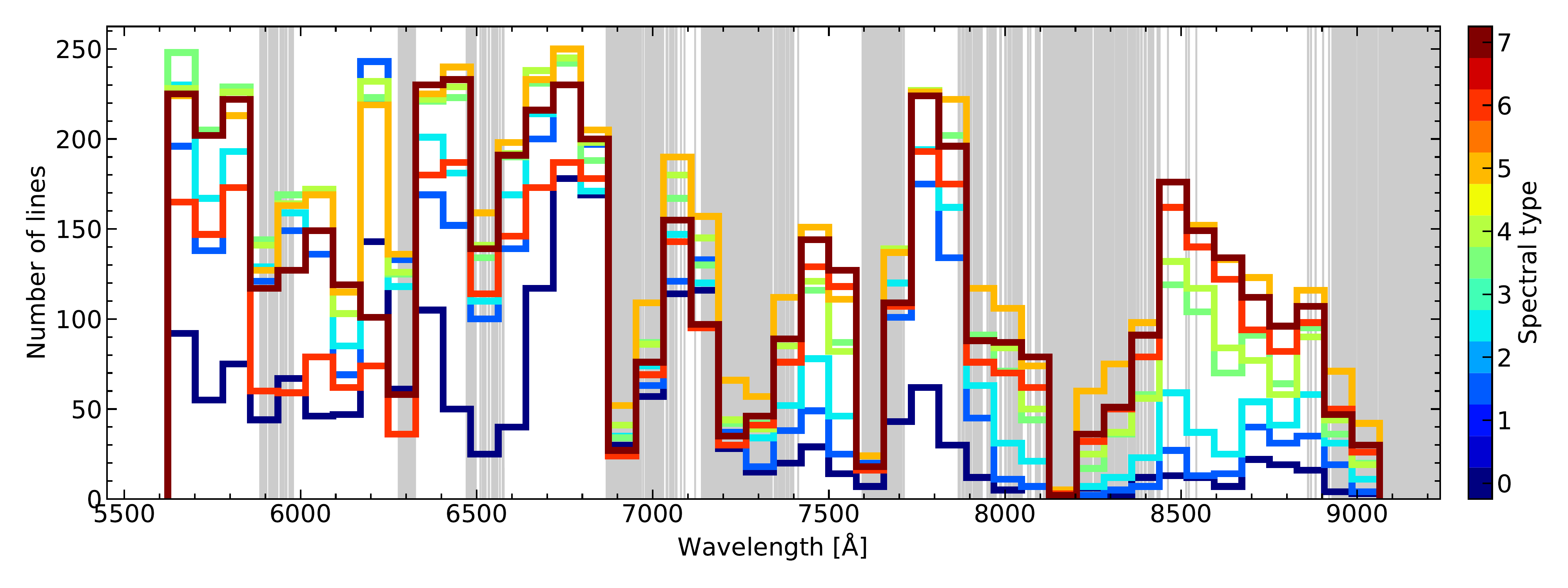}
\end{subfigure}
\\
\begin{subfigure}[b]{\linewidth}
\includegraphics[width=\textwidth]{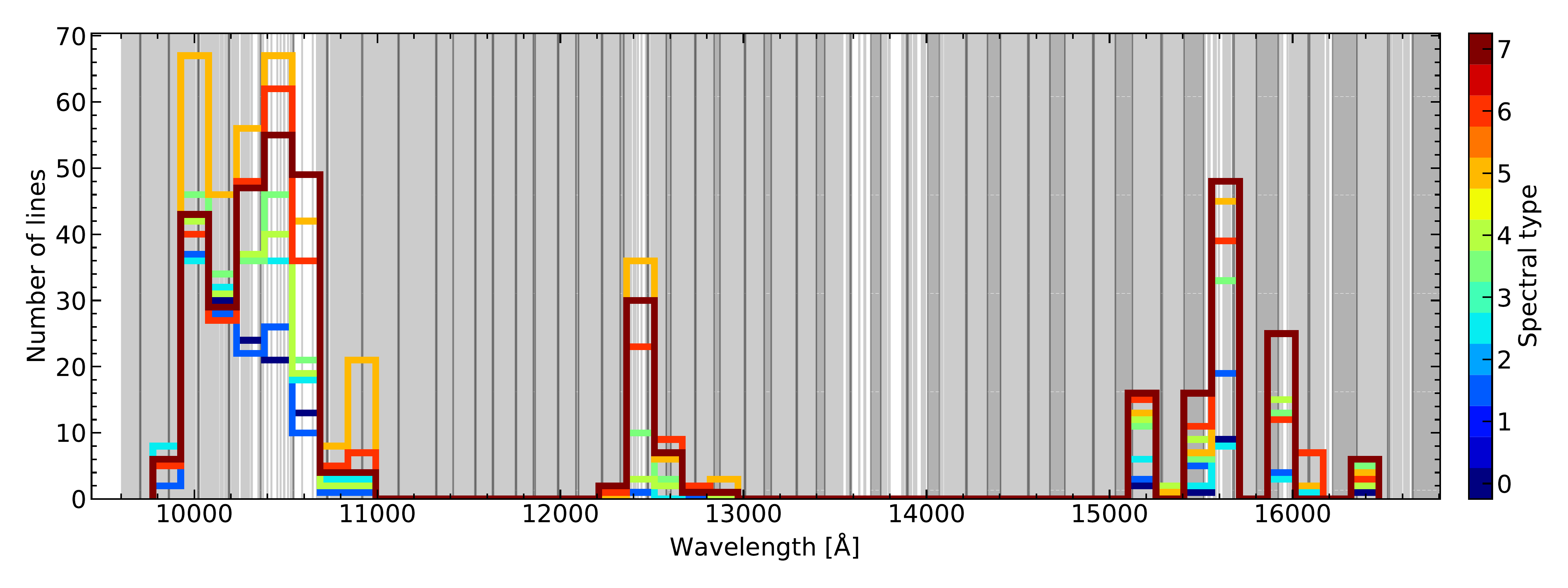}
\end{subfigure}
\caption{Histogram of the line wavelengths of the masks used to compute the CCFs of the low \vsini targets for the VIS ({\em top}) and NIR ({\em bottom}) channels. Different colours correspond to the different spectral types of the targets used to build the masks (see targets and number of lines in Table \ref{tab:maskcarmenestargets}). Light grey areas correspond to regions affected by telluric features, which are broadened by $\pm30\,\kms$.
Dark grey hatched areas correspond to regions without spectral information in the CARMENES data, corresponding to gaps between consecutive spectral orders (redwards of ~11500 $\AA$), and intra-order gaps in the NIR detector mosaic.
}
\label{fig:masks}
\end{figure*}

For each spectral subtype (M0.0\,V to M0.5\,V, M1.0\,V to M1.5\,V, etc.), we selected the objects with the highest S/N to build the best possible template. This generally corresponds to the brightest objects with the highest number of observations.
We restricted this search to targets with low \vsini ($\vsini \sim 2\,\kms$) and low levels of activity.
We did this because our early tests show that for targets with larger \vsini, we obtained better results (CCFs with smoother profiles and smaller RV dispersions) when using masks built from observations of a star with also larger rotational velocity, rather than using one of the standard masks obtained with observations of a slowly-rotating star.
The masks were built using the selection parameters shown in Table \ref{tab:maskcarmenesparams}.

Table \ref{tab:maskcarmenestargets} lists the targets used to create the masks together with the number of co-added observations, the S/N of the template, and the total number of lines of the masks for the CARMENES VIS and NIR spectral ranges.
Figure \ref{fig:masks} shows a histogram of the number of lines of the masks listed in Table \ref{tab:maskcarmenestargets} as a function of wavelength. 

We analysed the performance of the different VIS masks on targets of different spectral types. We did this for several targets with low \vsini, low activity levels, and more than 100 observations when possible, since most late spectral type targets are very active and have not been observed that many times.
We computed the CCF of these targets with the different masks and studied the CCF profiles and the parameters time series.

Although they have a different number of lines, we find that for most of the early and mid spectral type stars, from M0 to M6--M7, masks built from early and mid spectral type targets result in smooth CCF profiles and RV time series with similar scatter and modulation.
Therefore, we are able to obtain reliable CCF profiles and parameters with a single mask for all early and mid spectral type targets. However, we find that using more than one mask for different spectral subtypes slightly improves the precision of the parameters.
We note that different masks result in RV time series with different absolute offsets, with differences of the order of a few tens of \ms, thus showing the difficulty in computing the absolute displacement of a mask.

Regarding the other CCF parameters, they also show similar dispersions and modulations for masks of different spectral types, and their absolute values also seem to depend on the mask used. FWHM mean values can vary by some hundreds of \ms for masks of a different spectral type showing very similar RV scatters.
The same happens with the contrast and the bisector. Contrast mean values can vary by about 10\% and bisectors can fluctuate by some tens of \ms.
This shows that the same mask must be used if we want to compare CCF parameters.

% --------------------------------------

For early M dwarfs (M0\,V-M1\,V), the VIS channel CCF has the shape of an absorption line and rather flat wings. The same shape is found in G and K stars \citep[see e.g. Fig. 1 in][]{mayor1995Jup51Peg}.
As we move towards later-type stars, the flux level of the CCF continuum decreases and two humps appear at either side of the CCF (such as in Fig. \ref{fig:ccfparams}).
The appearance of the two humps and the decreasing continuum are a consequence of the very large density of spectral features in cooler stars. The mask lines are chosen to represent those spectral features that are best defined, that is, showing no obvious blends and asymmetries as well as with a clearer continuum. As the mask slides through the spectrum at different wavelength positions, the density of stellar lines is so high that a number of the mask lines randomly coincide with stellar absorption features. This means that the CCF continuum is depressed. The first CCF flux maximum occurs just before the mask lines enter the stellar absorption lines because all mask lines receive flux from the continuum as the mask is only composed of lines that are well behaved. The second maximum occurs for the same reason just after the mask lines leave the stellar lines. This explains the characteristic shape of the CCF for stars of spectral types later than M1 in the case of the CARMENES VIS spectra. Interestingly, we do not observe this phenomenon in the NIR channel CCFs as the lines in the spectrum are more widely spaced.

% --------------------------------------

\subsection{Fast-rotating stars}

The spectra of stars with high \vsini values are quite different from those of low \vsini stars of similar types because their lines are much broader and most of them become heavily blended. Moreover, line profiles can be affected by distortions due to stellar activity, which is generally stronger in fast rotators. 
Using one of the `standard' masks made from a low \vsini and inactive target for a star with larger \vsini (from $\sim 3\,\kms$ up to $\sim 10-15\,\kms$) still delivers a smooth CCF profile. However, the RV dispersion is larger than if we were to use a mask made from an object of \vsini close to the one of the target star.\ This would be the case even if the spectral type is not a very good match.

For even faster rotators, using the standard masks results in a CCF profile that shows a global minimum close to the RV value of the target, but mostly this contains noise, with the CCF showing bumps across its profile. The profile becomes noisier as the \vsini of the star increases up to a point where we are not able to distinguish any minima for the stars with the most extreme \vsini values ($\vsini \gtrsim 40\,\kms$).
In using a mask that is also made from a large \vsini target, the CCF profile becomes less noisy, but we cannot obtain smooth profiles as is the case for slower rotators.
When building such a mask, in order to be able to select any line, the FWHM cuts need to be larger than in the low \vsini case. The resulting masks have fewer lines, only a few hundred, which also impedes obtaining a smooth CCF because of the lower S/N that results from only considering a small number of lines.
Another problem is that most of the fast-rotating stars are the ones with the latest spectral types and, therefore,  they are fainter and deliver poorer S/N observations. Also, in general, fewer observations are available. All of this implies that the templates have a low S/N. In some extreme cases, we smoothed the templates applying a convolution with a Gaussian profile in order to identify local minima due to bona-fide absorption lines instead of noise features. For stars with $\vsini \gtrsim 50-60\,\kms$, it becomes unfeasible to identify a large number of lines following our methodology.
In all circumstances, and because of the numerous blends, CCF parameters of fast-rotating stars are much less reliable than for slow rotators.

% --------------------------------------

\subsection{Final masks}

In light of these results, we used a grid of masks to compute the CCFs of the CARMENES survey sample.
For targets with $\vsini \leq 4\,\kms$, we used the standard masks listed in Table \ref{tab:maskcarmenestargets}.
The procedure is to employ the mask corresponding to the spectral type closest to that of the target.
For faster rotators, we used a different set of masks made from targets with larger \vsini values.
We divided the targets according to their \vsini in bins of approximately $10\,\kms$, up to $\vsini \sim 60\,\kms$ in the VIS, and $\vsini \sim 30\,\kms$ in the NIR.
For each bin, we selected targets of different spectral types for which we can obtain templates with a sufficiently high S/N and build the masks. We finally obtain one to three masks per \vsini bin covering different spectral types, depending on target availability.

%--------------------------------------------------------------------

\subsection{Comparison between CCF and template-matching RVs}
\label{sec:ccfVSserval}

\begin{figure*}[]
\centering
\begin{subfigure}[b]{0.49\linewidth}
\includegraphics[width=\textwidth]{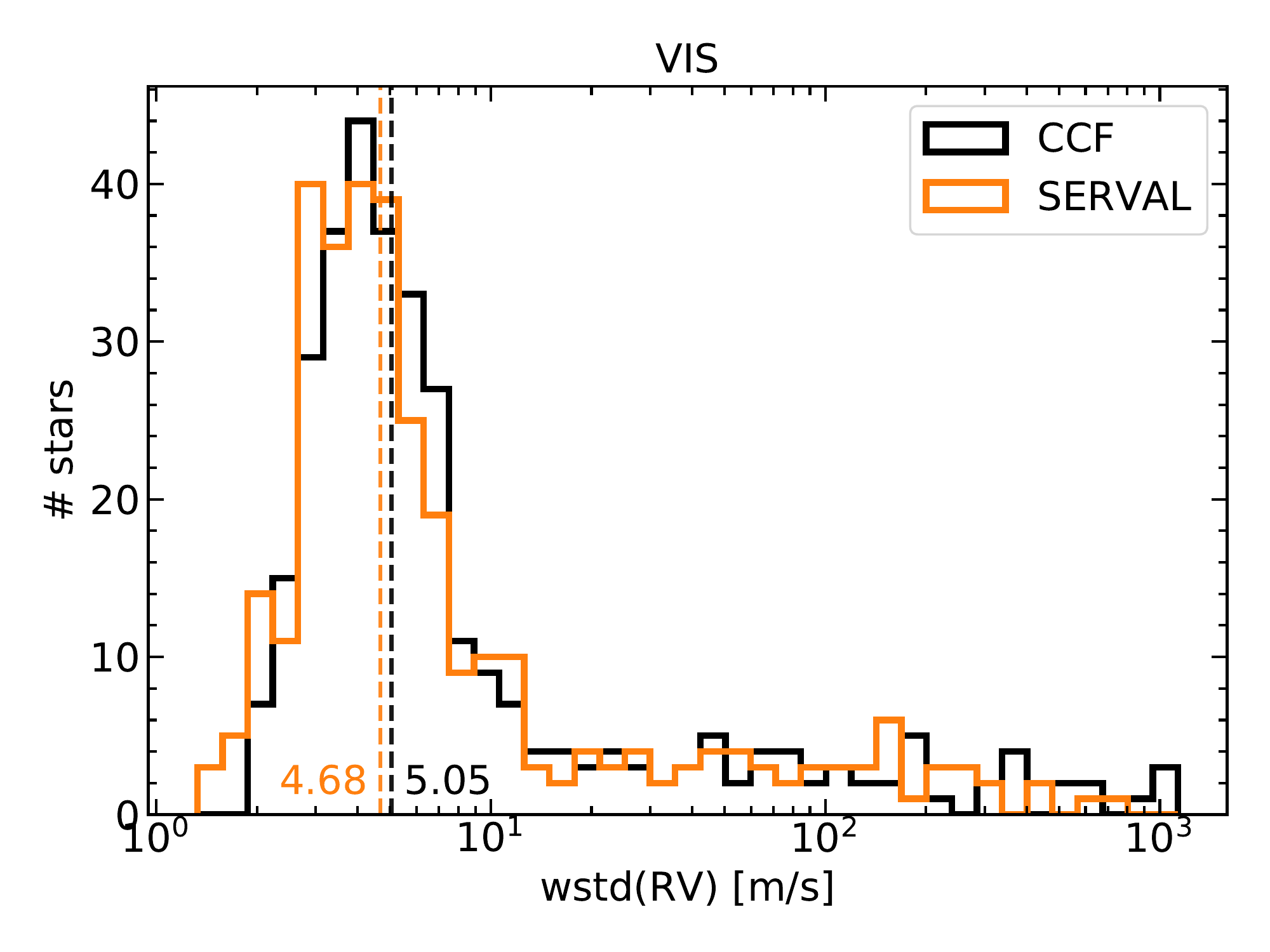}
\end{subfigure}
~
\begin{subfigure}[b]{0.49\linewidth}
\includegraphics[width=\textwidth]{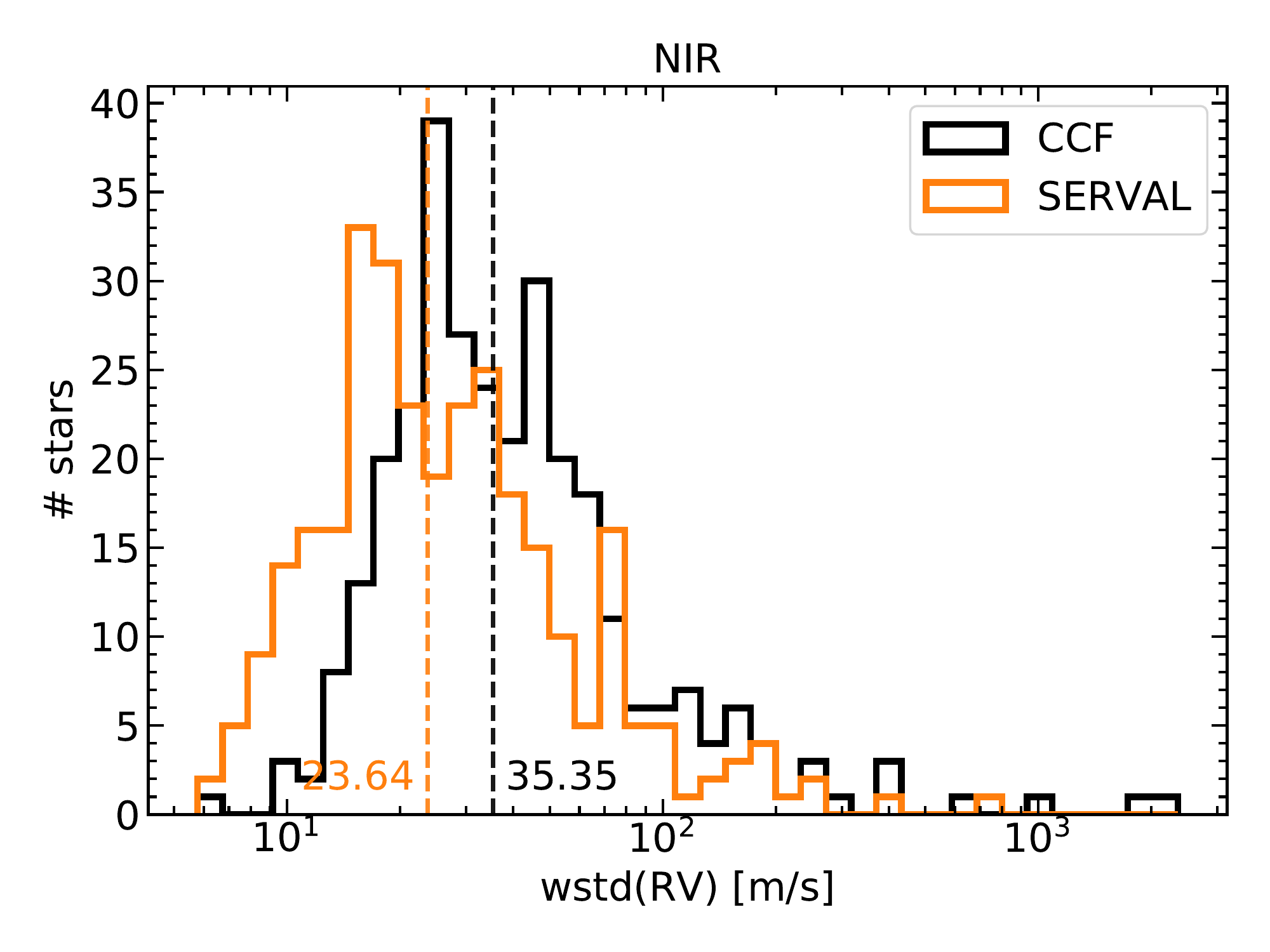}
\end{subfigure}
\caption{Time series RV dispersions (weighted standard deviations) obtained with the CCF (black) and with SERVAL (orange) in the VIS ({\em left}) and NIR ({\em right}) channels. The dashed lines show the median of each distribution.}
\label{fig:comparion_wstdccfrv_wstdservalrv_hist}
\end{figure*}

\begin{figure*}[]
\centering
\begin{subfigure}[b]{0.49\linewidth}
\includegraphics[width=\linewidth]{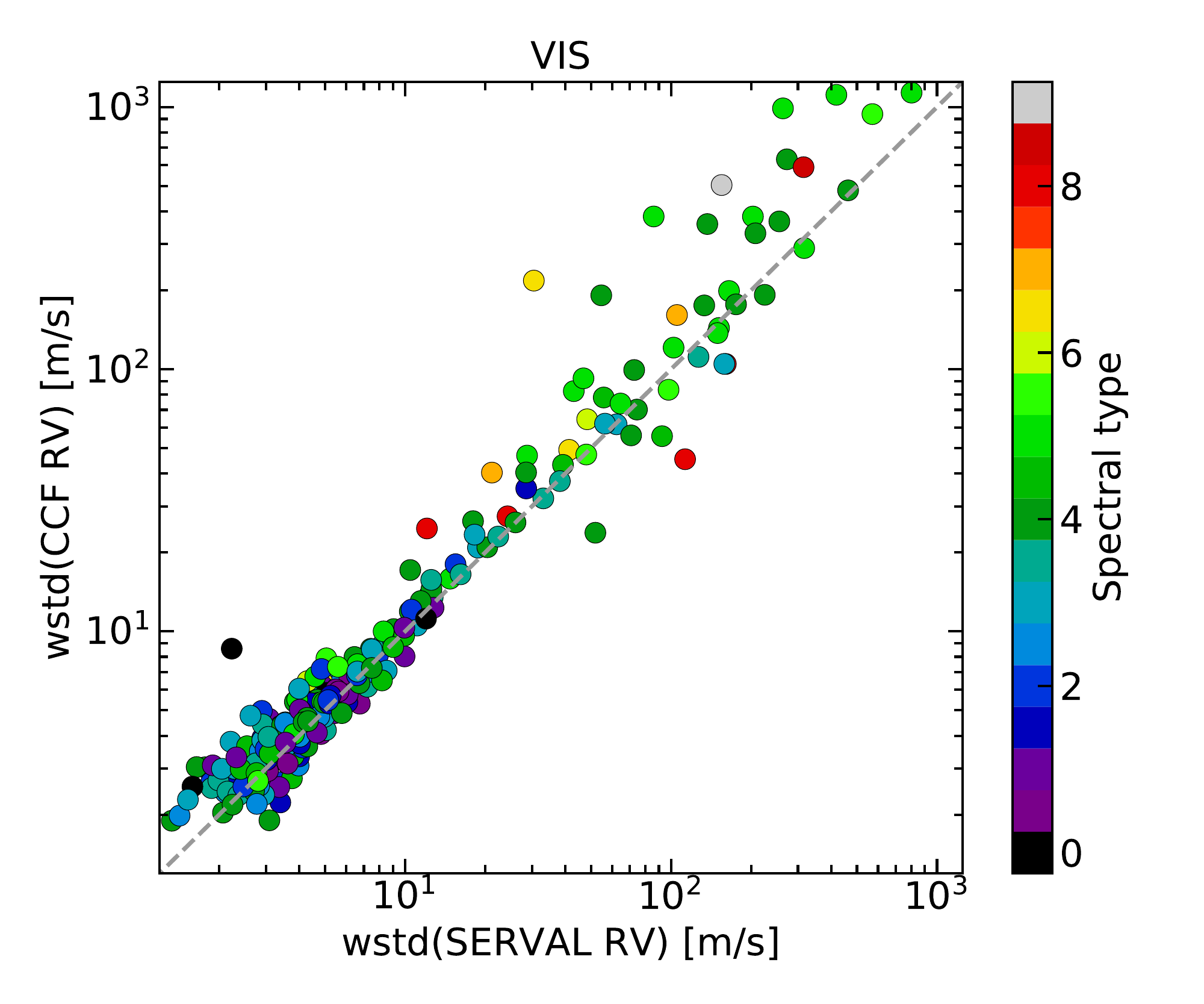}
\end{subfigure}
~
\begin{subfigure}[b]{0.49\linewidth}
\includegraphics[width=\linewidth]{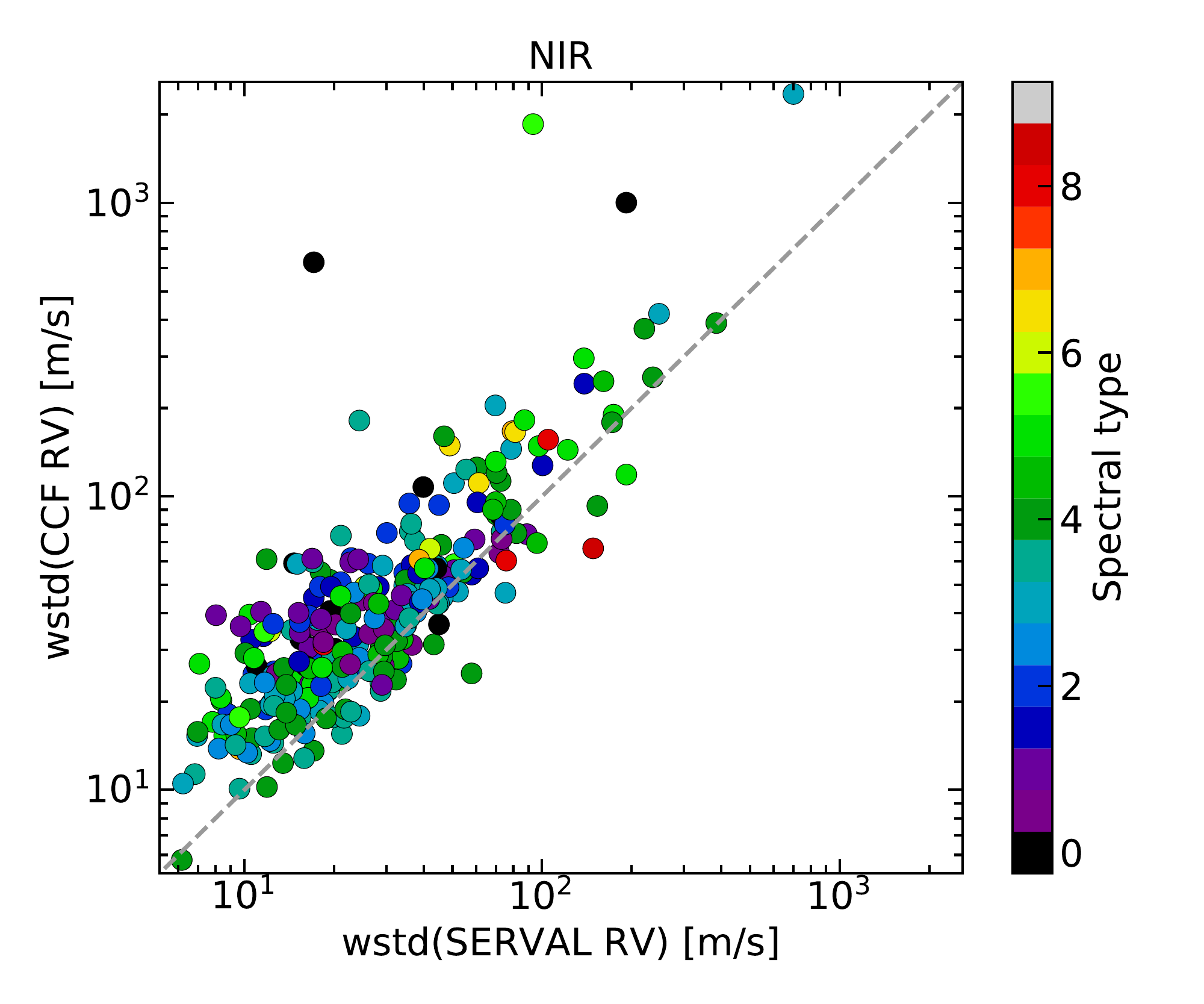}
\end{subfigure}
\caption{Comparison of the time series RV dispersions (weighted standard deviation) obtained with the CCF and SERVAL methods, in the VIS ({\em left}) and NIR ({\em right}), which are colour-coded with spectral type. The grey dashed line indicates the one-to-one relation.}
\label{fig:comparion_wstdccfrv_wstdservalrv_corr}
\end{figure*}

\begin{figure*}[]
\centering
\begin{subfigure}[b]{0.49\linewidth}
\includegraphics[width=\linewidth]{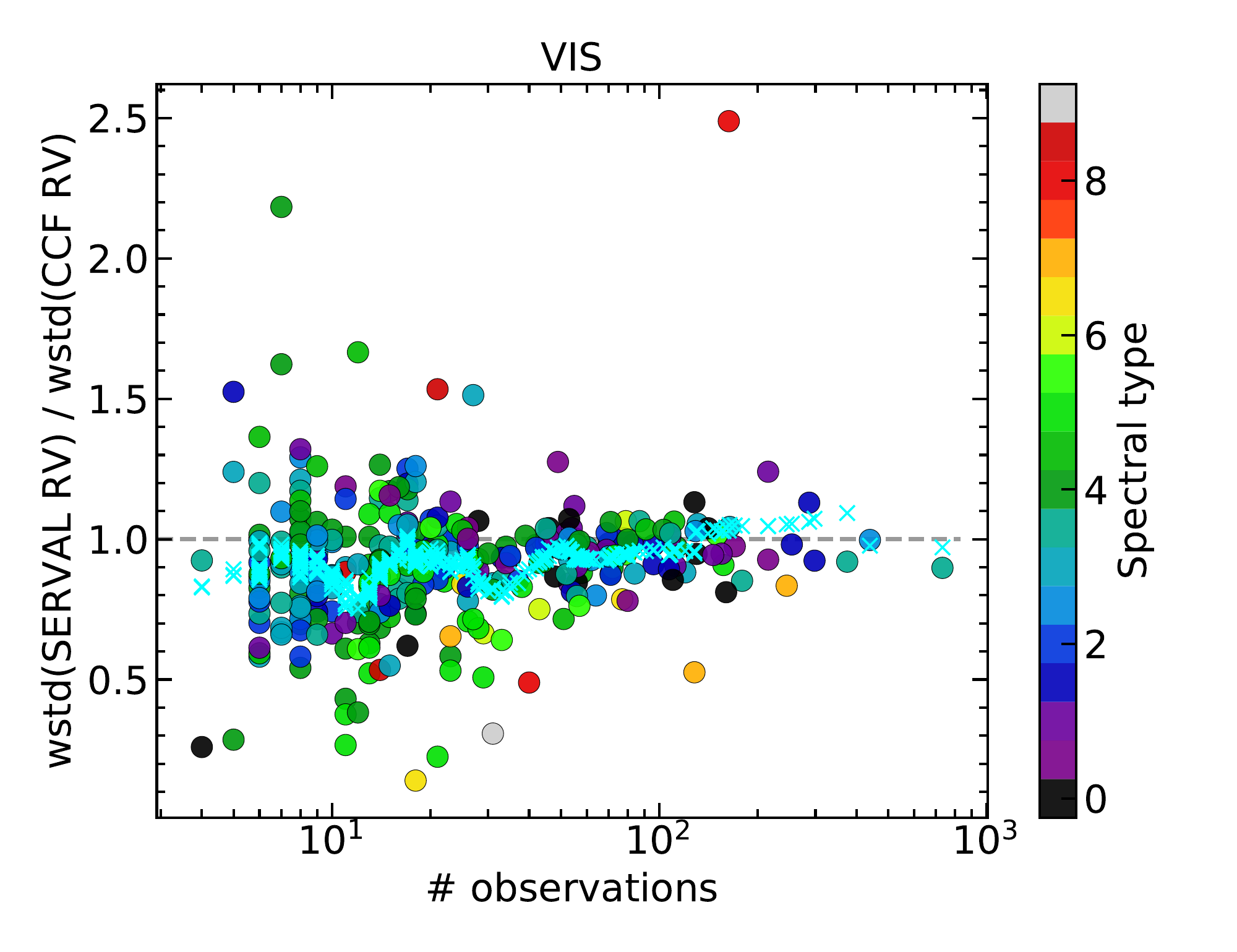}
\end{subfigure}
~
\begin{subfigure}[b]{0.49\linewidth}
\includegraphics[width=\linewidth]{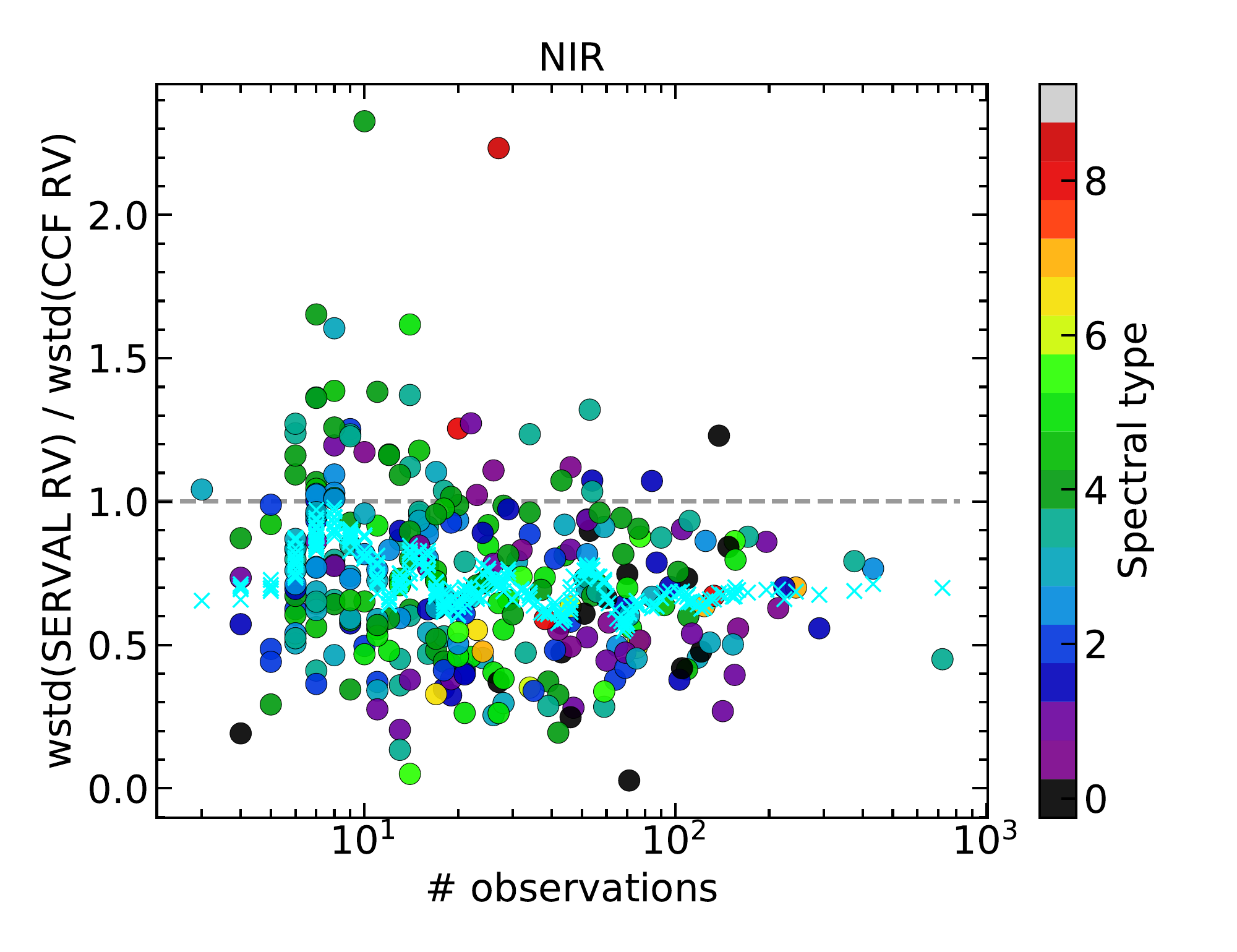}
\end{subfigure}
\caption{Ratio of the SERVAL and CCF RV dispersions (weighted standard deviations) as a function of the number of observations co-added by SERVAL to build the template, in the VIS ({\em left}) and NIR ({\em right}) channels, which are colour-coded with the stellar spectral type. The cyan crosses show the moving average of the data computed over windows of 20 data points.}
\label{fig:wstdccfrv_wstdservalrv_nobs}
\end{figure*}

To analyse the performance of the CCF method regarding the precision of the resulting RVs, we compared them with the velocities computed using the standard CARMENES RV pipeline SERVAL \citep{zechmeister2018serval}.
As mentioned above, SERVAL computes the RV of a series of observations by fitting a high S/N template made from the observations themselves. It matches the template to the observations order-by-order, and the final RV is a weighted average of the RVs of the individual orders.
Since it uses a complete spectrum template, it extracts more information than the CCF method, which only uses a subset of the absorption lines.

By default, SERVAL builds the template from the observations of the star that is being analysed, but it is also possible to use a different template built from observations of another star. Using another template gives better results in the case of faint stars or stars with a small number of observations.
For the CCFs, we only used masks that were created from high S/N templates, except for a few fast rotators where the templates obtained have an S/N lower than the standard masks for slow rotators.
Here, we use the default SERVAL results where a template has been created for each star with its own observations.
All of the analysed stars have been observed at least five times.

Aside from using templates of a different S/N, another difference between both methods is the use of different orders to compute the final RV.
The RV depends on the spectral region used since there is a chromatic dependence.
Also, the RV dispersion appears to be larger in the blue region of the spectral range, where the activity effects are expected to be stronger \cite[see e.g.][]{tal-or2018carmenesRVloud}.
Moreover, ground observations, such as the CARMENES ones, show regions contaminated with tellurics. Some orders are more affected than others, and they may introduce additional scatter due to unidentified telluric features.
In using different sets of orders and in specially removing blue orders in the case of active stars may reduce the RV dispersion.
SERVAL uses the VIS orders 108 to 67, except in some faint and late spectral type targets, where about the 20 to 30 bluest orders are omitted.
In the CCF VIS data, we used orders 108 to 68.\ However, between 5 and 10 orders were removed because the masks do not have enough lines to provide a reliable CCF. This is mainly due to the presence of tellurics. We also removed the bluest orders in the case of faint and late-type targets. Using these regions allowed us to obtain better results.
In the case of the NIR arm, SERVAL mainly uses about 20 orders, which are not contaminated by tellurics, in the {\it Y} and {\it H} bands.
We chose to use the same orders in the NIR CCFs to avoid telluric contamination.

In Fig. \ref{fig:comparion_wstdccfrv_wstdservalrv_hist} we plotted the distribution of the RV time series dispersions obtained with the CCF method and with SERVAL's template matching for each of the survey stars analysed in both VIS and NIR channels. We measured the dispersion as the weighted standard deviation of the RVs, where we used the squared inverse of the RV uncertainties as weights.
We corrected the RVs for an instrumental nightly zero-point drift \citep{trifonov2018carmenes1st,tal-or2019systematichires}. These nightly zero-points were calculated using the SERVAL RVs of stars with small variability observed in each night. We used the SERVAL measurements instead of the CCF ones because they are more precise in general.
Figure \ref{fig:comparion_wstdccfrv_wstdservalrv_corr} shows the correlation of the RV dispersions obtained with the two methods (the same quantities as in Fig. \ref{fig:comparion_wstdccfrv_wstdservalrv_hist}) for the analysed stars. 

In the VIS channel, the RV dispersions of all the stars show a similar distribution for RVs calculated with SERVAL and CCFs. Although, in general, the scatter is slightly larger in the latter case.
This is what we expected because the template matching technique used by SERVAL takes all of the spectral regions into account, rather than the regions around a set of some selected lines as the CCF does.\ Thus SERVAL is able to extract more RV information from the spectra.

For the NIR data, we obtain much larger dispersions than SERVAL in general. This may be caused by the low number of mask lines used (see Table \ref{tab:maskcarmenestargets}), which in turn is due to a large number of regions containing tellurics in the NIR range.
Not having enough lines in the mask results in CCFs with a low S/N and less precise parameters.
We note that the NIR CCF results are not as optimised as in the VIS case, which is the one we used to perform most of our tests due to its better precision compared to the NIR.
A different choice of parameters, both in the mask creation (different line selection values) and the CCF computation (different pixel and order masking) processes may lead to improved results.
In the same vein, we expect the RV precision to be improved by using telluric-corrected spectra as presented in \citet{nagel2019tdtm}.

There are some cases in which the dispersion of the CCF RVs is smaller than for SERVAL.
This occurs for stars where the template has a poorer S/N, that is, in the case of faint stars or stars with a small number of observations.
In Fig. \ref{fig:wstdccfrv_wstdservalrv_nobs}, we plotted the ratio of the RV dispersions obtained with SERVAL and with the CCFs as a function of the number of observations co-added to build the SERVAL template for each star.
We see that for the VIS data, the ratio is larger (which means lower dispersions for the CCF RVs compared to SERVAL RVs) in the case of stars with less than 10--20 observations.
Late spectral type targets, which are faint, also show a large ratio.
In these cases, instead of using a template made from the observations of the target itself, which is the default in SERVAL, we used a high S/N template built from observations of a similar star.\ We obtain RVs that are more precise and that show smaller scatter compared to the results of the default template.

%--------------------------------------------------------------------

\subsection{Activity analysis}\label{sec:carmenesindicators}

\begin{figure*}[]
\centering
\includegraphics[width=\linewidth]{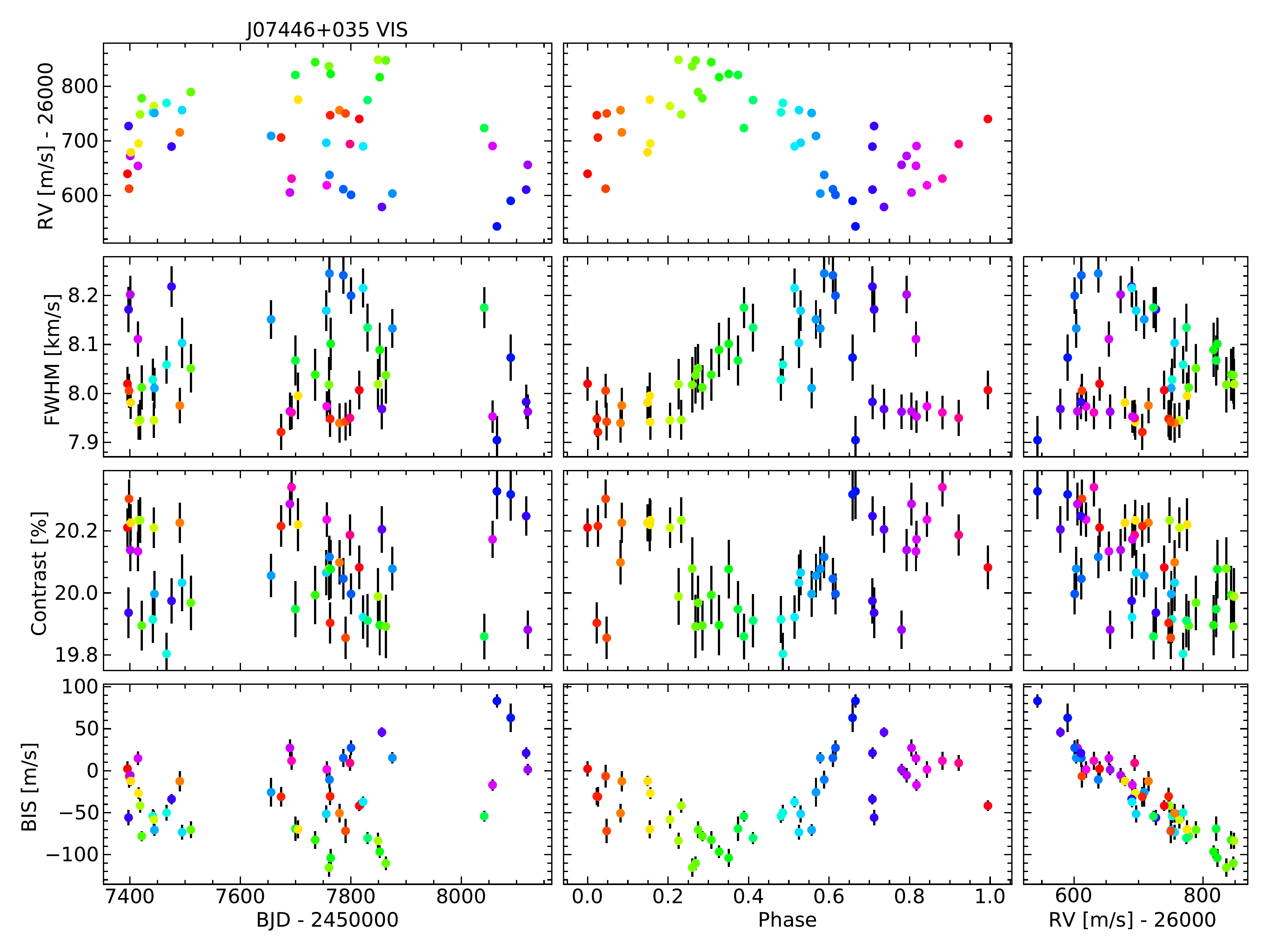}
\caption{
Time series of the four parameters derived from the CCF for the star YZ~CMi in the VIS channel.
We show RV, FWHM, contrast, and BIS (from top to bottom) as a function of barycentric Julian Date ({\em left}), phase of the rotation period ({\em middle}), and also plot the correlation of the activity indicators with the RV ({\em right}). 
The data points are coloured according to the phase.}
\label{fig:ccfpar_yzcmi_vis}
\end{figure*}

\begin{figure*}[]
\includegraphics[width=\linewidth]{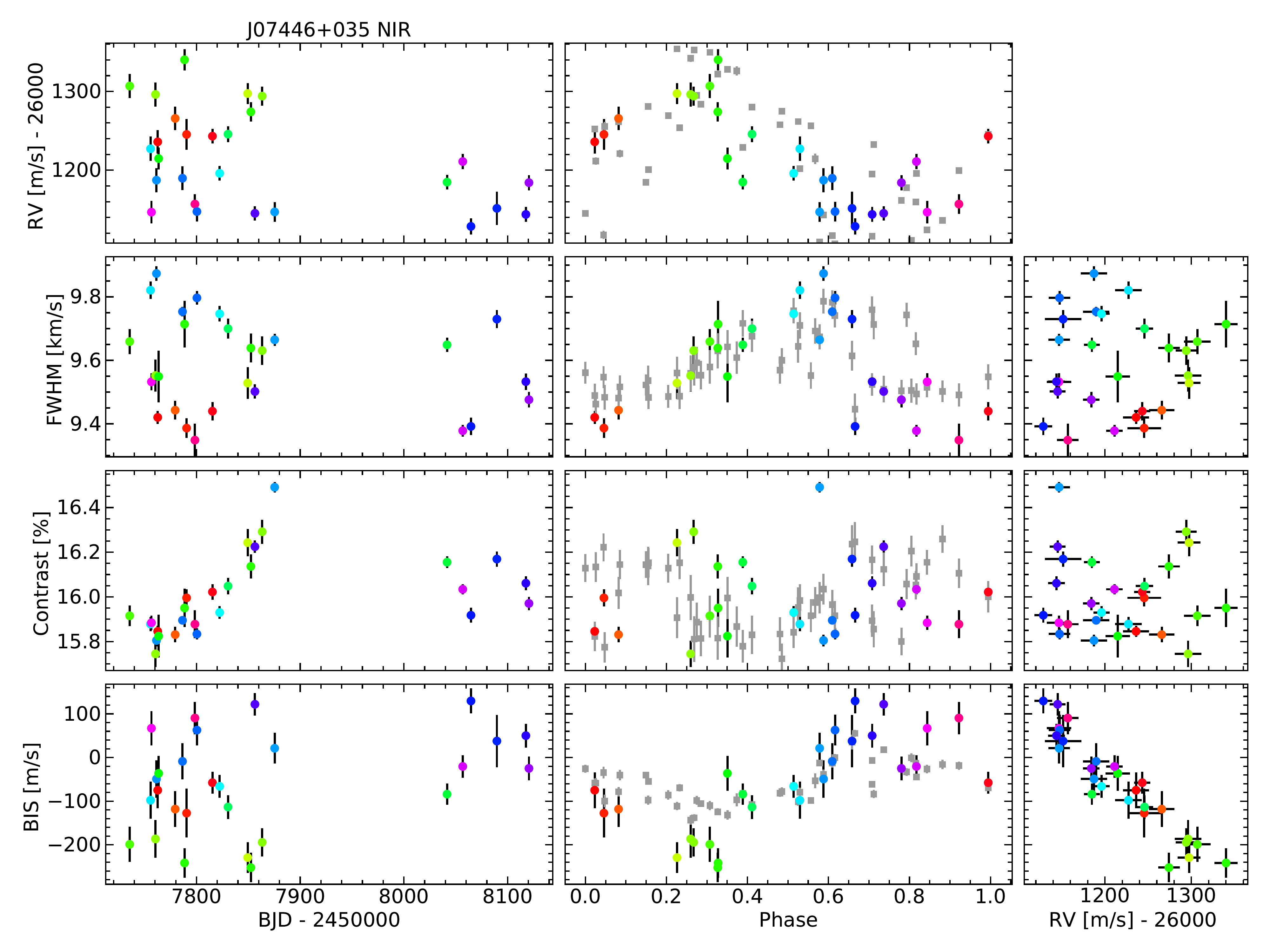}
\caption{Same as Figure \ref{fig:ccfpar_yzcmi_vis}, but for the NIR channel. In the phase-folded plots (central panels), we also plotted the VIS data as grey squares, which are shifted to the mean of the corresponding NIR parameter, to compare the amplitude of the modulation.}
\label{fig:ccfpar_yzcmi_nir}
\end{figure*}

To show an example of the CCF activity indicators (FWHM, contrast, BIS), in Figs. \ref{fig:ccfpar_yzcmi_vis} and \ref{fig:ccfpar_yzcmi_nir}, we plotted the RV time series, together with the three activity indicators, for an active star in the CARMENES survey sample, YZ CMi (GJ~285, Karmn J07446+035), corresponding to observations obtained with the VIS and NIR channels, respectively. YZ CMi is an active mid M dwarf \citep[spectral type M4.5\,V][]{reid1995pmsu} with a rotational velocity of $\vsini \simeq 4\,\kms$ \citep{reiners2018carmenes324} and a rotation period of $P_{\rm rot}\simeq 2.78$\,d \citep{diezalonso2018carmenesRotPhot}.

It shows a large-amplitude modulation in the RV of a few hundreds of \ms peak-to-peak in both channels due to stellar activity.
Both VIS and NIR data show similar behaviour.
The BIS is the indicator that shows the clearest correlation with the RVs. Both parameters show a similar modulation in the phase-folded plot but in opposite phase, and the correlation plot shows a strong anti-correlation, with a Pearson's correlation coefficient of about --0.9.
This anti-correlation was first observed by \citet{queloz2001noplanet} and has since  been  observed in several stars for which the RV modulation contains a signal caused by stellar activity \citep[e.g.][reported correlations in several M dwarfs]{bonfils2013harpsMdwarfsample}.
We also observed such anti-correlations in other active M dwarf stars from the CARMENES sample.
The behaviour of the BIS is similar to the one observed for the chromatic index of YZ CMi, which is a measure of the variation of the RV with the wavelength \citep{zechmeister2018serval, tal-or2018carmenesRVloud}.

The FWHM and the contrast also show a modulation at the stellar rotation period; although, it is not as clear as the BIS one.
The correlation of these two parameters with RV is not very clear either, but there is some loop shape in which the data points follow the stellar rotation phase.
This was already observed for another activity indicator, the differential line width, which is a parameter that measures variations in the stellar line width compared with a stellar template \citep[again see][]{zechmeister2018serval}.

This example shows the value of the CCF parameters to be reliable activity indicators.
These activity proxies, together with other photospheric and chromospheric indicators of activity, are being used in several projects that include CARMENES observations to study stellar activity and to help distinguish between planetary and activity-related signals in RV time series. These projects include both M-dwarf data taken under the CARMENES survey program \citep{kaminski2018carmenesplanetHD180617, luque2018carmenes2warmSEarth, reiners2018carmenes1stneptune, lalitha2019carmenes2planets, nagel2019carmenesGJ4276, perger2019carmeneshadesGl49, schofer2019carmenes4stars} as well as open time observations of other cool stars \citep{luque2018planetK2-292, palle2019epic4}.
We observe that the indicators in different stars display notable differences depending on the spectral type and activity level of the stars. A study of such variation is, however, out of the scope of this paper, and it will be discussed in forthcoming articles.

%--------------------------------------------------------------------

\subsection{Absolute RVs}\label{sec:rvabs}

We estimated absolute RVs of the CARMENES survey targets by using the RV values derived from the CCFs. Absolute RVs are necessary to determine stellar space motions and to carry out studies of galactic kinematics and dynamics.
Since our template masks are calibrated against PHOENIX spectra, our CCF RVs are appropriate to determine the true offset of the wavelengths from their reference values.
The radial velocity that we measured from the wavelength shifts of the absorption lines is displaced from the true motion of the stellar barycentre due to several effects \citep{lindegren2003definitionRV}.
For cool dwarfs, the dominant effects are the gravitational redshift that the photons experience as they escape the potential well, which is created by the mass of the star, and the convective shift due to the motion of hot and cold material in the stellar photosphere.
Therefore, apart from the RV that was directly derived from the CCF measurements, we also need to take these effects into account.
In Appendix \ref{sec:app_absrv}, we describe the values used to compute the absolute RVs and the associated errors in detail, and we show the values obtained in Table \ref{tab:rvabs}.

%--------------------------------------------------------------------
\section{Summary}\label{sec:conclusion}

We present a procedure to build weighted binary masks, compute CCFs of high resolution echelle spectra with these masks, and derive RVs and the following standard CCF activity indicators: FWHM, contrast, and BIS.
In order to build the masks, we used a stellar spectrum template built from observations and we selected the mask lines based on the profile of the local minima present in it.

% ----------------------------

We applied these methods to the CARMENES survey sample, which is a set of more than 300 nearby M dwarfs.
By using CARMENES observations of the brightest and most frequently observed targets, we created several masks covering the different spectral subtypes and rotational velocities of the stars in the sample.
We then used these masks to compute CCFs of the whole sample.
Our tests show that the CCF parameters depend on the mask used, and they tend to be more precise if the mask was built from observations of a star with similar characteristics as the target.
We found that the most important parameter to select a mask was the rotational velocity; results were more sensitive to this than to the spectral subtype of the mask.
For fast rotators, the absorption lines in the spectrum become severely blended, so the number of mask lines that we were able to select decreases from a few thousand of the lowest \vsini masks to only a few hundred, making the CCF and its parameters much less reliable.
Moreover, fast rotators tend to be active stars, so their line profiles are affected by deformations due to activity features on the stellar surface, which add noise to the results.

% ----------------------------

We compared the RVs derived from the CCF method with the ones obtained with the standard CARMENES pipeline SERVAL, which performs a fit of the observation with a spectrum template obtained by co-adding the observations themselves.
As expected, SERVAL generally performs better than the CCF because it is able to exploit more information from the whole spectrum, while the CCF only uses the region around a set of a few selected lines.
However, in some cases, the scatter was smaller in the CCF RVs than in SERVAL. This happened in faint stars or stars with few observations, where the template employed by SERVAL has a low S/N, while the CCF uses masks built from high S/N templates.

% ----------------------------

We show the behaviour of the CCF activity indicators of YZ~CMi, wich is an active mid-type M dwarf that displays a large modulation due to activity in its RV time series.
We found that the BIS shows a clear anti-correlation with the RVs in a similar way as the chromatic index.
On the other hand, the FWHM and contrast show a more complicated correlation with the RVs, but if we consider the stellar rotation phase of the signals, both parameters show phase-shifted correlations.

% ----------------------------

Finally, we estimated the absolute RVs of the CARMENES sample stars using the CCF RVs and discuss the main sources of shifts that affect the M-dwarf spectra.\ This includes the redshift caused by the gravitational potential of the star and the shift due to convective motions in the photosphere.

% ----------------------------

In conclusion, even though other algorithms allow us to obtain more precise RVs in the case of cool stars, the CCF method still offers a fast and straightforward way of obtaining RVs by simply cross-correlating the observed spectrum with a mask, which is a simple model of the absorption lines in the spectrum.
Moreover, we can compute different indicators related to stellar activity from the CCF profile, which have proven to be valuable when studying activity-related signals in RV time series.

%--------------------------------------------------------------------

\begin{acknowledgements}
We thank the anonymous referee for the constructive report.
CARMENES is an instrument for the Centro Astron\'omico Hispano-Alem\'an de
Calar Alto (CAHA, Almer\'{\i}a, Spain).
CARMENES is funded by the German Max-Planck-Gesellschaft (MPG),
the Spanish Consejo Superior de Investigaciones Cient\'{\i}ficas (CSIC),
the European Union through FEDER/ERF FICTS-2011-02 funds,
and the members of the CARMENES Consortium
(Max-Planck-Institut f\"ur Astronomie,
Instituto de Astrof\'{\i}sica de Andaluc\'{\i}a,
Landessternwarte K\"onigstuhl,
Institut de Ci\`encies de l'Espai,
Institut f\"ur Astrophysik G\"ottingen,
Universidad Complutense de Madrid,
Th\"uringer Landessternwarte Tautenburg,
Instituto de Astrof\'{\i}sica de Canarias,
Hamburger Sternwarte,
Centro de Astrobiolog\'{\i}a and
Centro Astron\'omico Hispano-Alem\'an),
with additional contributions by the Spanish Ministry of Economy,
the German Science Foundation through the Major Research Instrumentation
Programme and DFG Research Unit FOR2544 ``Blue Planets around Red Stars'',
the Klaus Tschira Stiftung,
the states of Baden-W\"urttemberg and Niedersachsen,
and by the Junta de Andaluc\'{\i}a.
Based on data from the CARMENES data archive at CAB (INTA-CSIC).
\end{acknowledgements}


% for the bibliography, at the end
\bibliographystyle{aa} % style aa.bst
\bibliography{lafarga.bib} % your references Yourfile.bib

\begin{thebibliography}{65}
\expandafter\ifx\csname natexlab\endcsname\relax\def\natexlab#1{#1}\fi

\bibitem[{Allende~Prieto {et~al.}(2013)Allende~Prieto, Koesterke, Ludwig,
  Freytag, \& Caffau}]{allendeprieto2013ConvectiveshiftGaia}
Allende~Prieto, C., Koesterke, L., Ludwig, H.-G., Freytag, B., \& Caffau, E.
  2013, A\&A, 550, A103

\bibitem[{{Alonso-Floriano} {et~al.}(2015){Alonso-Floriano}, Morales,
  Caballero, Montes, Klutsch, Mundt, {Cort{\'e}s-Contreras}, Ribas, Reiners,
  Amado, Quirrenbach, \& Jeffers}]{alonsofloriano2015carmenesinlowres}
{Alonso-Floriano}, F.~J., Morales, J.~C., Caballero, J.~A., {et~al.} 2015,
  A\&A, 577, A128

\bibitem[{{Anglada-Escud{\'e}} \& Butler(2012)}]{anglada-escude2012harpsTerra}
{Anglada-Escud{\'e}}, G. \& Butler, R.~P. 2012, The Astrophysical Journal
  Supplement Series, 200, 15

\bibitem[{{Astudillo-Defru} {et~al.}(2015){Astudillo-Defru}, Bonfils, Delfosse,
  S{\'e}gransan, Forveille, Bouchy, Gillon, Lovis, Mayor, Neves, Pepe, Perrier,
  Queloz, Rojo, Santos, \& Udry}]{astudillo-defru2015harps3Mdwarfs}
{Astudillo-Defru}, N., Bonfils, X., Delfosse, X., {et~al.} 2015, A\&A, 575,
  A119

\bibitem[{Baluev \& Shaidulin(2015)}]{baluev2015rm}
Baluev, R.~V. \& Shaidulin, V.~S. 2015, Mon. Not. R. Astron. Soc., 454, 4379

\bibitem[{Baranne {et~al.}(1979)Baranne, Mayor, \& Poncet}]{baranne1979Coravel}
Baranne, A., Mayor, M., \& Poncet, J.~L. 1979, Vistas in Astronomy, 23, 279

\bibitem[{Baranne {et~al.}(1996)Baranne, Queloz, Mayor, Adrianzyk, Knispel,
  Kohler, Lacroix, Meunier, Rimbaud, \& Vin}]{baranne1996elodie}
Baranne, A., Queloz, D., Mayor, M., {et~al.} 1996, Astronomy and Astrophysics
  Supplement Series, 119, 373

\bibitem[{Baroch {et~al.}(2018)Baroch, Morales, Ribas, {Tal-Or}, Zechmeister,
  Reiners, Caballero, Quirrenbach, Amado, Dreizler, Lalitha, Jeffers, Lafarga,
  B{\'e}jar, Colom{\'e}, {Cort{\'e}s-Contreras}, {D{\'i}ez-Alonso},
  {Galad{\'i}-Enr{\'i}quez}, Guenther, Hagen, Henning, Herrero, K{\"u}rster,
  Montes, Nagel, Passegger, Perger, Rosich, Schweitzer, \&
  Seifert}]{baroch2018carmenesBinaries}
Baroch, D., Morales, J.~C., Ribas, I., {et~al.} 2018, A\&A, 619, A32

\bibitem[{Bauer {et~al.}(2018)Bauer, Reiners, Beeck, \&
  Jeffers}]{bauer2018ConvBlueshift}
Bauer, F.~F., Reiners, A., Beeck, B., \& Jeffers, S.~V. 2018, A\&A, 610, A52

\bibitem[{Berdi{\~n}as {et~al.}(2016)Berdi{\~n}as, Amado, {Anglada-Escud{\'e}},
  {Rodr{\'i}guez-L{\'o}pez}, \& Barnes}]{berdinas2016systematicsccf}
Berdi{\~n}as, Z.~M., Amado, P.~J., {Anglada-Escud{\'e}}, G.,
  {Rodr{\'i}guez-L{\'o}pez}, C., \& Barnes, J. 2016, Mon. Not. R. Astron. Soc.,
  459, 3551

\bibitem[{Boisse {et~al.}(2011)Boisse, Bouchy, H{\'e}brard, Bonfils, Santos, \&
  Vauclair}]{boisse2011disentangling}
Boisse, I., Bouchy, F., H{\'e}brard, G., {et~al.} 2011, A\&A, 528, A4

\bibitem[{Boisse {et~al.}(2010)Boisse, Eggenberger, Santos, Lovis, Bouchy,
  H{\'e}brard, Arnold, Bonfils, Delfosse, Desort, D{\'i}az, Ehrenreich,
  Forveille, Gallenne, Lagrange, Moutou, Udry, Pepe, Perrier, Perruchot, Pont,
  Queloz, Santerne, S{\'e}gransan, \& {Vidal-Madjar}}]{boisse2010sophieJupiter}
Boisse, I., Eggenberger, A., Santos, N.~C., {et~al.} 2010, A\&A, 523, A88

\bibitem[{Bonfils {et~al.}(2013)Bonfils, Delfosse, Udry, Forveille, Mayor,
  Perrier, Bouchy, Gillon, Lovis, Pepe, Queloz, Santos, S{\'e}gransan, \&
  Bertaux}]{bonfils2013harpsMdwarfsample}
Bonfils, X., Delfosse, X., Udry, S., {et~al.} 2013, Astronomy and Astrophysics,
  549, A109

\bibitem[{Bouchy {et~al.}(2001)Bouchy, Pepe, \& Queloz}]{bouchy2001photonnoise}
Bouchy, F., Pepe, F., \& Queloz, D. 2001, A\&A, 374, 733

\bibitem[{Butler {et~al.}(1996)Butler, Marcy, Williams, McCarthy, Dosanjh, \&
  Vogt}]{butler1996dopplerprecision}
Butler, R.~P., Marcy, G.~W., Williams, E., {et~al.} 1996, Publications of the
  Astronomical Society of the Pacific, 108, 500

\bibitem[{Butler {et~al.}(2004)Butler, Vogt, Marcy, Fischer, Wright, Henry,
  Laughlin, \& Lissauer}]{butler2004gj436}
Butler, R.~P., Vogt, S.~S., Marcy, G.~W., {et~al.} 2004, Astrophys. J., 617,
  580

\bibitem[{{Cort{\'e}s-Contreras} {et~al.}(2017){Cort{\'e}s-Contreras},
  B{\'e}jar, Caballero, Gauza, Montes, {Alonso-Floriano}, Jeffers, Morales,
  Reiners, Ribas, Sch{\"o}fer, Quirrenbach, Amado, Mundt, \&
  Seifert}]{cortescontreras2017carmenesinhighres}
{Cort{\'e}s-Contreras}, M., B{\'e}jar, V. J.~S., Caballero, J.~A., {et~al.}
  2017, A\&A, 597, A47

\bibitem[{Cosentino {et~al.}(2012)Cosentino, Lovis, Pepe, Collier~Cameron,
  Latham, Molinari, Udry, Bezawada, Black, Born, Buchschacher, Charbonneau,
  Figueira, Fleury, Galli, Gallie, Gao, Ghedina, Gonzalez, Gonzalez, Guerra,
  Henry, Horne, Hughes, Kelly, Lodi, Lunney, Maire, Mayor, Micela, Ordway,
  Peacock, Phillips, Piotto, Pollacco, Queloz, Rice, Riverol, Riverol,
  San~Juan, Sasselov, Segransan, Sozzetti, Sosnowska, Stobie, Szentgyorgyi,
  Vick, \& Weber}]{cosentino2012harpsn}
Cosentino, R., Lovis, C., Pepe, F., {et~al.} 2012, Ground-Based Airborne
  Instrum. Astron. IV, 8446, 84461V

\bibitem[{Cretignier {et~al.}(2020)Cretignier, Dumusque, Allart, Pepe, \&
  Lovis}]{cretignier2019indivline}
Cretignier, M., Dumusque, X., Allart, R., Pepe, F., \& Lovis, C. 2020,
  Astronomy and Astrophysics, 633, A76

\bibitem[{Delfosse {et~al.}(1998)Delfosse, Forveille, Mayor, Perrier, Naef, \&
  Queloz}]{delfosse1998gj876}
Delfosse, X., Forveille, T., Mayor, M., {et~al.} 1998, A\&A, 338, L67

\bibitem[{D{\'i}ez~Alonso {et~al.}(2019)D{\'i}ez~Alonso, Caballero, Montes, {de
  Cos Juez}, Dreizler, Dubois, Jeffers, Lalitha, Naves, Reiners, Ribas,
  Vanaverbeke, Amado, B{\'e}jar, {Cort{\'e}s-Contreras}, Herrero, Hidalgo,
  K{\"u}rster, Logie, Quirrenbach, Rau, Seifert, Sch{\"o}fer, \&
  {Tal-Or}}]{diezalonso2018carmenesRotPhot}
D{\'i}ez~Alonso, E., Caballero, J.~A., Montes, D., {et~al.} 2019, A\&A, 621,
  A126

\bibitem[{Dumusque(2018)}]{dumusque2018indivline}
Dumusque, X. 2018, A\&A, 620, A47

\bibitem[{Figueira {et~al.}(2013)Figueira, Santos, Pepe, Lovis, \&
  Nardetto}]{figueira2013lineprofilevariations}
Figueira, P., Santos, N.~C., Pepe, F., Lovis, C., \& Nardetto, N. 2013, A\&A,
  557, A93

\bibitem[{Gray(2009)}]{gray2009thirdsignatureconvection}
Gray, D.~F. 2009, ApJ, 697, 1032

\bibitem[{Husser {et~al.}(2013)Husser, {Wende-von Berg}, Dreizler, Homeier,
  Reiners, Barman, \& Hauschildt}]{husser2013phoenix}
Husser, T.-O., {Wende-von Berg}, S., Dreizler, S., {et~al.} 2013, A\&A, 553, A6

\bibitem[{Kaminski {et~al.}(2018)Kaminski, Trifonov, Caballero, Quirrenbach,
  Ribas, Reiners, Amado, Zechmeister, Dreizler, Perger, {Tal-Or}, Bonfils,
  Mayor, {Astudillo-Defru}, Bauer, B{\'e}jar, Cifuentes, Colom{\'e},
  {Cort{\'e}s-Contreras}, Delfosse, {D{\'i}ez-Alonso}, Forveille, Guenther,
  Hatzes, Henning, Jeffers, K{\"u}rster, Lafarga, Luque, Mandel, Montes,
  Morales, Passegger, Pedraz, Reffert, Sadegi, Schweitzer, Seifert, Stahl, \&
  Udry}]{kaminski2018carmenesplanetHD180617}
Kaminski, A., Trifonov, T., Caballero, J.~A., {et~al.} 2018, A\&A, 618, A115

\bibitem[{K{\"u}rster {et~al.}(2003)K{\"u}rster, Endl, Rouesnel, Els, Kaufer,
  Brillant, Hatzes, Saar, \& Cochran}]{kuerster2003barnard}
K{\"u}rster, M., Endl, M., Rouesnel, F., {et~al.} 2003, A\&A, 403, 1077

\bibitem[{Lalitha {et~al.}(2019)Lalitha, Baroch, Morales, Passegger, Bauer,
  Guill{\'e}n, Dreizler, Oshagh, Reiners, Ribas, Caballero, Quirrenbach, Amado,
  B{\'e}jar, Colom{\'e}, {Cort{\'e}s-Contreras}, {Galad{\'i}-Enr{\'i}quez},
  {Gonz{\'a}lez-Cuesta}, Guenther, Hagen, Henning, Herrero, Husser, Jeffers,
  Kaminski, K{\"u}rster, Lafarga, Lodieu, {L{\'o}pez-Gonz{\'a}lez}, Montes,
  Perger, Rosich, Rodr{\'i}guez, {Rodr{\'i}guez-L{\'o}pez}, Schmitt, {Tal-Or},
  \& Zechmeister}]{lalitha2019carmenes2planets}
Lalitha, S., Baroch, D., Morales, J.~C., {et~al.} 2019, A\&A, 627, A116

\bibitem[{Lanza {et~al.}(2018)Lanza, Malavolta, Benatti, Desidera, Bignamini,
  Bonomo, Esposito, Figueira, Gratton, Scandariato, Damasso, Sozzetti, Biazzo,
  Claudi, Cosentino, Covino, Maggio, Masiero, Micela, Molinari, Pagano, Piotto,
  Poretti, Smareglia, Affer, Boccato, Borsa, Boschin, Giacobbe, Knapic, Leto,
  Maldonado, Mancini, Martinez~Fiorenzano, Messina, Nascimbeni, Pedani, \&
  Rainer}]{lanza2018lineprofilediagnostics}
Lanza, A.~F., Malavolta, L., Benatti, S., {et~al.} 2018, A\&A, 616, A155

\bibitem[{Lindegren \& Dravins(2003)}]{lindegren2003definitionRV}
Lindegren, L. \& Dravins, D. 2003, A\&A, 401, 1185

\bibitem[{Lisogorskyi {et~al.}(2019)Lisogorskyi, Jones, \&
  Feng}]{lisogoroskyi2019activityalphacenb}
Lisogorskyi, M., Jones, H. R.~A., \& Feng, F. 2019, Mon. Not. R. Astron. Soc.,
  485, 4804

\bibitem[{{L{\"o}hner-B{\"o}ttcher} {et~al.}(2019){L{\"o}hner-B{\"o}ttcher},
  Schmidt, Schlichenmaier, Steinmetz, \&
  Holzwarth}]{lohner-bottcher2019ConvBlueSunVis}
{L{\"o}hner-B{\"o}ttcher}, J., Schmidt, W., Schlichenmaier, R., Steinmetz, T.,
  \& Holzwarth, R. 2019, A\&A, 624, A57

\bibitem[{Luque {et~al.}(2019)Luque, Nowak, Pall{\'e}, Dai, Kaminski, Nagel,
  Hidalgo, Bauer, Lafarga, Livingston, Barrag{\'a}n, Hirano, Fridlund,
  Gandolfi, Justesen, Hjorth, Eylen, Winn, Esposito, Morales, Albrecht, Alonso,
  Amado, Beck, Caballero, Cabrera, Cochran, Csizmadia, Deeg, Eigm{\"u}ller,
  Endl, Erikson, Fukui, Grziwa, Guenther, Hatzes, Knudstrup, Korth, Lam, Lund,
  Mathur, {Monta{\~n}es-Rodr{\'i}guez}, Narita, Nespral, Niraula, P{\"a}tzold,
  Persson, {Prieto-Arranz}, Quirrenbach, Rauer, Redfield, Reiners, Ribas, \&
  Smith}]{luque2018planetK2-292}
Luque, R., Nowak, G., Pall{\'e}, E., {et~al.} 2019, A\&A, 623, A114

\bibitem[{Luque {et~al.}(2018)Luque, Nowak, Pall{\'e}, Kossakowski, Trifonov,
  Zechmeister, B{\'e}jar, Guill{\'e}n, {Tal-Or}, Hidalgo, Ribas, Reiners,
  Caballero, Amado, Quirrenbach, Aceituno, {Cort{\'e}s-Contreras},
  {D{\'i}ez-Alonso}, Dreizler, Guenther, Henning, Jeffers, Kaminski,
  K{\"u}rster, Lafarga, Montes, Morales, Passegger, Schmitt, \&
  Schweitzer}]{luque2018carmenes2warmSEarth}
Luque, R., Nowak, G., Pall{\'e}, E., {et~al.} 2018, A\&A, 620, A171

\bibitem[{Marcy {et~al.}(2001)Marcy, Butler, Fischer, Vogt, Lissauer, \&
  Rivera}]{marcy2001gj876}
Marcy, G.~W., Butler, R.~P., Fischer, D., {et~al.} 2001, Astrophys. J., 556,
  296

\bibitem[{Marcy {et~al.}(1998)Marcy, Butler, Vogt, Fischer, \&
  Lissauer}]{marcy1998gj876}
Marcy, G.~W., Butler, R.~P., Vogt, S.~S., Fischer, D., \& Lissauer, J.~J. 1998,
  ApJ, 505, L147

\bibitem[{{Mart{\'i}nez-Rodr{\'i}guez}
  {et~al.}(2019){Mart{\'i}nez-Rodr{\'i}guez}, Caballero, Cifuentes, Piro, \&
  Barnes}]{martinez-rodriguez2019exomoons}
{Mart{\'i}nez-Rodr{\'i}guez}, H., Caballero, J.~A., Cifuentes, C., Piro, A.~L.,
  \& Barnes, R. 2019, ArXiv E-Prints, arXiv:1910.12054

\bibitem[{Mayor {et~al.}(2003)Mayor, Pepe, Queloz, Bouchy, Rupprecht, Lo~Curto,
  Avila, Benz, Bertaux, Bonfils, Dall, Dekker, Delabre, Eckert, Fleury,
  Gilliotte, Gojak, Guzman, Kohler, Lizon, Longinotti, Lovis, Megevand,
  Pasquini, Reyes, Sivan, Sosnowska, Soto, Udry, {van Kesteren}, Weber, \&
  Weilenmann}]{mayor2003harps}
Mayor, M., Pepe, F., Queloz, D., {et~al.} 2003, The Messenger, 114, 20

\bibitem[{Mayor \& Queloz(1995)}]{mayor1995Jup51Peg}
Mayor, M. \& Queloz, D. 1995, Nature, 378, 355

\bibitem[{Meunier {et~al.}(2017)Meunier, Lagrange, Mbemba~Kabuiku, Alex,
  Mignon, \& Borgniet}]{meunier2017VariabilityGranualtionConvective}
Meunier, N., Lagrange, A.-M., Mbemba~Kabuiku, L., {et~al.} 2017, A\&A, 597, A52

\bibitem[{Nagel {et~al.}(2019{\natexlab{a}})Nagel, Czesla, Kaminski,
  Zechmeister, {Tal-Or}, Schmitt, Ribas, Reiners, Quirrenbach, Amado,
  Caballero, Alacid, Bauer, Bejar, {Cortes-Contreras}, Dreizler, Hatzes,
  Jeffers, K{\"u}rster, Lafarga, Montes, Montes, \& Pedraz}]{nagel2019tdtm}
Nagel, E., Czesla, S., Kaminski, A., {et~al.} 2019{\natexlab{a}}, A\&A,
  submitted

\bibitem[{Nagel {et~al.}(2019{\natexlab{b}})Nagel, Czesla, Schmitt, Dreizler,
  {Anglada-Escud{\'e}}, Rodr{\'i}guez, Ribas, Reiners, Quirrenbach, Amado,
  Caballero, Aceituno, B{\'e}jar, {Cort{\'e}s-Contreras},
  {Gonz{\'a}lez-Cuesta}, Guenther, Henning, Jeffers, Kaminski, K{\"u}rster,
  Lafarga, {L{\'o}pez-Gonz{\'a}lez}, Montes, Morales, Passegger,
  {Rodr{\'i}guez-L{\'o}pez}, Schweitzer, \&
  Zechmeister}]{nagel2019carmenesGJ4276}
Nagel, E., Czesla, S., Schmitt, J. H. M.~M., {et~al.} 2019{\natexlab{b}}, A\&A,
  622, A153

\bibitem[{Nardetto {et~al.}(2006)Nardetto, Mourard, Kervella, Mathias,
  M{\'e}rand, \& Bersier}]{nardetto2006lineasymmetry}
Nardetto, N., Mourard, D., Kervella, P., {et~al.} 2006, A\&A, 453, 309

\bibitem[{Pall{\'e} {et~al.}(2019)Pall{\'e}, Nowak, Luque, Hidalgo,
  Barrag{\'a}n, {Prieto-Arranz}, Hirano, Fridlund, Gandolfi, Livingston, Dai,
  Morales, Lafarga, Albrecht, Alonso, Amado, Caballero, Cabrera, Cochran,
  Csizmadia, Deeg, Eigm{\"u}ller, Endl, Erikson, Fukui, Guenther, Grziwa,
  Hatzes, Korth, K{\"u}rster, Kuzuhara, Rodr{\'i}guez, Murgas, Narita, Nespral,
  P{\"a}tzold, Persson, Quirrenbach, Rauer, Redfield, Reiners, Ribas, Smith,
  Eylen, Winn, \& Zechmeister}]{palle2019epic4}
Pall{\'e}, E., Nowak, G., Luque, R., {et~al.} 2019, A\&A, 623, A41

\bibitem[{Pepe {et~al.}(2002)Pepe, Mayor, Galland, Naef, Queloz, Santos, Udry,
  \& Burnet}]{pepe2002coralieSaturns}
Pepe, F., Mayor, M., Galland, F., {et~al.} 2002, A\&A, 388, 632

\bibitem[{Perger {et~al.}(2019)Perger, Scandariato, Ribas, Morales, Affer,
  Azzaro, Amado, {Anglada-Escud{\'e}}, Baroch, Barrado, Bauer, B{\'e}jar,
  Caballero, {Cort{\'e}s-Contreras}, Damasso, Dreizler, {Gonz{\'a}lez-Cuesta},
  Hern{\'a}ndez, Guenther, Henning, Herrero, Jeffers, Kaminski, K{\"u}rster,
  Lafarga, Leto, {L{\'o}pez-Gonz{\'a}lez}, Maldonado, Micela, Montes,
  Pinamonti, Quirrenbach, Rebolo, Reiners, Rodr{\'i}guez,
  {Rodr{\'i}guez-L{\'o}pez}, Schmitt, Sozzetti, Mascare{\~n}o,
  {Toledo-Padr{\'o}n}, S{\'a}nchez, Osorio, \&
  Zechmeister}]{perger2019carmeneshadesGl49}
Perger, M., Scandariato, G., Ribas, I., {et~al.} 2019, A\&A, 624, A123

\bibitem[{Queloz(1995)}]{queloz1995spectroscopylowSNR}
Queloz, D. 1995, New Dev. Array Technol. Appl., 167, 221

\bibitem[{Queloz {et~al.}(2001)Queloz, Henry, Sivan, Baliunas, Beuzit, Donahue,
  Mayor, Naef, Perrier, \& Udry}]{queloz2001noplanet}
Queloz, D., Henry, G.~W., Sivan, J.~P., {et~al.} 2001, A\&A, 379, 279

\bibitem[{Quirrenbach {et~al.}(2016)Quirrenbach, Amado, Caballero, Mundt,
  Reiners, Ribas, Seifert, Abril, Aceituno, {Alonso-Floriano},
  {Anwand-Heerwart}, Azzaro, Bauer, Barrado, Becerril, Bejar, Benitez,
  Berdinas, Brinkm{\"o}ller, Cardenas, Casal, Claret, Colom{\'e},
  {Cortes-Contreras}, Czesla, Doellinger, Dreizler, Feiz, Fernandez, Ferro,
  Fuhrmeister, Galadi, Gallardo, {G{\'a}lvez-Ortiz}, {Garcia-Piquer}, Garrido,
  Gesa, Galera, Hern{\'a}ndez, Peinado, Gr{\"o}zinger, Gu{\`a}rdia, Guenther,
  de~Guindos, Hagen, Hatzes, Hauschildt, Helmling, Henning, Hermann, Arabi,
  Casta{\~n}o, Hernando, Herrero, Huber, Huber, Huke, Jeffers, de~Juan,
  Kaminski, Kehr, Kim, Klein, Kl{\"u}ter, K{\"u}rster, Lafarga, Lara, Lamert,
  Laun, Launhardt, Lemke, Lenzen, Llamas, del Fresno, {L{\'o}pez-Puertas},
  {L{\'o}pez-Santiago}, Salas, Madinabeitia, Mall, Mandel, Mancini, Molina,
  Fern{\'a}ndez, Mart{\'i}n, {Mart{\'i}n-Ruiz}, Marvin, Mathar, Mirabet,
  Montes, Morales, Mu{\~n}oz, Nagel, Naranjo, Nowak, Pall{\'e}, Panduro,
  Passegger, Pavlov, Pedraz, Perez, {P{\'e}rez-Medialdea}, Perger, Pluto,
  Ram{\'o}n, Rebolo, Redondo, Reffert, Reinhart, Rhode, Rix, Rodler,
  Rodr{\'i}guez, L{\'o}pez, Rohloff, Rosich, Carrasco, {Sanz-Forcada}, Sarkis,
  Sarmiento, Sch{\"a}fer, Schiller, Schmidt, Schmitt, Sch{\"o}fer, Schweitzer,
  Shulyak, Solano, Stahl, Storz, Tabernero, Tala, {Tal-Or}, Ulbrich, Veredas,
  Linares, Vilardell, Wagner, Winkler, Osorio, Zechmeister, Eiff,
  {Anglada-Escud{\'e}}, del Burgo, {Garcia-Vargas}, Klutsch, Lizon,
  {Lopez-Morales}, Ofir, {P{\'e}rez-Calpena}, Perryman, {S{\'a}nchez-Blanco},
  Strachan, St{\"u}rmer, Su{\'a}rez, Trifonov, Tulloch, \&
  Xu}]{quirrenbach2016CARMENES}
Quirrenbach, A., Amado, P.~J., Caballero, J.~A., {et~al.} 2016, in Ground-Based
  and {{Airborne Instrumentation}} for {{Astronomy VI}}, Vol. 9908
  ({International Society for Optics and Photonics}), 990812

\bibitem[{Quirrenbach {et~al.}(2018)Quirrenbach, Amado, Ribas, Reiners,
  Caballero, Seifert, Aceituno, Azzaro, Baroch, Barrado, Bauer, Becerril,
  B{\`e}jar, Ben{\'i}tez, Brinkm{\"o}ller, Guill{\'e}n, Cifuentes, Colom{\'e},
  {Cort{\'e}s-Contreras}, Czesla, Dreizler, Fr{\"o}lich, Fuhrmeister,
  {Galad{\'i}-Enr{\'i}quez}, Hern{\'a}ndez, Peinado, Guenther, de~Guindos,
  Hagen, Hatzes, Hauschildt, Helmling, Henning, Herbort, Casta{\~n}o, Herrero,
  Hintz, Jeffers, Johnson, de~Juan, Kaminski, Klahr, K{\"u}rster, Lafarga,
  Sairam, Lamp{\'o}n, Lara, Launhardt, del Fresno, {L{\'o}pez-Puertas}, Luque,
  Mandel, Marfil, Mart{\'i}n, {Mart{\'i}n-Ruiz}, Mathar, Montes, Morales,
  Nagel, Nortmann, Nowak, Pall{\'e}, Passegger, Pavlov, Pedraz,
  {P{\'e}rez-Medialdea}, Perger, Rebolo, Reffert, Rodr{\'i}guez, L{\'o}pez,
  Rosich, Sabotta, Sadegi, Salz, {S{\'a}nchez-L{\'o}pez}, {Sanz-Forcada},
  Sarkis, Sch{\"a}fer, Schiller, Schmitt, Sch{\"o}fer, Schweitzer, Shulyak,
  Solano, Stahl, Pinto, Trifonov, Osorio, Yan, Zechmeister, Abell{\'a}n, Abril,
  {Alonso-Floriano}, Eiff, {Anglada-Escud{\'e}}, {Anwand-Heerwart},
  {Arroyo-Torres}, Berdi{\~n}as, Bergondy, Bl{\"u}mcke, del Burgo, Cano, Carro,
  C{\'a}rdenas, Casal, Claret, {D{\'i}ez-Alonso}, Doellinger, Dorda, Feiz,
  Fern{\'a}ndez, Ferro, Gaisn{\'e}, Gallardo, {G{\'a}lvez-Ortiz},
  {Garc{\'i}a-Piquer}, {Garc{\'i}a-Vargas}, Garrido, Gesa, Galera,
  {Gonz{\'a}lez-{\'A}lvarez}, {Gonz{\'a}lez-Cuesta}, Grohnert, Gr{\"o}zinger,
  Gu{\`a}rdia, Guijarro, Hedrosa, Hermann, Hermelo, Arab{\'i}, Hernando,
  Hidalgo, Holgado, Huber, Huber, Huke, Kehr, Kim, Klein, Kl{\"u}ter, Klutsch,
  Labarga, Labiche, Lamert, Laun, L{\'a}zaro, Lemke, Lenzen, Llamas, Lizon,
  Lodieu, Gonz{\'a}lez, {L{\'o}pez-Morales}, Salas, {L{\'o}pez-Santiago},
  Madinabeitia, Mall, Mancini, Molina, {Mart{\'i}nez-Rodr{\'i}guez},
  Fern{\'a}ndez, Marvin, Mirabet, {Moreno-Raya}, Moya, Mundt, Naranjo, Panduro,
  Pascual, {P{\'e}rez-Calpena}, Perryman, Pluto, Ram{\'o}n, Redondo, Reinhart,
  Rhode, Rix, Rodler, Rohloff, {S{\'a}nchez-Blanco}, Carrasco, Sarmiento,
  Schmidt, Storz, Strachan, St{\"u}rmer, Su{\'a}rez, Tabernero, {Tal-Or},
  Tulloch, Ulbrich, Veredas, Linares, {Vidal-Dasilva}, Vilardell, Wagner,
  Winkler, Wolthoff, Xu, \& Zhao}]{quirrenbach2018CARMENES}
Quirrenbach, A., Amado, P.~J., Ribas, I., {et~al.} 2018, in Ground-Based and
  {{Airborne Instrumentation}} for {{Astronomy VII}}, Vol. 10702
  ({International Society for Optics and Photonics}), 107020W

\bibitem[{Rajpaul {et~al.}(2020)Rajpaul, Aigrain, \&
  Buchhave}]{rajpaul2020rvmethod}
Rajpaul, V.~M., Aigrain, S., \& Buchhave, L.~A. 2020, Monthly Notices of the
  Royal Astronomical Society

\bibitem[{Reid {et~al.}(1995)Reid, Hawley, \& Gizis}]{reid1995pmsu}
Reid, I.~N., Hawley, S.~L., \& Gizis, J.~E. 1995, Astron. J., 110, 1838

\bibitem[{Reiners {et~al.}(2018{\natexlab{a}})Reiners, Ribas, Zechmeister,
  Caballero, Trifonov, Dreizler, Morales, {Tal-Or}, Lafarga, Quirrenbach,
  Amado, Kaminski, Jeffers, Aceituno, B{\'e}jar, Gu{\`a}rdia, Guenther, Hagen,
  Montes, Passegger, Seifert, Schweitzer, {Cort{\'e}s-Contreras}, Abril,
  {Alonso-Floriano}, Eiff, Antona, {Anglada-Escud{\'e}}, {Anwand-Heerwart},
  {Arroyo-Torres}, Azzaro, Baroch, Barrado, Bauer, Becerril, Ben{\'i}tez,
  Berdi{\~n}as, Bergond, Bl{\"u}mcke, Brinkm{\"o}ller, {del Burgo}, Cano,
  C{\'a}rdenas~V{\'a}zquez, Casal, Cifuentes, Claret, Colom{\'e}, Czesla,
  {D{\'i}ez-Alonso}, Feiz, Fern{\'a}ndez, Ferro, Fuhrmeister,
  {Galad{\'i}-Enr{\'i}quez}, {Garcia-Piquer}, Garc{\'i}a~Vargas, Gesa,
  G{\'o}mez~Galera, Gonz{\'a}lez~Hern{\'a}ndez, {Gonz{\'a}lez-Peinado},
  Gr{\"o}zinger, Grohnert, Guijarro, {de Guindos}, {Guti{\'e}rrez-Soto},
  Hatzes, Hauschildt, Hedrosa, Helmling, Henning, Hermelo,
  Hern{\'a}ndez~Arab{\'i}, Hern{\'a}ndez~Casta{\~n}o, Hern{\'a}ndez~Hernando,
  Herrero, Huber, Huke, Johnson, {de Juan}, Kim, Klein, Kl{\"u}ter, Klutsch,
  K{\"u}rster, Labarga, Lamert, Lamp{\'o}n, Lara, Laun, Lemke, Lenzen,
  Launhardt, {L{\'o}pez del Fresno}, {L{\'o}pez-Gonz{\'a}lez},
  {L{\'o}pez-Puertas}, L{\'o}pez~Salas, {L{\'o}pez-Santiago}, Luque,
  Mag{\'a}n~Madinabeitia, Mall, Mancini, Mandel, Marfil, Mar{\'i}n~Molina,
  Maroto~Fern{\'a}ndez, Mart{\'i}n, {Mart{\'i}n-Ruiz}, Marvin, Mathar, Mirabet,
  {Moreno-Raya}, Moya, Mundt, Nagel, Naranjo, Nortmann, Nowak, Ofir, Oreiro,
  Pall{\'e}, Panduro, Pascual, Pavlov, Pedraz, {P{\'e}rez-Calpena},
  P{\'e}rez~Medialdea, Perger, Perryman, Pluto, Rabaza, Ram{\'o}n, Rebolo,
  Redondo, Reffert, Reinhart, Rhode, Rix, Rodler, Rodr{\'i}guez,
  {Rodr{\'i}guez-L{\'o}pez}, Rodr{\'i}guez~Trinidad, Rohloff, Rosich, Sadegi,
  {S{\'a}nchez-Blanco}, S{\'a}nchez~Carrasco, {S{\'a}nchez-L{\'o}pez},
  {Sanz-Forcada}, Sarkis, Sarmiento, Sch{\"a}fer, Schmitt, Schiller,
  Sch{\"o}fer, Solano, Stahl, Strachan, St{\"u}rmer, Su{\'a}rez, Tabernero,
  Tala, Tulloch, Ulbrich, Veredas, Vico~Linares, Vilardell, Wagner, Winkler,
  Wolthoff, Xu, Yan, \& Zapatero~Osorio}]{reiners2018carmenes1stneptune}
Reiners, A., Ribas, I., Zechmeister, M., {et~al.} 2018{\natexlab{a}}, A\&A,
  609, L5

\bibitem[{Reiners {et~al.}(2018{\natexlab{b}})Reiners, Zechmeister, Caballero,
  Ribas, Morales, Jeffers, Sch{\"o}fer, {Tal-Or}, Quirrenbach, Amado, Kaminski,
  Seifert, Abril, Aceituno, {Alonso-Floriano}, {Ammler-von Eiff}, Antona,
  {Anglada-Escud{\'e}}, {Anwand-Heerwart}, {Arroyo-Torres}, Azzaro, Baroch,
  Barrado, Bauer, Becerril, B{\'e}jar, Ben{\'i}tez, Berdi{\~n}as, Bergond,
  Bl{\"u}mcke, Brinkm{\"o}ller, {del Burgo}, Cano, C{\'a}rdenas~V{\'a}zquez,
  Casal, Cifuentes, Claret, Colom{\'e}, {Cort{\'e}s-Contreras}, Czesla,
  {D{\'i}ez-Alonso}, Dreizler, Feiz, Fern{\'a}ndez, Ferro, Fuhrmeister,
  {Galad{\'i}-Enr{\'i}quez}, {Garcia-Piquer}, Garc{\'i}a~Vargas, Gesa, Galera,
  Gonz{\'a}lez~Hern{\'a}ndez, {Gonz{\'a}lez-Peinado}, Gr{\"o}zinger, Grohnert,
  Gu{\`a}rdia, Guenther, Guijarro, de~Guindos, {Guti{\'e}rrez-Soto}, Hagen,
  Hatzes, Hauschildt, Hedrosa, Helmling, Henning, Hermelo,
  Hern{\'a}ndez~Arab{\'i}, Hern{\'a}ndez~Casta{\~n}o, Hern{\'a}ndez~Hernando,
  Herrero, Huber, Huke, Johnson, de~Juan, Kim, Klein, Kl{\"u}ter, Klutsch,
  K{\"u}rster, Lafarga, Lamert, Lamp{\'o}n, Lara, Laun, Lemke, Lenzen,
  Launhardt, {L{\'o}pez del Fresno}, {L{\'o}pez-Gonz{\'a}lez},
  {L{\'o}pez-Puertas}, L{\'o}pez~Salas, {L{\'o}pez-Santiago}, Luque,
  Mag{\'a}n~Madinabeitia, Mall, Mancini, Mandel, Marfil, Mar{\'i}n~Molina,
  Maroto~Fern{\'a}ndez, Mart{\'i}n, {Mart{\'i}n-Ruiz}, Marvin, Mathar, Mirabet,
  Montes, {Moreno-Raya}, Moya, Mundt, Nagel, Naranjo, Nortmann, Nowak, Ofir,
  Oreiro, Pall{\'e}, Panduro, Pascual, Passegger, Pavlov, Pedraz,
  {P{\'e}rez-Calpena}, Medialdea, Perger, Perryman, Pluto, Rabaza, Ram{\'o}n,
  Rebolo, Redondo, Reffert, Reinhart, Rhode, Rix, Rodler, Rodr{\'i}guez,
  {Rodr{\'i}guez-L{\'o}pez}, Rodr{\'i}guez~Trinidad, Rohloff, Rosich, Sadegi,
  {S{\'a}nchez-Blanco}, S{\'a}nchez~Carrasco, {S{\'a}nchez-L{\'o}pez},
  {Sanz-Forcada}, Sarkis, Sarmiento, Sch{\"a}fer, Schmitt, Schiller,
  Schweitzer, Solano, Stahl, Strachan, St{\"u}rmer, Su{\'a}rez, Tabernero,
  Tala, Trifonov, Tulloch, Ulbrich, Veredas, Vico~Linares, Vilardell, Wagner,
  Winkler, Wolthoff, Xu, Yan, \& Zapatero~Osorio}]{reiners2018carmenes324}
Reiners, A., Zechmeister, M., Caballero, J.~A., {et~al.} 2018{\natexlab{b}},
  A\&A, 612, A49

\bibitem[{Sch{\"o}fer {et~al.}(2019)Sch{\"o}fer, Jeffers, Reiners, Zechmeister,
  Fuhrmeister, Lafarga, Ribas, Quirrenbach, Amado, Caballero,
  {Anglada-Escud{\'e}}, Bauer, B{\'e}jar, {Cort{\'e}s-Contreras}, D{\'i}ez,
  Dreizler, Guenther, Herbort, Johnson, Kaminski, K{\"u}rster, Montes, Morales,
  Pedraz, \& {Tal-Or}}]{schofer2019carmenes4stars}
Sch{\"o}fer, P., Jeffers, S.~V., Reiners, A., {et~al.} 2019, A\&A, submitted

\bibitem[{Schweitzer {et~al.}(2019)Schweitzer, Passegger, Cifuentes, B{\'e}jar,
  {Cort{\'e}s-Contreras}, Caballero, del Burgo, Czesla, K{\"u}rster, Montes,
  Osorio, Ribas, Reiners, Quirrenbach, Amado, Aceituno, {Anglada-Escud{\'e}},
  Bauer, Dreizler, Jeffers, Guenther, Henning, Kaminski, Lafarga, Marfil,
  Morales, Schmitt, Seifert, Solano, Tabernero, \&
  Zechmeister}]{schweitzer2019carmenesMR}
Schweitzer, A., Passegger, V.~M., Cifuentes, C., {et~al.} 2019, A\&A, 625, A68

\bibitem[{Simola {et~al.}(2019)Simola, Dumusque, \&
  {Cisewski-Kehe}}]{simola2018ccfskewnormal}
Simola, U., Dumusque, X., \& {Cisewski-Kehe}, J. 2019, A\&A, 622, A131

\bibitem[{Skrutskie {et~al.}(2006)Skrutskie, Cutri, Stiening, Weinberg,
  Schneider, Carpenter, Beichman, Capps, Chester, Elias, Huchra, Liebert,
  Lonsdale, Monet, Price, Seitzer, Jarrett, Kirkpatrick, Gizis, Howard, Evans,
  Fowler, Fullmer, Hurt, Light, Kopan, Marsh, McCallon, Tam, Van~Dyk, \&
  Wheelock}]{skrutskie20062mass}
Skrutskie, M.~F., Cutri, R.~M., Stiening, R., {et~al.} 2006, Astron. J., 131,
  1163

\bibitem[{{Tal-Or} {et~al.}(2019){Tal-Or}, Trifonov, Zucker, Mazeh, \&
  Zechmeister}]{tal-or2019systematichires}
{Tal-Or}, L., Trifonov, T., Zucker, S., Mazeh, T., \& Zechmeister, M. 2019,
  Mon. Not. R. Astron. Soc., 484, L8

\bibitem[{{Tal-Or} {et~al.}(2018){Tal-Or}, Zechmeister, Reiners, Jeffers,
  Sch{\"o}fer, Quirrenbach, Amado, Ribas, Caballero, Aceituno, Bauer,
  B{\'e}jar, Czesla, Dreizler, Fuhrmeister, Hatzes, Johnson, K{\"u}rster,
  Lafarga, Montes, Morales, Reffert, Sadegi, Seifert, \&
  Shulyak}]{tal-or2018carmenesRVloud}
{Tal-Or}, L., Zechmeister, M., Reiners, A., {et~al.} 2018, A\&A, 614, A122

\bibitem[{Trifonov {et~al.}(2018)Trifonov, K{\"u}rster, Zechmeister, {Tal-Or},
  Caballero, Quirrenbach, Amado, Ribas, Reiners, Reffert, Dreizler, Hatzes,
  Kaminski, Launhardt, Henning, Montes, B{\'e}jar, Mundt, Pavlov, Schmitt,
  Seifert, Morales, Nowak, Jeffers, {Rodr{\'i}guez-L{\'o}pez}, {del Burgo},
  {Anglada-Escud{\'e}}, {L{\'o}pez-Santiago}, Mathar, {Ammler-von Eiff},
  Guenther, Barrado, Gonz{\'a}lez~Hern{\'a}ndez, Mancini, St{\"u}rmer, Abril,
  Aceituno, {Alonso-Floriano}, Antona, {Anwand-Heerwart}, {Arroyo-Torres},
  Azzaro, Baroch, Bauer, Becerril, Ben{\'i}tez, Berdi{\~n}as, Bergond,
  Bl{\"u}mcke, Brinkm{\"o}ller, Cano, C{\'a}rdenas~V{\'a}zquez, Casal,
  Cifuentes, Claret, Colom{\'e}, {Cort{\'e}s-Contreras}, Czesla,
  {D{\'i}ez-Alonso}, Feiz, Fern{\'a}ndez, Ferro, Fuhrmeister,
  {Galad{\'i}-Enr{\'i}quez}, {Garcia-Piquer}, Garc{\'i}a~Vargas, Gesa,
  G{\'o}mez~Galera, {Gonz{\'a}lez-Peinado}, Gr{\"o}zinger, Grohnert,
  Gu{\`a}rdia, Guijarro, {de Guindos}, {Guti{\'e}rrez-Soto}, Hagen, Hauschildt,
  Hedrosa, Helmling, Hermelo, Hern{\'a}ndez~Arab{\'i},
  Hern{\'a}ndez~Casta{\~n}o, Hern{\'a}ndez~Hernando, Herrero, Huber, Huke,
  Johnson, {de Juan}, Kim, Klein, Kl{\"u}ter, Klutsch, Lafarga, Lamp{\'o}n,
  Lara, Laun, Lemke, Lenzen, {L{\'o}pez del Fresno}, {L{\'o}pez-Gonz{\'a}lez},
  {L{\'o}pez-Puertas}, L{\'o}pez~Salas, Luque, Mag{\'a}n~Madinabeitia, Mall,
  Mandel, Marfil, Mar{\'i}n~Molina, Maroto~Fern{\'a}ndez, Mart{\'i}n,
  {Mart{\'i}n-Ruiz}, Marvin, Mirabet, Moya, {Moreno-Raya}, Nagel, Naranjo,
  Nortmann, Ofir, Oreiro, Pall{\'e}, Panduro, Pascual, Passegger, Pedraz,
  {P{\'e}rez-Calpena}, P{\'e}rez~Medialdea, Perger, Perryman, Pluto, Rabaza,
  Ram{\'o}n, Rebolo, Redondo, Reinhardt, Rhode, Rix, Rodler, Rodr{\'i}guez,
  Rodr{\'i}guez~Trinidad, Rohloff, Rosich, Sadegi, {S{\'a}nchez-Blanco},
  S{\'a}nchez~Carrasco, {S{\'a}nchez-L{\'o}pez}, {Sanz-Forcada}, Sarkis,
  Sarmiento, Sch{\"a}fer, Schiller, Sch{\"o}fer, Schweitzer, Solano, Stahl,
  Strachan, Su{\'a}rez, Tabernero, Tala, Tulloch, Veredas, Vico~Linares,
  Vilardell, Wagner, Winkler, Wolthoff, Xu, Yan, \&
  Zapatero~Osorio}]{trifonov2018carmenes1st}
Trifonov, T., K{\"u}rster, M., Zechmeister, M., {et~al.} 2018, A\&A, 609, A117

\bibitem[{{van Kerkwijk} {et~al.}(1995){van Kerkwijk}, {van Paradijs},
  Zuiderwijk, {Hammerschlag-Hensberge}, Kaper, \&
  Sterken}]{vankerkwijk1995hd77581vela}
{van Kerkwijk}, M.~H., {van Paradijs}, J., Zuiderwijk, E.~J., {et~al.} 1995,
  Astronomy and Astrophysics, 303, 483

\bibitem[{Wise {et~al.}(2018)Wise, {Dodson-Robinson}, Bevenour, \&
  Provini}]{wise2018activelines}
Wise, A.~W., {Dodson-Robinson}, S.~E., Bevenour, K., \& Provini, A. 2018,
  Astron. J., 156, 180

\bibitem[{Zechmeister {et~al.}(2014)Zechmeister, {Anglada-Escud{\'e}}, \&
  Reiners}]{zechmeister2014fox}
Zechmeister, M., {Anglada-Escud{\'e}}, G., \& Reiners, A. 2014, A\&A, 561, A59

\bibitem[{Zechmeister {et~al.}(2018)Zechmeister, Reiners, Amado, Azzaro, Bauer,
  B{\'e}jar, Caballero, Guenther, Hagen, Jeffers, Kaminski, K{\"u}rster,
  Launhardt, Montes, Morales, Quirrenbach, Reffert, Ribas, Seifert, {Tal-Or},
  \& Wolthoff}]{zechmeister2018serval}
Zechmeister, M., Reiners, A., Amado, P.~J., {et~al.} 2018, A\&A, 609, A12

\end{thebibliography}

\begin{appendix}
        
\section{Absolute RVs}\label{sec:app_absrv}

To compute the absolute RVs of the CARMENES survey stars, $\mathrm{RV_{abs}}$, we followed
\begin{equation}\label{eq:rvabs}
\mathrm{RV_{abs}} = \mathrm{RV_{CCF}} + v_{\rm grav} + v_{\rm conv},
\end{equation}
where $\mathrm{RV_{CCF}}$ is the RV directly derived from the CCF measurements, $v_{\rm grav}$ is the gravitational shift of the star, and $v_{\rm conv}$ is the convective shift of the stellar photosphere.

We computed $\mathrm{RV_{CCF}}$ as the weighted mean of the RV time series of each target.
Ideally, we would add in the model of each time series signal from companions and stellar activity on top of the RV zero point.
For simplicity, we ignored those effects, which are mostly short-term (and therefore averaged out in long time series) and at the \ms level.
Exceptions include single-lined spectroscopic binaries with long periods where the system velocity has not been established yet. Double-lined spectroscopic binaries in the sample were not considered here; \cite{baroch2018carmenesBinaries} already provided absolute RVs for them.
We used the nightly zero-point corrected VIS data and removed the observations without drift correction (except in stars without any instrumental drift estimation).
We estimated the uncertainty arising from the dispersion of the different RV measurements of each star, $\sigma_{\rm CCF}$, as their weighted standard deviation divided by the square root of the number of measurements.

A systematic error source when estimating the absolute RV shift is the zero-point of the CCF masks.
As explained in Sect. \ref{sec:maskrvabs}, in order to transform the masks to an absolute reference, we cross-correlated them with PHOENIX synthetic models, which we assumed to have a zero offset, and we also used the RV derived from the best-fit Gaussian to the CCF to correct the mask shift.
There is uncertainty as to the mask zero-point, $\sigma_{\rm mask}$, because of mismatches between the mask and the model lines. To measure this uncertainty, we used the formal error of the Gaussian fit, which has typical values of $\sim 30\,\ms$ for a mask made from low \vsini targets, and it increases by some tens of \ms for the fastest rotators.

% --------------------------------------

The gravitational redshift of a photon escaping from the surface of a star of mass $M$ and radius $R$ is given by
\begin{equation}\label{eq:grav}
v_{\mathrm{grav}} = \frac{G M }{c R},
\end{equation}
where $G$ is the gravitational constant and $c$ is the speed of light.
We used this equation with stellar mass and radius values from \cite{schweitzer2019carmenesMR} to estimate the gravitational redshift, and we assumed the uncertainties derived by error propagation from Eq. \ref{eq:grav}.

% --------------------------------------

The net velocity shift due to convective motions in the photosphere is more difficult to quantify.
In the Sun, the effect of the upward motion of the granules and the downward motion of the intergranular lanes results in a net average blueshift of about $400\,\ms$.
However, the exact value depends on the characteristics of the line used, such as the line strength, the excitation potential, or the formation depth in the atmosphere.
There is also a limb effect: The value of the net shift changes from the centre to the limb of the solar disc \citep{lohner-bottcher2019ConvBlueSunVis}.
For other cool stars, 3D hydrodynamical atmosphere models that compute the convective shift using several lines in a specific wavelength range \citep{allendeprieto2013ConvectiveshiftGaia}, as well as observations of specific spectral lines \citep{meunier2017VariabilityGranualtionConvective}, also show a dependence of the convective shift on the line or lines used. In particular, lines at shorter wavelengths experience larger blueshifts than those at longer wavelengths, and shallower lines (formed deeper in the atmosphere) are also more blueshifted than stronger lines.
The effective temperature of the star also affects the convective shift \citep{bauer2018ConvBlueshift} from blueshift values of $~500\,\ms$ for early G-type stars decreasing until values of $~100\,\ms$ for late K-type stars.
The blueshift also decreases with the metallicity, the surface gravity, and the activity level of the star.

We found neither theoretical nor observational studies that delved into the effects of convective blueshift in M dwarfs.
Therefore, we extrapolated the results obtained for earlier spectral type stars.
We expect the convective shift in M dwarfs to be small since they have lower effective temperatures and can have large levels of activity. There even seems to be evidence of a reversed convection pattern, showing a convective redshift instead of a blueshift \citep{kuerster2003barnard}.
Due to the difficulty in estimating a value for the convective shift $v_{\rm conv}$ for our target stars, we assumed it to be zero with an uncertainty $\sigma_{\rm conv}$ of 100 \ms.

% --------------------------------------

The final absolute RV values, $\mathrm{RV_{abs}}$, obtained using Eq.~\ref{eq:rvabs}, are shown in Table \ref{tab:rvabs}.
The $\mathrm{RV_{abs}}$ uncertainties are the quadratic sum of the following four sources of error that we considered: the error of the RV measurements dispersion $\sigma_{\rm CCF}$, the error from the gravitational redshift $ \sigma_{\rm grav}$, 100 \ms to account for the uncorrected convective shift $\sigma_{\rm conv}$ (which for most targets is the dominating one), and the error of the zero-point of the mask $\sigma_{\rm mask}$
\begin{equation}
\sigma_{\rm abs}^2 = \sigma_{\rm CCF}^2 + \sigma_{\rm grav}^2 + \sigma_{\rm conv}^2 + \sigma_{\rm mask}^2.
\end{equation}
The same table also contains the average RV from the CCF measurements, $\mathrm{RV_{CCF}}$ (which already includes the correction of the mask zero-point), together with the combined uncertainty of the RV scatter $\sigma_{\rm CCF}$ and the mask zero-point $\sigma_{\rm mask}$ given by their quadratic sum.
We also list these two uncertainties separately.
Finally, we also give the gravitational redshift estimates together with their uncertainties.

\longtab[1]{
\begin{longtable}{lccccc}
\caption{Absolute RV ($\mathrm{RV_{abs}}$), weighted mean RV from the VIS CCF measurements ($\mathrm{RV_{CCF}}$) and gravitational redshift ($v_{\rm grav}$).\tablefootmark{a}}\\
\hline\hline
Karmn & $\mathrm{RV_{abs}}$ & $\mathrm{RV_{CCF}}$ & $\sigma_{\rm CCF}$ & $\sigma_{\rm mask}$ & $v_{\rm grav}$ \\
 & [\ms] & [\ms] & [\ms] & [\ms] & [\ms] \\
\hline
\endfirsthead
\caption{continued.}\\
\toprule
Karmn & $\mathrm{RV_{abs}}$ & $\mathrm{RV_{CCF}}$ & $\sigma_{\rm CCF}$ & $\sigma_{\rm mask}$ & $v_{\rm grav}$ \\
& [\ms] & [\ms] & [\ms] & [\ms] & [\ms] \\
\hline
\endhead
\hline
\endfoot
J00051+457  &    $-1203\pm 107$ &     $-561\pm 23$ &                  0.52 &                 23 &   $642\pm 31$ \\
J00067-075  &   $-40679\pm 113$ &   $-40123\pm 20$ &                  0.38 &                 20 &   $556\pm 48$ \\
J00162+198E &    $-2242\pm 109$ &    $-1625\pm 24$ &                  0.69 &                 24 &   $617\pm 35$ \\
J00183+440  &    $10849\pm 107$ &    $11481\pm 23$ &                  0.20 &                 23 &   $632\pm 31$ \\
J00184+440  &    $10089\pm 111$ &    $10676\pm 25$ &                  0.19 &                 25 &   $587\pm 40$ \\
J00286-066  &   $-13342\pm 108$ &   $-12711\pm 24$ &                  0.51 &                 24 &   $631\pm 33$ \\
J00389+306  &    $-1261\pm 108$ &     $-626\pm 25$ &                  0.53 &                 25 &   $634\pm 32$ \\
J00570+450  &     $5644\pm 108$ &     $6271\pm 25$ &                  0.65 &                 25 &   $627\pm 33$ \\
J01013+613  &     $6606\pm 108$ &     $7237\pm 23$ &                  1.07 &                 23 &   $631\pm 32$ \\
J01019+541  &    $-5788\pm 168$ &   $-5203\pm 128$ &                 87.71 &                 93 &   $585\pm 43$ \\
J01025+716  &      $606\pm 108$ &     $1246\pm 25$ &                  0.29 &                 25 &   $640\pm 32$ \\
J01026+623  &    $-6945\pm 107$ &    $-6304\pm 23$ &                  0.56 &                 23 &   $642\pm 30$ \\
J01033+623  &    $-7132\pm 131$ &    $-6521\pm 77$ &                 60.38 &                 47 &   $610\pm 38$ \\
J01048-181  &    $10799\pm 111$ &    $11373\pm 20$ &                  1.12 &                 20 &   $574\pm 44$ \\
J01125-169  &    $27271\pm 112$ &    $27846\pm 23$ &                  0.31 &                 23 &   $574\pm 44$ \\
J01339-176  &     $5504\pm 109$ &     $6120\pm 24$ &                  1.87 &                 24 &   $616\pm 35$ \\
J01352-072  &     $8685\pm 237$ &    $9335\pm 212$ &                190.67 &                 93 &   $650\pm 33$ \\
J01433+043  &   $-26866\pm 108$ &   $-26232\pm 23$ &                  1.69 &                 23 &   $633\pm 33$ \\
J01518+644  &   $-13703\pm 255$ &   $-13063\pm 25$ &                  0.97 &                 25 &  $640\pm 233$ \\
J02002+130  &   $-29411\pm 111$ &   $-28832\pm 25$ &                  1.93 &                 25 &   $579\pm 42$ \\
J02015+637  &   $-85028\pm 109$ &   $-84391\pm 25$ &                  0.80 &                 25 &   $637\pm 35$ \\
J02070+496  &    $18109\pm 108$ &    $18735\pm 25$ &                  1.12 &                 25 &   $625\pm 33$ \\
J02088+494  &   $-10368\pm 123$ &    $-9735\pm 64$ &                 26.87 &                 58 &   $633\pm 33$ \\
J02123+035  &    $-3526\pm 107$ &    $-2889\pm 23$ &                  0.46 &                 23 &   $637\pm 31$ \\
J02222+478  &   $-39154\pm 106$ &   $-38511\pm 18$ &                  0.63 &                 18 &   $643\pm 30$ \\
J02336+249  &    $-7267\pm 109$ &    $-6662\pm 24$ &                  2.32 &                 24 &   $605\pm 37$ \\
J02358+202  &       $19\pm 108$ &      $661\pm 23$ &                  0.91 &                 23 &   $642\pm 33$ \\
J02362+068  &    $25356\pm 109$ &    $25971\pm 24$ &                  0.51 &                 24 &   $615\pm 35$ \\
J02442+255  &    $29675\pm 108$ &    $30304\pm 25$ &                  0.37 &                 25 &   $629\pm 33$ \\
J02519+224  &     $8501\pm 131$ &     $9142\pm 78$ &                 51.95 &                 58 &   $640\pm 32$ \\
J02530+168  &    $67889\pm 118$ &    $68417\pm 28$ &                  0.28 &                 28 &   $528\pm 56$ \\
J02565+554W &    $75667\pm 107$ &    $76313\pm 23$ &                  1.32 &                 23 &   $645\pm 29$ \\
J03133+047  &    $27572\pm 110$ &    $28156\pm 20$ &                  0.76 &                 20 &   $585\pm 42$ \\
J03181+382  &    $-5201\pm 107$ &    $-4557\pm 23$ &                  0.65 &                 23 &   $644\pm 29$ \\
J03213+799  &   $-14220\pm 108$ &   $-13586\pm 23$ &                  0.62 &                 23 &   $634\pm 32$ \\
J03217-066  &    $25486\pm 107$ &    $26125\pm 23$ &                  1.54 &                 23 &   $639\pm 32$ \\
J03463+262  &    $34941\pm 105$ &    $35585\pm 18$ &                  0.55 &                 18 &   $644\pm 28$ \\
J03473-019  &    $17235\pm 107$ &    $17876\pm 21$ &                  6.19 &                 20 &   $641\pm 32$ \\
J03531+625  &  $-120947\pm 108$ &  $-120319\pm 25$ &                  0.74 &                 25 &   $628\pm 33$ \\
J04153-076  &   $-44197\pm 109$ &   $-43578\pm 24$ &                  1.23 &                 24 &   $619\pm 36$ \\
J04198+425  &    $20940\pm 132$ &    $21486\pm 70$ &                 23.46 &                 66 &   $546\pm 49$ \\
J04219+213  &    $18174\pm 106$ &    $18819\pm 19$ &                  5.97 &                 18 &   $645\pm 31$ \\
J04225+105  &    $36078\pm 108$ &    $36715\pm 25$ &                  0.70 &                 25 &   $637\pm 32$ \\
J04290+219  &   $-36205\pm 105$ &   $-35558\pm 14$ &                  0.35 &                 14 &   $647\pm 30$ \\
J04311+589  &    $27810\pm 109$ &    $28429\pm 24$ &                  0.94 &                 24 &   $619\pm 36$ \\
J04376+528  &    $33268\pm 105$ &    $33913\pm 14$ &                  0.45 &                 14 &   $645\pm 28$ \\
J04376-110  &    $-7749\pm 107$ &    $-7111\pm 23$ &                  0.65 &                 23 &   $638\pm 31$ \\
J04429+189  &    $25248\pm 107$ &    $25889\pm 23$ &                  0.96 &                 23 &   $641\pm 32$ \\
J04429+214  &     $2171\pm 108$ &     $2803\pm 25$ &                  1.19 &                 25 &   $632\pm 33$ \\
J04472+206  &    $23345\pm 330$ &   $23976\pm 312$ &                297.85 &                 93 &   $631\pm 35$ \\
J04520+064  &    $-9747\pm 108$ &    $-9118\pm 25$ &                  0.91 &                 25 &   $629\pm 34$ \\
J04538-177  &   $-15460\pm 107$ &   $-14826\pm 23$ &                  0.71 &                 23 &   $634\pm 32$ \\
J04588+498  &   $-34999\pm 106$ &   $-34354\pm 18$ &                  1.46 &                 18 &   $645\pm 28$ \\
J05019+011  &    $19060\pm 110$ &    $19706\pm 31$ &                 23.35 &                 21 &   $645\pm 33$ \\
J05019-069  &    $41378\pm 110$ &    $41967\pm 24$ &                  1.26 &                 24 &   $589\pm 40$ \\
J05033-173  &    $14575\pm 109$ &    $15190\pm 25$ &                  1.05 &                 25 &   $616\pm 35$ \\
J05062+046  &    $18442\pm 130$ &    $19083\pm 76$ &                 48.50 &                 58 &   $641\pm 33$ \\
J05084-210  &    $21563\pm 253$ &   $22214\pm 230$ &                222.38 &                 58 &   $650\pm 35$ \\
J05127+196  &   $-26001\pm 108$ &   $-25365\pm 23$ &                  0.62 &                 23 &   $636\pm 32$ \\
J05280+096  &    $59697\pm 109$ &    $60305\pm 25$ &                  0.78 &                 25 &   $608\pm 36$ \\
J05314-036  &     $7685\pm 114$ &     $8329\pm 23$ &                  0.48 &                 23 &   $644\pm 51$ \\
J05348+138  &    $36788\pm 108$ &    $37416\pm 25$ &                  0.57 &                 25 &   $629\pm 33$ \\
J05360-076  &    $28462\pm 108$ &    $29086\pm 24$ &                  0.68 &                 24 &   $625\pm 34$ \\
J05365+113  &    $20869\pm 106$ &    $21514\pm 18$ &                  1.08 &                 18 &   $645\pm 29$ \\
J05366+112  &    $20912\pm 109$ &    $21530\pm 24$ &                  2.68 &                 24 &   $618\pm 35$ \\
J05394+406  &    $-6877\pm 118$ &    $-6338\pm 29$ &                  8.98 &                 28 &   $539\pm 55$ \\
J05415+534  &     $1176\pm 107$ &     $1818\pm 23$ &                  0.60 &                 23 &   $642\pm 29$ \\
J05421+124  &   $105065\pm 109$ &   $105673\pm 24$ &                  0.34 &                 24 &   $608\pm 36$ \\
J06000+027  &    $29294\pm 109$ &    $29903\pm 25$ &                  6.74 &                 24 &   $609\pm 37$ \\
J06011+595  &     $1073\pm 109$ &     $1687\pm 25$ &                  0.34 &                 25 &   $614\pm 35$ \\
J06024+498  &    $19687\pm 111$ &    $20259\pm 20$ &                  0.70 &                 20 &   $572\pm 44$ \\
J06103+821  &    $-2582\pm 107$ &    $-1947\pm 23$ &                  0.60 &                 23 &   $635\pm 32$ \\
J06105-218  &     $3539\pm 108$ &     $4182\pm 18$ &                  0.60 &                 18 &   $642\pm 36$ \\
J06246+234  &   $-12244\pm 111$ &   $-11656\pm 24$ &                  1.06 &                 24 &   $587\pm 41$ \\
J06318+414  &     $3717\pm 162$ &    $4352\pm 123$ &                 79.83 &                 93 &   $635\pm 35$ \\
J06371+175  &   $-59392\pm 106$ &   $-58754\pm 18$ &                  0.70 &                 18 &   $638\pm 31$ \\
J06396-210  &    $-8039\pm 117$ &    $-7426\pm 24$ &                  3.77 &                 24 &   $613\pm 55$ \\
J06421+035  &    $81557\pm 108$ &    $82187\pm 25$ &                  0.50 &                 25 &   $630\pm 33$ \\
J06548+332  &    $22025\pm 108$ &    $22654\pm 25$ &                  0.22 &                 25 &   $629\pm 33$ \\
J06574+740  &    $-3328\pm 182$ &   $-2688\pm 105$ &                 87.98 &                 58 &  $641\pm 110$ \\
J06594+193  &   $-30072\pm 112$ &   $-29510\pm 20$ &                  0.76 &                 20 &   $562\pm 47$ \\
J07033+346  &     $1415\pm 109$ &     $2029\pm 24$ &                  4.42 &                 24 &   $613\pm 36$ \\
J07044+682  &   $-51400\pm 108$ &   $-50766\pm 25$ &                  0.94 &                 25 &   $634\pm 33$ \\
J07274+052  &    $17274\pm 109$ &    $17895\pm 25$ &                  0.11 &                 25 &   $622\pm 35$ \\
J07287-032  &      $548\pm 108$ &     $1180\pm 25$ &                  0.84 &                 25 &   $632\pm 32$ \\
J07319+362N &    $-1705\pm 108$ &    $-1071\pm 25$ &                  1.02 &                 25 &   $634\pm 33$ \\
J07353+548  &   $-15901\pm 108$ &   $-15270\pm 23$ &                  1.22 &                 23 &   $631\pm 32$ \\
J07386-212  &   $-29804\pm 109$ &   $-29183\pm 25$ &                  1.66 &                 25 &   $621\pm 34$ \\
J07393+021  &    $18884\pm 106$ &    $19529\pm 18$ &                  0.65 &                 18 &   $645\pm 29$ \\
J07403-174  &   $-29340\pm 119$ &   $-28811\pm 32$ &                  1.25 &                 32 &   $528\pm 55$ \\
J07446+035  &    $26077\pm 111$ &    $26707\pm 32$ &                 11.35 &                 30 &   $630\pm 36$ \\
J07472+503  &   $-15387\pm 116$ &   $-14771\pm 47$ &                  6.74 &                 47 &   $615\pm 35$ \\
J07558+833  &     $7027\pm 119$ &     $7642\pm 53$ &                 24.29 &                 47 &   $615\pm 36$ \\
J07582+413  &   $-21958\pm 109$ &   $-21342\pm 25$ &                  0.42 &                 25 &   $616\pm 35$ \\
J08119+087  &    $13833\pm 111$ &    $14416\pm 24$ &                  0.67 &                 24 &   $583\pm 42$ \\
J08126-215  &     $8065\pm 109$ &     $8680\pm 24$ &                  0.86 &                 24 &   $615\pm 35$ \\
J08161+013  &    $61229\pm 107$ &    $61866\pm 23$ &                  0.51 &                 23 &   $637\pm 32$ \\
J08293+039  &    $21863\pm 108$ &    $22504\pm 25$ &                  0.91 &                 25 &   $641\pm 31$ \\
J08298+267  &     $9372\pm 121$ &     $9936\pm 48$ &                 10.06 &                 47 &   $564\pm 49$ \\
J08315+730  &   $-91460\pm 109$ &   $-90841\pm 24$ &                  1.16 &                 24 &   $618\pm 35$ \\
J08358+680  &    $18234\pm 108$ &    $18865\pm 25$ &                  0.82 &                 25 &   $630\pm 33$ \\
J08402+314  &    $66074\pm 109$ &    $66692\pm 25$ &                  1.05 &                 25 &   $618\pm 34$ \\
J08409-234  &    $86913\pm 108$ &    $87547\pm 26$ &                  7.84 &                 25 &   $635\pm 33$ \\
J08413+594  &     $5560\pm 112$ &     $6122\pm 20$ &                  3.89 &                 20 &   $561\pm 47$ \\
J08526+283  &    $26561\pm 109$ &    $27176\pm 24$ &                  1.39 &                 24 &   $614\pm 36$ \\
J08536-034  &    $-7712\pm 157$ &   $-7225\pm 100$ &                 90.68 &                 42 &   $487\pm 68$ \\
J09003+218  &     $7551\pm 132$ &     $8114\pm 71$ &                 53.57 &                 47 &   $564\pm 49$ \\
J09005+465  &     $1387\pm 110$ &     $1989\pm 24$ &                  0.98 &                 24 &   $601\pm 38$ \\
J09028+680  &   $-25629\pm 109$ &   $-25014\pm 24$ &                  1.32 &                 24 &   $615\pm 35$ \\
J09033+056  &   $-28696\pm 118$ &   $-28118\pm 44$ &                 33.75 &                 28 &   $578\pm 45$ \\
J09133+688  &    $13001\pm 159$ &    $13647\pm 25$ &                  1.01 &                 25 &  $646\pm 121$ \\
J09140+196  &    $12245\pm 109$ &    $12885\pm 29$ &                 14.98 &                 25 &   $640\pm 32$ \\
J09143+526  &    $10041\pm 108$ &    $10686\pm 18$ &                  0.91 &                 18 &   $645\pm 35$ \\
J09144+526  &    $11253\pm 107$ &    $11898\pm 18$ &                  0.55 &                 18 &   $645\pm 33$ \\
J09161+018  &   $-13372\pm 118$ &   $-12750\pm 53$ &                 23.68 &                 47 &   $622\pm 35$ \\
J09163-186  &    $13927\pm 107$ &    $14566\pm 23$ &                  1.59 &                 23 &   $639\pm 32$ \\
J09307+003  &    $45245\pm 109$ &    $45867\pm 25$ &                  1.22 &                 25 &   $622\pm 34$ \\
J09360-216  &   $-35459\pm 108$ &   $-34833\pm 25$ &                  0.79 &                 25 &   $627\pm 34$ \\
J09411+132  &    $10519\pm 107$ &    $11158\pm 23$ &                  0.74 &                 23 &   $639\pm 31$ \\
J09423+559  &    $14366\pm 108$ &    $14996\pm 25$ &                  1.64 &                 25 &   $631\pm 33$ \\
J09425+700  &     $6058\pm 108$ &     $6699\pm 23$ &                  1.36 &                 23 &   $641\pm 32$ \\
J09428+700  &     $5718\pm 108$ &     $6354\pm 25$ &                  1.62 &                 25 &   $636\pm 33$ \\
J09439+269  &    $33547\pm 108$ &    $34179\pm 25$ &                  3.84 &                 25 &   $632\pm 33$ \\
J09447-182  &     $7720\pm 109$ &     $8337\pm 24$ &                  0.44 &                 24 &   $618\pm 35$ \\
J09449-123  &    $16332\pm 379$ &   $16959\pm 364$ &                352.21 &                 93 &   $627\pm 36$ \\
J09468+760  &   $-28735\pm 107$ &   $-28094\pm 23$ &                  0.94 &                 23 &   $641\pm 30$ \\
J09511-123  &    $61213\pm 106$ &    $61853\pm 18$ &                  0.81 &                 18 &   $641\pm 30$ \\
J09561+627  &    $14273\pm 106$ &    $14918\pm 18$ &                  0.81 &                 18 &   $645\pm 29$ \\
J10023+480  &   $-10758\pm 107$ &   $-10113\pm 23$ &                  1.38 &                 23 &   $645\pm 29$ \\
J10122-037  &     $6958\pm 107$ &     $7600\pm 23$ &                  0.65 &                 23 &   $642\pm 31$ \\
J10125+570  &    $-4405\pm 108$ &    $-3782\pm 25$ &                  0.96 &                 25 &   $623\pm 34$ \\
J10167-119  &   $-11454\pm 108$ &   $-10813\pm 25$ &                  0.99 &                 25 &   $640\pm 32$ \\
J10196+198  &    $11650\pm 107$ &    $12286\pm 21$ &                  4.56 &                 20 &   $636\pm 32$ \\
J10251-102  &    $20621\pm 107$ &    $21262\pm 23$ &                  1.16 &                 23 &   $641\pm 30$ \\
J10289+008  &     $7370\pm 108$ &     $8005\pm 23$ &                  0.31 &                 23 &   $636\pm 32$ \\
J10350-094  &    $13693\pm 108$ &    $14324\pm 25$ &                  0.90 &                 25 &   $631\pm 32$ \\
J10360+051  &    $20185\pm 108$ &    $20813\pm 25$ &                  4.91 &                 25 &   $628\pm 34$ \\
J10396-069  &     $2335\pm 108$ &     $2973\pm 25$ &                  1.35 &                 25 &   $639\pm 32$ \\
J10416+376  &    $-2467\pm 109$ &    $-1861\pm 24$ &                  1.86 &                 24 &   $606\pm 37$ \\
J10482-113  &     $1183\pm 117$ &     $1740\pm 32$ &                  0.71 &                 32 &   $557\pm 51$ \\
J10504+331  &   $-60280\pm 154$ &  $-59646\pm 112$ &                109.38 &                 24 &   $634\pm 33$ \\
J10508+068  &    $-1958\pm 109$ &    $-1344\pm 24$ &                  0.36 &                 24 &   $614\pm 36$ \\
J10564+070  &    $18874\pm 116$ &    $19442\pm 32$ &                  0.69 &                 32 &   $568\pm 48$ \\
J10584-107  &    $10176\pm 109$ &    $10778\pm 20$ &                  2.72 &                 20 &   $602\pm 38$ \\
J11000+228  &     $2256\pm 243$ &     $2888\pm 25$ &                  0.27 &                 25 &  $632\pm 220$ \\
J11026+219  &   $-14881\pm 107$ &   $-14238\pm 23$ &                  1.55 &                 23 &   $643\pm 29$ \\
J11033+359  &   $-85644\pm 112$ &   $-85016\pm 23$ &                  0.16 &                 23 &   $629\pm 44$ \\
J11054+435  &    $67948\pm 108$ &    $68578\pm 23$ &                  0.36 &                 23 &   $631\pm 32$ \\
J11055+435  &    $68605\pm 114$ &    $69163\pm 23$ &                  1.35 &                 23 &   $558\pm 49$ \\
J11110+304W &   $-16331\pm 107$ &   $-15688\pm 23$ &                  0.64 &                 23 &   $643\pm 29$ \\
J11126+189  &    $30575\pm 107$ &    $31218\pm 23$ &                  1.15 &                 23 &   $643\pm 31$ \\
J11201-104  &    $11382\pm 107$ &    $12026\pm 23$ &                  4.00 &                 23 &   $643\pm 31$ \\
J11289+101  &    $36185\pm 109$ &    $36802\pm 25$ &                  0.90 &                 25 &   $617\pm 35$ \\
J11302+076  &     $-436\pm 108$ &      $200\pm 25$ &                  0.78 &                 25 &   $637\pm 32$ \\
J11306-080  &    $15804\pm 108$ &    $16432\pm 25$ &                  0.65 &                 25 &   $628\pm 33$ \\
J11417+427  &    $-9970\pm 108$ &    $-9341\pm 24$ &                  2.82 &                 24 &   $629\pm 33$ \\
J11421+267  &     $8699\pm 108$ &     $9335\pm 25$ &                  1.11 &                 25 &   $636\pm 32$ \\
J11467-140  &    $19555\pm 108$ &    $20196\pm 25$ &                  0.54 &                 25 &   $641\pm 31$ \\
J11474+667  &   $-10029\pm 112$ &    $-9404\pm 36$ &                 29.74 &                 20 &   $625\pm 36$ \\
J11476+002  &     $5310\pm 109$ &     $5938\pm 25$ &                  5.84 &                 24 &   $627\pm 35$ \\
J11476+786  &  $-112578\pm 109$ &  $-111965\pm 25$ &                  0.44 &                 25 &   $613\pm 36$ \\
J11477+008  &   $-31968\pm 110$ &   $-31375\pm 24$ &                  0.48 &                 24 &   $593\pm 40$ \\
J11509+483  &   $-36201\pm 110$ &   $-35613\pm 24$ &                  0.99 &                 24 &   $588\pm 40$ \\
J11511+352  &     $-694\pm 107$ &      $-56\pm 23$ &                  0.29 &                 23 &   $638\pm 31$ \\
J12054+695  &     $4638\pm 108$ &     $5260\pm 24$ &                  1.49 &                 24 &   $621\pm 34$ \\
J12100-150  &    $79580\pm 108$ &    $80210\pm 25$ &                  0.77 &                 25 &   $630\pm 33$ \\
J12111-199  &   $-10005\pm 108$ &    $-9376\pm 25$ &                  0.85 &                 25 &   $629\pm 33$ \\
J12123+544S &   $-18313\pm 106$ &   $-17668\pm 18$ &                  0.36 &                 18 &   $645\pm 29$ \\
J12156+526  &   $-10192\pm 171$ &   $-9548\pm 134$ &                 96.72 &                 93 &   $644\pm 33$ \\
J12189+111  &     $5105\pm 121$ &     $5681\pm 53$ &                 24.47 &                 47 &   $576\pm 44$ \\
J12230+640  &     $8223\pm 108$ &     $8863\pm 25$ &                  0.91 &                 25 &   $640\pm 31$ \\
J12248-182  &    $50238\pm 108$ &    $50853\pm 23$ &                  0.68 &                 23 &   $615\pm 34$ \\
J12312+086  &    $17945\pm 106$ &    $18589\pm 18$ &                  0.70 &                 18 &   $643\pm 32$ \\
J12350+098  &    $32375\pm 108$ &    $33014\pm 25$ &                  1.44 &                 25 &   $639\pm 31$ \\
J12373-208  &     $8457\pm 108$ &     $9092\pm 24$ &                  0.87 &                 24 &   $634\pm 33$ \\
J12388+116  &    $-5129\pm 108$ &    $-4491\pm 25$ &                  1.51 &                 25 &   $638\pm 32$ \\
J12428+418  &    $-5316\pm 108$ &    $-4684\pm 24$ &                  3.05 &                 24 &   $632\pm 33$ \\
J12479+097  &    $18149\pm 109$ &    $18772\pm 25$ &                  0.47 &                 25 &   $623\pm 34$ \\
J13005+056  &   $-25926\pm 119$ &   $-25330\pm 50$ &                 16.10 &                 47 &   $596\pm 40$ \\
J13102+477  &   $-13033\pm 110$ &   $-12446\pm 20$ &                  1.26 &                 20 &   $587\pm 41$ \\
J13196+333  &   $-12669\pm 107$ &   $-12025\pm 23$ &                  0.85 &                 23 &   $643\pm 29$ \\
J13209+342  &   $-36398\pm 238$ &   $-35756\pm 23$ &                  0.54 &                 23 &  $642\pm 214$ \\
J13229+244  &   $-20177\pm 109$ &   $-19560\pm 24$ &                  0.67 &                 24 &   $617\pm 35$ \\
J13283-023W &   $-40370\pm 108$ &   $-39734\pm 25$ &                  1.29 &                 25 &   $637\pm 33$ \\
J13293+114  &    $27343\pm 108$ &    $27980\pm 25$ &                  1.55 &                 25 &   $637\pm 33$ \\
J13299+102  &    $13357\pm 106$ &    $13997\pm 18$ &                  0.22 &                 18 &   $640\pm 30$ \\
J13427+332  &     $5834\pm 109$ &     $6449\pm 25$ &                  0.73 &                 25 &   $615\pm 35$ \\
J13450+176  &    $19625\pm 106$ &    $20265\pm 18$ &                  0.85 &                 18 &   $640\pm 29$ \\
J13457+148  &    $14832\pm 108$ &    $15471\pm 23$ &                  0.21 &                 23 &   $639\pm 34$ \\
J13458-179  &     $4008\pm 108$ &     $4634\pm 25$ &                  0.88 &                 25 &   $627\pm 34$ \\
J13536+776  &    $-8208\pm 109$ &    $-7591\pm 24$ &                 10.59 &                 21 &   $617\pm 36$ \\
J13582+125  &   $-10737\pm 109$ &   $-10131\pm 25$ &                  1.35 &                 25 &   $606\pm 37$ \\
J13591-198  &   $-17102\pm 109$ &   $-16488\pm 25$ &                  5.50 &                 24 &   $615\pm 36$ \\
J14010-026  &   $-26785\pm 107$ &   $-26144\pm 23$ &                  0.71 &                 23 &   $641\pm 30$ \\
J14082+805  &     $6416\pm 107$ &     $7061\pm 23$ &                  1.01 &                 23 &   $645\pm 31$ \\
J14152+450  &    $13247\pm 108$ &    $13884\pm 25$ &                  1.24 &                 25 &   $637\pm 32$ \\
J14173+454  &     $2489\pm 123$ &     $3108\pm 62$ &                 40.20 &                 47 &   $619\pm 36$ \\
J14251+518  &   $-11013\pm 108$ &   $-10380\pm 25$ &                  0.74 &                 25 &   $633\pm 32$ \\
J14257+236E &     $7315\pm 106$ &     $7960\pm 18$ &                  0.52 &                 18 &   $645\pm 28$ \\
J14257+236W &     $8236\pm 105$ &     $8882\pm 18$ &                  0.57 &                 18 &   $646\pm 28$ \\
J14294+155  &     $6871\pm 107$ &     $7512\pm 23$ &                  1.65 &                 23 &   $640\pm 31$ \\
J14307-086  &   $-23181\pm 105$ &   $-22534\pm 18$ &                  0.39 &                 18 &   $647\pm 28$ \\
J14310-122  &    $-2709\pm 108$ &    $-2085\pm 25$ &                  1.65 &                 25 &   $624\pm 34$ \\
J14321+081  &   $-22601\pm 114$ &   $-22021\pm 34$ &                 10.17 &                 32 &   $580\pm 44$ \\
J14342-125  &    $-2361\pm 109$ &    $-1742\pm 24$ &                  0.43 &                 24 &   $618\pm 35$ \\
J14524+123  &     $4760\pm 108$ &     $5403\pm 23$ &                  2.08 &                 23 &   $643\pm 33$ \\
J14544+355  &   $-41692\pm 108$ &   $-41063\pm 25$ &                  0.88 &                 25 &   $629\pm 33$ \\
J15013+055  &    $-6899\pm 108$ &    $-6271\pm 25$ &                  0.56 &                 25 &   $628\pm 33$ \\
J15095+031  &   $-33423\pm 108$ &   $-32786\pm 25$ &                  1.28 &                 25 &   $637\pm 32$ \\
J15194-077  &   $-10284\pm 109$ &    $-9662\pm 25$ &                  1.32 &                 25 &   $622\pm 34$ \\
J15218+209  &     $6071\pm 107$ &     $6716\pm 21$ &                  4.82 &                 20 &   $645\pm 31$ \\
J15305+094  &      $856\pm 121$ &     $1434\pm 52$ &                 22.09 &                 47 &   $578\pm 45$ \\
J15369-141  &     $1475\pm 108$ &     $2098\pm 24$ &                  0.31 &                 24 &   $623\pm 34$ \\
J15499+796  &   $-16902\pm 145$ &   $-16285\pm 98$ &                 78.70 &                 58 &   $617\pm 38$ \\
J15598-082  &   $-18112\pm 107$ &   $-17472\pm 23$ &                  1.45 &                 23 &   $640\pm 32$ \\
J16028+205  &     $5570\pm 109$ &     $6182\pm 24$ &                  1.06 &                 24 &   $612\pm 36$ \\
J16092+093  &   $-45766\pm 108$ &   $-45135\pm 25$ &                  1.42 &                 25 &   $631\pm 32$ \\
J16102-193  &    $-7655\pm 109$ &    $-7002\pm 31$ &                 20.08 &                 23 &   $653\pm 32$ \\
J16167+672N &   $-19418\pm 108$ &   $-18780\pm 25$ &                  0.35 &                 25 &   $639\pm 33$ \\
J16167+672S &   $-20011\pm 106$ &   $-19364\pm 18$ &                  0.43 &                 18 &   $647\pm 30$ \\
J16254+543  &   $-14003\pm 108$ &   $-13379\pm 23$ &                  0.52 &                 23 &   $624\pm 33$ \\
J16303-126  &   $-22180\pm 109$ &   $-21560\pm 25$ &                  0.40 &                 25 &   $621\pm 35$ \\
J16313+408  &   $-23752\pm 116$ &   $-23156\pm 43$ &                 36.36 &                 23 &   $596\pm 41$ \\
J16327+126  &   $-33704\pm 108$ &   $-33071\pm 25$ &                  1.12 &                 25 &   $633\pm 32$ \\
J16462+164  &    $17908\pm 108$ &    $18544\pm 25$ &                  0.42 &                 25 &   $636\pm 32$ \\
J16554-083N &    $14895\pm 110$ &    $15498\pm 25$ &                  0.69 &                 25 &   $604\pm 37$ \\
J16555-083  &    $13863\pm 117$ &    $14409\pm 28$ &                  3.74 &                 28 &   $546\pm 53$ \\
J16570-043  &    $-4575\pm 116$ &    $-3963\pm 47$ &                  5.69 &                 47 &   $612\pm 36$ \\
J16581+257  &     $3350\pm 107$ &     $3991\pm 23$ &                  0.54 &                 23 &   $642\pm 31$ \\
J17033+514  &    $36589\pm 110$ &    $37182\pm 24$ &                  0.99 &                 24 &   $593\pm 39$ \\
J17052-050  &    $33847\pm 107$ &    $34485\pm 23$ &                  0.55 &                 23 &   $639\pm 31$ \\
J17071+215  &   $-51697\pm 108$ &   $-51064\pm 25$ &                  1.27 &                 25 &   $633\pm 33$ \\
J17115+384  &   $-45409\pm 108$ &   $-44776\pm 25$ &                  0.48 &                 25 &   $632\pm 33$ \\
J17166+080  &   $-31559\pm 107$ &   $-30925\pm 23$ &                  1.59 &                 23 &   $634\pm 32$ \\
J17198+417  &   $-20404\pm 108$ &   $-19774\pm 25$ &                  1.03 &                 25 &   $629\pm 33$ \\
J17303+055  &   $-13669\pm 106$ &   $-13026\pm 18$ &                  0.48 &                 18 &   $643\pm 29$ \\
J17338+169  &   $-22504\pm 306$ &  $-21873\pm 287$ &                271.31 &                 93 &   $631\pm 38$ \\
J17355+616  &   $-16082\pm 106$ &   $-15440\pm 18$ &                  0.82 &                 18 &   $643\pm 29$ \\
J17378+185  &   $-10488\pm 108$ &    $-9853\pm 23$ &                  0.32 &                 23 &   $636\pm 32$ \\
J17542+073  &   $-29236\pm 109$ &   $-28614\pm 24$ &                  1.85 &                 24 &   $622\pm 35$ \\
J17578+046  &  $-111156\pm 110$ &  $-110567\pm 20$ &                  0.15 &                 20 &   $589\pm 41$ \\
J17578+465  &   $-32260\pm 108$ &   $-31626\pm 25$ &                  1.35 &                 25 &   $633\pm 33$ \\
J18022+642  &    $-2219\pm 119$ &    $-1627\pm 50$ &                 17.49 &                 47 &   $592\pm 41$ \\
J18027+375  &     $3091\pm 111$ &     $3668\pm 20$ &                  0.78 &                 20 &   $577\pm 43$ \\
J18051-030  &    $31706\pm 107$ &    $32344\pm 23$ &                  0.42 &                 23 &   $637\pm 31$ \\
J18075-159  &   $-33797\pm 110$ &   $-33200\pm 24$ &                  2.35 &                 24 &   $596\pm 40$ \\
J18131+260  &    $-9239\pm 138$ &    $-8612\pm 88$ &                 85.23 &                 23 &   $628\pm 34$ \\
J18165+048  &   $-54112\pm 110$ &   $-53521\pm 20$ &                  0.85 &                 20 &   $591\pm 40$ \\
J18174+483  &   $-24885\pm 107$ &   $-24240\pm 23$ &                  1.51 &                 23 &   $645\pm 31$ \\
J18180+387E &     $-397\pm 108$ &      $225\pm 25$ &                  1.24 &                 25 &   $622\pm 33$ \\
J18189+661  &     $4779\pm 119$ &     $5355\pm 49$ &                 12.78 &                 47 &   $576\pm 43$ \\
J18221+063  &   $-44585\pm 108$ &   $-43965\pm 24$ &                  0.57 &                 24 &   $619\pm 34$ \\
J18224+620  &   $-14556\pm 112$ &   $-13970\pm 24$ &                  0.78 &                 24 &   $586\pm 45$ \\
J18319+406  &   $-19955\pm 108$ &   $-19325\pm 25$ &                  1.48 &                 25 &   $630\pm 33$ \\
J18346+401  &    $11545\pm 108$ &    $12180\pm 25$ &                  0.48 &                 25 &   $635\pm 33$ \\
J18353+457  &   $-32697\pm 106$ &   $-32053\pm 18$ &                  1.25 &                 18 &   $645\pm 29$ \\
J18356+329  &      $117\pm 215$ &     $643\pm 183$ &                157.74 &                 93 &   $526\pm 51$ \\
J18363+136  &   $-45944\pm 109$ &   $-45321\pm 24$ &                  0.95 &                 24 &   $622\pm 35$ \\
J18409-133  &   $-33939\pm 107$ &   $-33297\pm 23$ &                  1.23 &                 23 &   $642\pm 30$ \\
J18419+318  &   $-32654\pm 108$ &   $-32027\pm 25$ &                  2.03 &                 25 &   $628\pm 33$ \\
J18480-145  &    $-5412\pm 108$ &    $-4780\pm 25$ &                  1.09 &                 25 &   $631\pm 32$ \\
J18482+076  &   $-35043\pm 110$ &   $-34455\pm 20$ &                  1.53 &                 20 &   $588\pm 42$ \\
J18498-238  &   $-11390\pm 110$ &   $-10797\pm 25$ &                  2.00 &                 25 &   $594\pm 39$ \\
J18580+059  &     $9229\pm 106$ &     $9873\pm 18$ &                  0.96 &                 18 &   $644\pm 29$ \\
J19070+208  &    $31441\pm 108$ &    $32064\pm 23$ &                  0.65 &                 23 &   $623\pm 33$ \\
J19072+208  &    $31207\pm 108$ &    $31830\pm 23$ &                  0.86 &                 23 &   $623\pm 33$ \\
J19084+322  &    $-2606\pm 108$ &    $-1979\pm 25$ &                  0.97 &                 25 &   $627\pm 33$ \\
J19098+176  &   $-14857\pm 110$ &   $-14262\pm 24$ &                  0.75 &                 24 &   $596\pm 39$ \\
J19169+051N &    $34973\pm 108$ &    $35612\pm 25$ &                  0.27 &                 25 &   $640\pm 31$ \\
J19169+051S &    $35179\pm 115$ &    $35731\pm 28$ &                  3.96 &                 28 &   $551\pm 48$ \\
J19216+208  &     $3911\pm 110$ &     $4519\pm 24$ &                  0.86 &                 24 &   $608\pm 39$ \\
J19251+283  &   $-41403\pm 267$ &   $-40773\pm 25$ &                  1.18 &                 25 &  $630\pm 247$ \\
J19346+045  &   $-59501\pm 105$ &   $-58857\pm 18$ &                  0.83 &                 18 &   $644\pm 27$ \\
J19422-207  &    $-2702\pm 111$ &    $-2094\pm 32$ &                 21.55 &                 23 &   $608\pm 37$ \\
J19511+464  &   $-13555\pm 122$ &   $-12942\pm 60$ &                 15.92 &                 58 &   $613\pm 36$ \\
J20093-012  &   $-54561\pm 111$ &   $-53970\pm 24$ &                 13.26 &                 20 &   $591\pm 41$ \\
J20260+585  &   $-60631\pm 110$ &   $-60047\pm 20$ &                  0.26 &                 20 &   $585\pm 42$ \\
J20305+654  &     $9694\pm 108$ &    $10326\pm 25$ &                  0.38 &                 25 &   $632\pm 34$ \\
J20336+617  &   $-21816\pm 108$ &   $-21184\pm 24$ &                  0.55 &                 24 &   $632\pm 33$ \\
J20405+154  &   $-60376\pm 110$ &   $-59782\pm 24$ &                  0.75 &                 24 &   $595\pm 39$ \\
J20450+444  &   $-25661\pm 107$ &   $-25026\pm 23$ &                  0.82 &                 23 &   $635\pm 31$ \\
J20525-169  &    $15396\pm 109$ &    $16005\pm 24$ &                  0.88 &                 24 &   $608\pm 36$ \\
J20533+621  &   $-18136\pm 107$ &   $-17493\pm 23$ &                  0.22 &                 23 &   $642\pm 29$ \\
J20556-140S &  $-142621\pm 110$ &  $-142042\pm 20$ &                  1.52 &                 20 &   $579\pm 43$ \\
J20567-104  &    $34345\pm 108$ &    $34984\pm 25$ &                  2.19 &                 25 &   $639\pm 32$ \\
J21019-063  &   $-15867\pm 108$ &   $-15229\pm 25$ &                  1.55 &                 25 &   $638\pm 32$ \\
J21152+257  &   $-16926\pm 108$ &   $-16284\pm 25$ &                  1.40 &                 25 &   $642\pm 34$ \\
J21164+025  &   $-22946\pm 108$ &   $-22310\pm 25$ &                  0.78 &                 25 &   $636\pm 32$ \\
J21221+229  &     $4713\pm 107$ &     $5353\pm 23$ &                  1.15 &                 23 &   $641\pm 30$ \\
J21348+515  &   $-14760\pm 108$ &   $-14122\pm 25$ &                  0.37 &                 25 &   $637\pm 32$ \\
J21463+382  &   $-83609\pm 110$ &   $-83013\pm 24$ &                  0.57 &                 24 &   $596\pm 38$ \\
J21466+668  &   $-10392\pm 108$ &    $-9771\pm 24$ &                  0.46 &                 24 &   $621\pm 34$ \\
J21466-001  &   $-29165\pm 109$ &   $-28548\pm 24$ &                  1.25 &                 24 &   $618\pm 35$ \\
J22012+283  &    $-3877\pm 175$ &   $-3255\pm 139$ &                103.35 &                 93 &   $622\pm 35$ \\
J22020-194  &   $-24227\pm 134$ &   $-23602\pm 25$ &                  1.24 &                 25 &   $625\pm 85$ \\
J22021+014  &    $17180\pm 106$ &    $17823\pm 18$ &                  0.66 &                 18 &   $643\pm 32$ \\
J22057+656  &   $-47534\pm 107$ &   $-46895\pm 23$ &                  0.39 &                 23 &   $639\pm 32$ \\
J22096-046  &   $-16231\pm 108$ &   $-15592\pm 25$ &                  2.41 &                 25 &   $639\pm 32$ \\
J22114+409  &   $-17462\pm 110$ &   $-16878\pm 20$ &                  0.85 &                 20 &   $583\pm 42$ \\
J22115+184  &   $-52315\pm 107$ &   $-51671\pm 23$ &                  0.72 &                 23 &   $644\pm 32$ \\
J22125+085  &     $7828\pm 108$ &     $8459\pm 25$ &                  0.58 &                 25 &   $631\pm 33$ \\
J22137-176  &   $-25077\pm 110$ &   $-24485\pm 24$ &                  0.82 &                 24 &   $592\pm 40$ \\
J22231-176  &    $-2564\pm 110$ &    $-1972\pm 24$ &                  2.20 &                 24 &   $591\pm 40$ \\
J22252+594  &     $3232\pm 108$ &     $3866\pm 24$ &                  0.66 &                 24 &   $634\pm 33$ \\
J22298+414  &     $1500\pm 109$ &     $2114\pm 24$ &                  1.22 &                 24 &   $613\pm 35$ \\
J22330+093  &    $-7473\pm 107$ &    $-6835\pm 23$ &                  1.28 &                 23 &   $638\pm 32$ \\
J22468+443  &     $-280\pm 110$ &      $347\pm 30$ &                  3.58 &                 30 &   $627\pm 34$ \\
J22503-070  &    $-6778\pm 106$ &    $-6135\pm 18$ &                  0.57 &                 18 &   $643\pm 29$ \\
J22518+317  &    $-3083\pm 117$ &    $-2445\pm 50$ &                 18.20 &                 47 &   $639\pm 32$ \\
J22532-142  &    $-2427\pm 111$ &    $-1801\pm 33$ &                 22.95 &                 24 &   $625\pm 34$ \\
J22559+178  &   $-32880\pm 107$ &   $-32238\pm 23$ &                  0.93 &                 23 &   $643\pm 29$ \\
J22565+165  &   $-28306\pm 107$ &   $-27664\pm 23$ &                  0.24 &                 23 &   $642\pm 30$ \\
J23064-050  &   $-53852\pm 118$ &   $-53323\pm 28$ &                  3.70 &                 28 &   $529\pm 57$ \\
J23216+172  &    $-7400\pm 108$ &    $-6769\pm 24$ &                  0.48 &                 24 &   $632\pm 33$ \\
J23245+578  &   $-34167\pm 107$ &   $-33524\pm 23$ &                  0.93 &                 23 &   $642\pm 30$ \\
J23340+001  &    $-5477\pm 108$ &    $-4842\pm 25$ &                  0.40 &                 25 &   $635\pm 32$ \\
J23351-023  &   $-41614\pm 113$ &   $-41056\pm 20$ &                  0.68 &                 20 &   $558\pm 48$ \\
J23381-162  &    $19840\pm 108$ &    $20471\pm 23$ &                  0.54 &                 23 &   $631\pm 32$ \\
J23419+441  &   $-78648\pm 111$ &   $-78073\pm 20$ &                  0.44 &                 20 &   $575\pm 44$ \\
J23431+365  &    $-3586\pm 109$ &    $-2981\pm 24$ &                  1.04 &                 24 &   $606\pm 37$ \\
J23492+024  &   $-72102\pm 107$ &   $-71469\pm 23$ &                  0.21 &                 23 &   $633\pm 31$ \\
J23505-095  &   $-22492\pm 109$ &   $-21871\pm 24$ &                  0.67 &                 24 &   $621\pm 35$ \\
J23548+385  &     $4746\pm 109$ &     $5372\pm 27$ &                 11.38 &                 24 &   $626\pm 34$ \\
\label{tab:rvabs}
\end{longtable}
\tablefoot{
\tablefoottext{a}{The $\mathrm{RV_{abs}}$ uncertainty is the quadratic sum of the uncertainties from the RV scatter ($\sigma_{\rm CCF}$), the mask zero-point ($\sigma_{\rm mask}$), the gravitational redshift ($\sigma_{\rm grav}$), and the convective blueshift ($\sigma_{\rm conv}= 100\,\ms$). The uncertainty in $\mathrm{RV_{CCF}}$ is the quadratic sum of the scatter and mask zero-points uncertainties, which are also listed separately. The $\sigma_{\rm grav}$ uncertainty comes from error propagation of Eq. \ref{eq:grav}. See Sect. \ref{sec:rvabs} for details.}
}
}
\end{appendix}

\end{document}